\documentclass[aps,pre,reprint,superscriptaddress,longbibliography,floatfix]{revtex4-2}
\usepackage[T1]{fontenc}
\usepackage{lmodern}
\usepackage{amsmath,amssymb,amsthm,mathtools,graphicx,booktabs,array,bm,mathrsfs,dsfont}

\usepackage[colorlinks=true,allcolors=blue]{hyperref}
\usepackage{physics}

\newcommand{\inprod}[2]{(#1\,|\,#2)}

\newcommand{\vbar}{\bar{v}}
\newcommand{\E}{\mathbb{E}}
\newcommand{\Lg}{\mathcal{L}}
\newcommand{\Fc}{\mathcal{F}}
\newcommand{\Qc}{\mathcal{Q}}

\newcommand{\He}{\mathrm{He}}

\newcommand{\ba}{\vb*{a}}
\newcommand{\bb}{\vb*{b}}
\newcommand{\bj}{\vb*{j}}
\newcommand{\bu}{\vb*{u}}
\newcommand{\bF}{\vb*{f}}

\newcommand{\bg}{\vb*{g}}

\newcommand{\bx}{\vb*{x}}
\newcommand{\by}{\vb*{y}}
\newcommand{\bv}{\vb*{v}}
\newcommand{\bz}{\vb*{z}}
\newcommand{\bw}{\vb*{w}}
\newcommand{\bW}{\vb*{W}}
\newcommand{\bZ}{\vb*{Z}}

\newcommand{\btheta}{\vb*{\vartheta}}
\newcommand{\bnu}{\vb*{\nu}}
\newcommand{\bh}{\vb*{h}}
\newcommand{\br}{\vb*{r}}
\newcommand{\bR}{\vb*{R}}
\newcommand{\bq}{\vb*{q}}

\newcommand{\D}{\mathsf{D}}
\newcommand{\T}{\mathsf{T}}
\newcommand{\M}{\mathsf{M}}
\newcommand{\Cm}{\mathsf{C}}
\newcommand{\Rm}{\mathsf{R}}
\newcommand{\Rc}{\mathcal{R}}

\newcommand{\Ac}{\mathcal{A}}
\newcommand{\Nc}{\mathcal{N}}

\newcommand{\Imat}{\mathsf{I}}
\newcommand{\jr}{\bar{\jmath}}
\newcommand{\st}{\mathrm{ss}}
\newcommand{\cond}{\,|\,}
\newcommand{\irr}{\mathrm{irr}}
\newcommand{\eps}{\epsilon}

\newcommand{\Vd}{\mathcal{V}_{\mathrm{dyn}}}
\newcommand{\Bform}{\mathscr{B}}
\newcommand{\Icost}{\mathcal{I}}
\newcommand{\Rgamma}{R_\gamma}
\newcommand{\hLg}{\widehat{\Lg}}
\newcommand{\hGam}{\widehat\Gamma}
\newcommand{\Hdyn}{\mathscr{H}}
\newcommand{\Cclass}{\mathscr{C}}

\DeclareMathOperator{\Var}{Var}
\DeclareMathOperator{\Cov}{Cov}

\providecommand{\erf}{\operatorname{erf}}

\begin{document}
\title{Exact finite-time fluctuation-response relations for Markov processes: A martingale framework}
\author{Tan Van Vu}
\email{tan.vu@yukawa.kyoto-u.ac.jp}
\affiliation{Center for Gravitational Physics and Quantum Information, Yukawa Institute for Theoretical Physics, Kyoto University, Kitashirakawa Oiwakecho, Sakyo-ku, Kyoto 606-8502, Japan}

\author{Van Tuan Vo}
\affiliation{Quantum AI \& Cyber Security Research Institute, FPT Corporation, 10 Pham Van Bach, Cau Giay, Ha Noi 100000, Vietnam}

\author{Ruicheng Bao}
\affiliation{Department of Physics, Graduate School of Science, The University of Tokyo, Hongo, Bunkyo-ku, Tokyo 113-0033, Japan}

\author{Keiji Saito}
\affiliation{Department of Physics, Kyoto University, Kyoto 606-8502, Japan}

\begin{abstract}
Fluctuations and responses of time-integrated observables are both experimentally measurable, and relations between them can reveal the dissipation that sustains a nonequilibrium state. Here we develop a martingale framework that gives exact finite-time fluctuation-response identities for additive observables of general Markov processes. We obtain an exact identity for the variance of an observable in terms of its Doob martingale, which splits the total variance into a part inherited from the initial ensemble and a part generated by noise during the observation. We show that, at fixed preparation, the response of the mean of the observable to any admissible perturbation is the covariance between the Doob martingale and a second martingale that represents the perturbation. For diffusions and jump processes, a single martingale amplitude determines both the fluctuations and the response of the observable. These identities yield inequalities, variational principles, and multivariate and frequency-domain bounds, whose costs are set by the corresponding physical perturbations. Applying the framework to underdamped dynamics, we obtain a friction-response thermodynamic uncertainty relation (TUR) that can be saturated at any finite time and thus gives a variational principle for the entropy production. We also demonstrate exponentially strong violations of the conventional TUR by velocity-resolved currents, even in linear Gaussian dynamics, and derive a weak-noise dissipation-coherence bound for limit cycles, which shows that inertial oscillators can beat the overdamped bound only if their frequency is weakly sensitive to friction. For overdamped dynamics and jump processes, the framework recovers and extends recent fluctuation-response identities and identifies when the corresponding uncertainty relations are saturated.
\end{abstract}
\maketitle

\section{Introduction}\label{sec:intro}
Fluctuations and responses are two complementary ways to probe a small system. The fluctuations of time-integrated currents, such as the displacement of a molecular motor or the heat exchanged with a bath, quantify how much transport varies from trajectory to trajectory, whereas their response quantifies how the mean changes when a force, a kinetic coefficient, or a transition rate is perturbed. In equilibrium, the two are tied together by the fluctuation-dissipation theorem, which expresses the linear response of an observable through spontaneous equilibrium fluctuations, namely its correlation with the variable to which the perturbation couples~\cite{Kubo.1991,Marconi.2008.PR}. Out of equilibrium, this connection breaks down. Generalized fluctuation-dissipation relations still express a response as a correlation function, but the correlated quantity is no longer fixed by the perturbation alone: it depends on the nonequilibrium state and on how the perturbation changes the dynamics~\cite{Agarwal.1972,Marconi.2008.PR,Baiesi.2009.PRL,Prost.2009.PRL,Seifert.2010.EPL}. Relating the responses and fluctuations of the same observable is of practical importance, because they carry information about the cost of maintaining a nonequilibrium state. In Langevin systems, for instance, the extent to which the fluctuation-dissipation theorem is violated determines the rate of energy dissipation~\cite{Harada.2005.PRL}. Since measuring such costs generally requires access to all forces and degrees of freedom involved, they are often difficult to determine directly, and relations among fluctuations, responses, and costs offer a way to infer them from measurable statistics~\cite{Li.2019.NC,Manikandan.2020.PRL,Vu.2020.PRE,Otsubo.2020.PRE,Seifert.2019.ARCMP}. Prominent examples include thermodynamic uncertainty relations (TURs), which bound the precision of currents by the entropy production~\cite{Barato.2015.PRL,Gingrich.2016.PRL,Pietzonka.2017.PRE,Horowitz.2017.PRE,Dechant.2018.JSM,Hasegawa.2019.PRE,Hasegawa.2019.PRL,Horowitz.2020.NP,Liu.2020.PRL,Vo.2022.JPA,Dieball.2023.PRL,Smith.2023.PRL,Vo.2025.PRL,Stutzer.2026.PRX}; kinetic uncertainty relations (KURs), which bound the precision of counting observables by the dynamical activity, i.e., the mean number of transitions~\cite{Garrahan.2017.PRE,Terlizzi.2019.JPA,Prech.2025.PRX,Macieszczak.2024.arxiv}; and fluctuation-response inequalities, which bound the response of an observable by its fluctuations and the cost of the perturbation~\cite{Dechant.2020.PNAS,Owen.2020.PRX,Ptaszynski.2024.PRL,Bao.2024.arxiv,Zheng.2025.PRE,Liu.2025.CP,Kwon.2025.PRL}. Beyond thermodynamic inference, these relations reveal universal trade-offs among precision, response, and cost that hold across broad classes of dynamics, regardless of microscopic details~\cite{Pietzonka.2018.PRL,Hartich.2021.PRL}.

However, these uncertainty relations and response bounds are inequalities: they leave open how close a bound is to equality and which observable and perturbation are most informative, and they do not provide an exact relation between the fluctuations of an observable and its responses, as the fluctuation-dissipation theorem does in equilibrium. Such exact relations have recently been obtained for Markov jump processes. In steady states and in the long-time limit, current covariances can be written as quadratic forms of the responses to perturbations of individual transition rates~\cite{Aslyamov.2025.PRL}, and analogous relations hold for state observables and state-current correlations~\cite{Ptaszynski.2026.PRE,Ptaszynski.2026.PRE2}. For a given perturbation, these identities bound the response by the fluctuations and the cost of the perturbation; kinetic and entropic perturbations, for instance, have different costs and therefore yield different bounds~\cite{Aslyamov.2025.PRL,Kwon.2025.PRL}. A dynamical fluctuation-response relation extends these identities to time-dependent driving and finite observation times~\cite{Aslyamov.2026.arxiv}. In this setting, the variance consists of two parts: one inherited from the initial distribution, and one generated during the observation, of which only the latter is expressed through the responses. Bounds based on the total variance therefore include fluctuations that are unrelated to the responses. Consequently, they become loose whenever the initial state accounts for a substantial part of the variance.

For diffusive dynamics, the picture is less complete. In overdamped Langevin dynamics, correlations of the empirical density and current yield a fluctuation-response identity in the long-time limit, whereas at finite times Cram\'er-Rao bounds give only inequalities for force, mobility, and temperature perturbations~\cite{Chun.2026.arxiv}. In steady states, frequency-domain relations connect susceptibilities and power spectra to thermodynamic and kinetic costs~\cite{Dechant.2026.PRL,Zheng.2026a.arxiv,Kwon.2026.arxiv,Gu.2026.arxiv}. Taken together, these results point to a common structure across jump and diffusive dynamics. Their derivations, however, rely on different tools, such as cumulant generating functions, information inequalities, and correlations of empirical fields, each suited to a particular class of dynamics. A unified formulation should provide exact finite-time identities for general Markovian dynamics and arbitrary initial distributions. The known inequalities and their equality conditions would then follow from these identities, with the contribution of the initial state separated out. Such a formulation would also clarify which perturbation yields which bound and how finite-time results reduce to stationary ones.

\begin{figure*}[t!]
 \centering\includegraphics[width=1.0\textwidth]{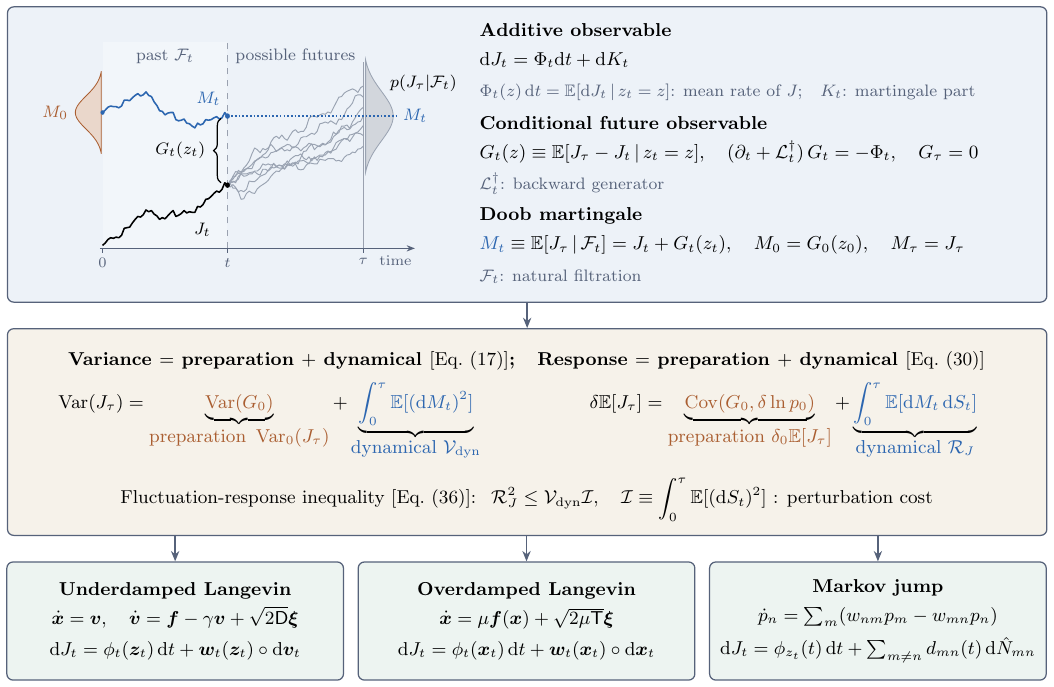}
 \caption{Structure of the martingale framework. Top: for an additive observable $J_t$, the expected future accumulation $G_t(z)$ defines the Doob martingale $M_t=J_t+G_t(z_t)=\E[J_\tau\cond\Fc_t]$, which evolves from $M_0=G_0(z_0)$ to $M_\tau=J_\tau$. Middle: the variance of $J_\tau$ splits into the preparation variance inherited from the initial ensemble (orange) and the dynamical variance generated by the martingale increments (blue). Likewise, the response of the mean splits into the initial response (orange) and the overlap of $M_t$ with the martingale $S_t$ representing the perturbation (blue). Applying the Cauchy-Schwarz inequality to the dynamical parts gives the fluctuation-response inequality. These identities and bounds hold for any Markov process, provided that the variances are finite and the perturbation is admissible. Bottom: specifying the dynamics makes them explicit for underdamped and overdamped Langevin dynamics and for Markov jump processes.}
 \label{fig:framework}
\end{figure*}

Underdamped dynamics poses distinct challenges for such a formulation. When inertia is non-negligible, as in levitated nanoparticles or other weakly damped systems, thermal noise acts only on the velocities, which are odd under time reversal, while the positions change only through the velocities. Changes of the force, friction, or temperature therefore inject probability current only in velocity space, so that the identities available for jump and overdamped dynamics do not extend directly. The same features also allow underdamped currents to violate the conventional TUR, which holds for steady-state currents in overdamped and jump dynamics~\cite{Fischer.2018.PRE,Vu.2019.PRE.UnderdampedTUR,Pietzonka.2022.PRL}. Several modified TURs restore a valid bound by accounting for inertia through auxiliary perturbations, additional dynamical costs, or response factors~\cite{Vu.2019.PRE.UnderdampedTUR,Lee.2021.PRE.TUR,Kwon.2022.NJP,Fu.2022.PRE,Dechant.2022.arxiv,Lyu.2024.PRE}. For a restricted class of currents, it has also been conjectured that the precision is bounded by that of free diffusion~\cite{Fischer.2020.PRE}. Despite this progress, these results remain inequalities for particular classes of observables or perturbations, and whether they can be saturated is largely unknown. In particular, an exact finite-time relation that expresses the variance of an observable through local responses is still missing. Such a relation would clarify which constraints survive in underdamped dynamics and why the conventional TUR can fail.

In this work, we develop a martingale framework that provides such a relation for general Markov processes (Fig.~\ref{fig:framework}). In stochastic thermodynamics, martingales have mainly been built from the entropy production, revealing generic properties of its fluctuations and universal statistics of its infima and stopping times~\cite{Neri.2017.PRX,Pigolotti.2017.PRL,Roldan.2023.AP}. Here, the central object is instead the Doob martingale~\cite{Doob.1953} of an additive observable: the accumulated value plus the expected future accumulation given the present state. Because its increments are uncorrelated with its initial value, the variance splits into a preparation variance, inherited from the initial ensemble, and a dynamical variance, generated by the noise during the observation and equal to the mean of the predictable quadratic variation. The same martingale governs the response: for any admissible perturbation at fixed preparation, the linear response of the mean is the covariance of this martingale with a second one representing the perturbation, whose squared norm is the cost of the perturbation. The dynamical variance and the response are thus a squared norm and an inner product in the same Hilbert space, and the Cauchy-Schwarz inequality, with its variational and matrix forms, turns each identity into a bound on the response in terms of the dynamical variance and the cost. For diffusions and jump processes, both quantities are expressed through a single martingale amplitude, the local response of the mean to a probability current injected at a given time and state. Standard martingale tools~\cite{Roldan.2023.AP,Cattiaux.2012} then give the long-time and frequency-domain results directly.

The framework unifies known fluctuation-response relations and yields new results for jump, overdamped, and underdamped dynamics. For jump processes, it reproduces the dynamical fluctuation-response relation of Ref.~\cite{Aslyamov.2026.arxiv} and, in the stationary limit, those of Refs.~\cite{Kwon.2025.PRL,Ptaszynski.2026.PRE,Ptaszynski.2026.PRE2}. For overdamped dynamics, it extends the long-time fluctuation-response identity of Ref.~\cite{Chun.2026.arxiv} to finite times and arbitrary initial distributions. Subtracting the contribution of the initial ensemble from the variance tightens the TUR for time-dependent driving~\cite{Koyuk.2020.PRL} in overdamped dynamics; in jump processes, it tightens the KUR, whereas the corresponding improved TUR was obtained in Ref.~\cite{Aslyamov.2026.arxiv}. We derive the equality conditions of these improved relations and show that, under strictly monotonic driving, they are saturated at any finite time by observables with suitably time-dependent weights; for jump currents, the saturated bound is the tighter one with the pseudo entropy production. In the frequency domain, the finite-frequency response identities and inequalities of Refs.~\cite{Dechant.2026.PRL,Kwon.2026.arxiv,Zheng.2026a.arxiv} follow as long-time limits of their finite-time counterparts.

The main new results concern underdamped dynamics, where such a relation has been missing. We obtain an exact finite-time fluctuation-response relation, valid for arbitrary initial distributions and any admissible perturbation. Applied to a change of friction, it gives a friction-response TUR that excludes the contribution of the initial ensemble and is saturated at every finite time by an explicitly constructed observable. This TUR thus becomes a variational principle: maximizing, over observables, the ratio of the squared friction response to the dynamical variance determines the entropy production exactly, so dissipation can be inferred from measured fluctuations and friction responses. The same TUR explains the arbitrarily strong violations of the conventional TUR by velocity-resolved currents in linear Gaussian dynamics. It also yields a weak-noise bound on the dissipation per coherent oscillation of a limit cycle in terms of the friction sensitivity of its frequency, which shows that inertial oscillators can fall below the overdamped dissipation-coherence bound only if their frequency is weakly sensitive to friction. The companion Letter~\cite{Vu.2026.arxiv.FR1} reports the underdamped results, whereas this paper develops the general framework, gives the full derivations and equality conditions, and establishes the connections to existing results.

The remainder of the paper is organized as follows. Section~\ref{sec:general} develops the general framework: the variance and response identities, the resulting bounds and their equality conditions, and their long-time and frequency-domain forms. Section~\ref{sec:under} applies the framework to underdamped dynamics, covering the friction-response TUR and its saturation, short-time behavior, multivariate bounds, the dissipation-coherence trade-off for limit cycles, and violations of the conventional TUR by velocity-resolved currents. Sections~\ref{sec:over} and \ref{sec:jump} treat overdamped dynamics and jump processes, respectively, identifying the known results recovered in each case and deriving the equality conditions of the improved TUR and KUR. Section~\ref{sec:discussion} summarizes the results and discusses open directions. Technical derivations are given in the appendices.

\section{General framework}\label{sec:general}
This section develops the martingale framework for fluctuations and responses of general Markov processes. The variance identity, the response identity, and the inequalities that follow from them require only the generator of the process and its martingale structure. We then specialize the results to diffusive and jump dynamics, where all quantities become explicit.

\subsection{Markov processes and observables}
We consider a Markov process whose state $z_t$ (or $\bz_t$ for multidimensional cases) at time $t$ contains all the information needed to predict future increments. For example, $\bz_t$ is a phase-space point $(\bx_t,\bv_t)$ for underdamped dynamics and a position $\bx_t$ for overdamped dynamics, whereas $z_t$ is a discrete state for Markov jump processes. Let $p_t(z)$ be the probability density (or, for discrete states, the probability) of finding the system in state $z$ at time $t$. It evolves according to the master equation
\begin{equation}
	\partial_t p_t=\Lg_t p_t,
\end{equation}
where $\Lg_t$ denotes the forward generator. Its adjoint, the backward generator $\Lg_t^\dagger$, acts on observables and satisfies $\inprod{\Lg_t^\dagger a}{p}=\inprod{a}{\Lg_t p}$, where the inner product $\inprod{\cdot}{\cdot}$ is an integral over a continuous state space and a sum over a discrete one. Throughout this work, we write $\ev{a}_t\equiv\inprod{a}{p_t}$ for an instantaneous average and $\E[\cdot]$ for an average over trajectories, and $\Var(X)\equiv\E[(X-\E[X])^2]$ and $\Cov(X,Y)\equiv\E[(X-\E[X])(Y-\E[Y])]$ for the variance and covariance of random variables $X$ and $Y$, respectively. Conditional versions, such as $\Var(X\cond z_0)$, are defined with $\E[\cdot\cond z_0]$ in place of $\E[\cdot]$. We also assume the usual regularity: the observables have finite variance, all functions involved are smooth enough for It\^o's formula, the martingale terms that appear are square integrable, derivatives with respect to perturbation parameters can be taken inside averages, and boundary terms in integrations by parts vanish. 

We focus on an additive observable $J_t$ that can accumulate both while the system resides in a state and when the state changes. In general, its stochastic increment can be written as
\begin{equation}
	\dd{J_t}=\Phi_t(z_{t^-})\dd{t}+\dd{K_t},\quad J_0=0,
\label{eq:general.additive}
\end{equation}
where $\Phi_t$ is the effective rate of the observable and $K_t$ is a martingale. The explicit forms of $\Phi_t$ and $K_t$ are determined once the dynamics and the observable are specified. A residence observable, such as an occupation time, contributes only through the term $\Phi_t\dd{t}$, whereas a current, such as a displacement or a heat flow, may contribute through both $\Phi_t$ and $K_t$. Throughout, $J_t$ is a functional of the state trajectory $z_{[0,t]}$, without measurement noise of its own, so that $K_t$ is driven by the same noise as $z_t$. The class of observables in Eq.~\eqref{eq:general.additive} covers all diffusion and jump observables considered in this work, as well as observables of processes with both diffusive and jump components.

To describe the noise in the observable and the noise in the state on the same footing, we treat the state and the accumulated observable as a single joint process. We consider Markov additive observables: given the history up to time $t$, the law of the future increments $(z_s,J_s-J_t)_{s\ge t}$ depends only on $t$ and $z_t$. All observables considered in this work have this property. Consequently, the joint variable $Z_t=(z_t,J_t)$ is again a Markov process. Let $y$ denote the value of the observable and $\hLg_t^\dagger$ the backward generator of $Z_t$, which acts on functions $f(z,y)$. Three properties follow from the additive structure. First, $\hLg_t^\dagger$ commutes with translations in $y$, because the future increments of $J_t$ do not depend on the value already accumulated. Second, it reduces to $\Lg_t^\dagger$ on functions of $z$ alone, because the observable does not act back on the state. Third, applied to the coordinate function $y$, it returns the effective rate, $\hLg_t^\dagger y=\Phi_t$, because, by Eq.~\eqref{eq:general.additive}, the increment $J_{t+\Delta t}-J_t$ has the conditional mean $\Phi_t(z)\Delta t+o(\Delta t)$ given $z_t=z$ and $J_t=y$. The joint fluctuations of the state and the observable are then characterized by the local covariance rate
\begin{equation}
	\hGam_t[f,g]\equiv\hLg_t^\dagger(fg)-f\,\hLg_t^\dagger g-g\,\hLg_t^\dagger f,
\label{eq:general.covrate}
\end{equation}
which is symmetric and bilinear in $f$ and $g$. Its meaning follows from a one-step expansion. For a short time step $\Delta t$, the backward generator gives $\E[f(Z_{t+\Delta t})\cond Z_t=Z]=f(Z)+(\hLg_t^\dagger f)(Z)\Delta t+o(\Delta t)$ for any function $f$ of the joint state in its domain. Applying this expansion to $f$, $g$, and $fg$, we obtain
\begin{align}
	&\Cov\qty[f(Z_{t+\Delta t}),g(Z_{t+\Delta t})\cond Z_t=Z]\notag\\
	&\quad=[\hLg_t^\dagger(fg)-f\,\hLg_t^\dagger g-g\,\hLg_t^\dagger f](Z)\Delta t+o(\Delta t)\notag\\
	&\quad=\hGam_t[f,g](Z)\Delta t+o(\Delta t),
\label{eq:general.onestep}
\end{align}
because the zeroth-order terms $f(Z)g(Z)$ cancel and the product of the two first-order corrections is of order $\Delta t^2$. Thus, $\hGam_t[f,g]\dd{t}$ is the conditional covariance of the increments of $f$ and $g$ over the next time step, and $\hGam_t[f,f]\ge0$ because it is a variance. In probability theory, $\hGam_t/2$ is known as the carr\'e du champ operator~\cite{Bakry.2014}; our convention omits the factor $1/2$.

By the same expansion, the increment of a function $f_t(Z_t)$ over a short time step splits into its conditional mean, $(\partial_t+\hLg_t^\dagger)f_t\dd{t}$, and a noise term with zero conditional mean. Accumulating only the noise term along the trajectory defines, for a time-dependent function $f_t$ in the domain of the extended generator, the process
\begin{equation}
	N^f_t\equiv f_t(Z_t)-f_0(Z_0)-\int_0^t(\partial_s+\hLg_s^\dagger)f_s(Z_s)\dd{s},
\label{eq:general.Nf}
\end{equation}
which is a martingale because the integral removes the conditional mean rate of change of $f_s(Z_s)$ at every time. The increments of $N^f$ and $N^g$ inherit the covariance~\eqref{eq:general.onestep},
\begin{equation}
	\E[\dd{N^f_t}\dd{N^g_t}\cond\Fc_{t^-}]=\hGam_t[f_t,g_t](Z_{t^-})\dd{t},
\label{eq:general.Ncov}
\end{equation}
as shown in Appendix~\ref{app:covariance}, where the explicit time dependence of $f_t$ and $g_t$ is also treated~\cite{Ethier.1986}. Here, $\Fc_t$ denotes the natural filtration generated by the trajectory $z_{[0,t]}$, i.e., the information available to an observer who has recorded the full state of the system from time $0$ up to time $t$. Note that $\dd{N^f_t}$ denotes the increment at time $t$, including a possible jump at $t$. Because the processes considered here have no jumps at deterministic times, $\Fc_{t^-}$ and $Z_{t^-}$ may be replaced by $\Fc_t$ and $Z_t$ at any fixed time without changing the result.

The observable itself provides the simplest example. For the coordinate function $f=y$, Eq.~\eqref{eq:general.Nf} with $\hLg_t^\dagger y=\Phi_t$ and $J_0=0$ gives $N^y_t=J_t-\int_0^t\Phi_s(z_s)\dd{s}$, which is exactly $K_t$ by Eq.~\eqref{eq:general.additive}. Equation~\eqref{eq:general.Ncov} then gives
\begin{equation}
	\E[(\dd{K_t})^2\cond\Fc_{t^-}]=\hGam_t[y,y](Z_{t^-})\dd{t}.
\label{eq:Knoise.int}
\end{equation}
This rate depends on $z$ only. To see this, apply Eq.~\eqref{eq:general.onestep} with $f=g=y$ at the joint state $(z,y_0)$; since subtracting the known value $J_t=y_0$ does not change a conditional variance,
\begin{align}
	&\hGam_t[y,y](z,y_0)\Delta t\notag\\
	&\quad=\Var[J_{t+\Delta t}-J_t\cond z_t=z,\,J_t=y_0]+o(\Delta t).
\label{eq:general.noiseK}
\end{align}
By Eq.~\eqref{eq:general.additive}, the increment $J_{t+\Delta t}-J_t$ is determined by the trajectory of $z$ on $[t,t+\Delta t]$, and the statistics of this trajectory given $z_t=z$ do not depend on $y_0$, because the observable does not act back on the state. The right-hand side of Eq.~\eqref{eq:general.noiseK}, and hence $\hGam_t[y,y]$, is therefore independent of $y_0$. Thus, $\hGam_t[y,y]$ is a function of $z$ alone and quantifies the intensity of the noise in the observable increment.

\subsection{Doob martingale and backward equation}\label{sec:general.doob}
The central object is the conditional future observable, i.e., the expected future accumulation of $J_t$. For a finite observation time $\tau$, it is defined, given the state at time $t$, as
\begin{equation}
	G_t(z)\equiv\E[J_\tau-J_t\cond z_t=z].
\label{eq:general.G}
\end{equation}
By the Markov property, the expected remaining contribution depends only on $z_t$, even though $J_t$ depends on the entire past trajectory. We then define the Doob martingale \cite{Doob.1953} of the observable as
\begin{equation}
	M_t\equiv J_t+G_t(z_t)=\E[J_\tau\cond\Fc_t].
\label{eq:general.M}
\end{equation}
The tower property gives $\E[M_t\cond\Fc_s]=M_s$ for $0\le s\le t$, so $M_t$ is a martingale, with endpoints $M_0=G_0(z_0)=\E[J_\tau\cond z_0]$ and $M_\tau=J_\tau$.

Splitting $J_\tau-J_t$ at $t+\Delta t$ and using the tower and Markov properties gives, as $\Delta t\to0^+$, $G_t(z)=\Phi_t(z)\Delta t+\E[G_{t+\Delta t}(z_{t+\Delta t})\cond z_t=z]+o(\Delta t)$. Expanding the second term as $G_t(z)+(\partial_t+\Lg_t^\dagger)G_t(z)\Delta t+o(\Delta t)$ yields the backward equation
\begin{equation}
	\partial_tG_t+\Lg_t^\dagger G_t=-\Phi_t
\label{eq:general.backward}
\end{equation}
with the terminal condition $G_\tau=0$. Equivalently, $G_t=\int_t^\tau\mathcal{U}^\dagger(s,t)\Phi_s\dd{s}$, where $\mathcal{U}(s,t)\equiv\mathcal{T}_{\leftarrow}\exp(\int_t^s\Lg_u\dd{u})$ is the forward propagator, $p_s=\mathcal{U}(s,t)p_t$, and its adjoint $\mathcal{U}^\dagger(s,t)\equiv\mathcal{T}_{\rightarrow}\exp(\int_t^s\Lg_u^\dagger\dd{u})$ propagates observables backward from $s$ to $t$. Here, $\mathcal{T}_{\leftarrow}$ ($\mathcal{T}_{\rightarrow}$) denotes the time-ordering operator that places later (earlier) times to the left. The detailed derivation of Eq.~\eqref{eq:general.backward} is given in Appendix~\ref{app:backward}.

The backward equation removes the drift of the martingale. With $H_t(z,y)\equiv y+G_t(z)$, we have $M_t=H_t(Z_t)$ and $(\partial_t+\hLg_t^\dagger)H_t=\Phi_t+\partial_tG_t+\Lg_t^\dagger G_t=0$, so that Eq.~\eqref{eq:general.Nf} gives $M_t=M_0+N^H_t$. Since $N^f$ is linear in $f$, separating the two parts of $H_t$ gives
\begin{equation}
	\dd{M_t}=\dd{K_t}+\dd{N^G_t},
\label{eq:general.dM}
\end{equation}
where $N^G$ is the martingale part of $G_t(z_t)$. The drift $\Phi_t\dd{t}$ of the observable increment is exactly compensated by the drift of the expected future accumulation. What remains is the noise in the observable increment and the noise-induced change of the expected future.

\subsection{Exact variance identity}
The martingale representation separates the variance of the observable into two parts: one inherited from the preparation, i.e., the initial distribution, and one generated by the noise during the observation. Recall that the endpoints of the martingale are $M_0=\E[J_\tau\cond z_0]$ and $M_\tau=J_\tau$, and the tower property gives $\E[M_0]=\E[J_\tau]$. The centered observable therefore splits into a preparation part and a martingale increment,
\begin{equation}
	J_\tau-\E[J_\tau]=(M_0-\E[M_0])+(M_\tau-M_0).
\label{eq:general.split}
\end{equation}
The first part is a function of $z_0$ alone, and the martingale property gives $\E[M_\tau-M_0\cond\Fc_0]=0$, so that
\begin{align}
	&\E[(M_0-\E[M_0])(M_\tau-M_0)]\notag\\
	&\quad=\E[(M_0-\E[M_0])\,\E[M_\tau-M_0\cond\Fc_0]]=0.
\label{eq:general.cross}
\end{align}
Thus, these two parts are uncorrelated. The same argument shows that $M_\tau-M_0$ is orthogonal to every square-integrable function of $z_0$. Squaring Eq.~\eqref{eq:general.split} and averaging therefore gives
\begin{equation}
	\Var(J_\tau)=\Var_0(J_\tau)+\E[(M_\tau-M_0)^2],
\label{eq:general.varstep}
\end{equation}
where $\Var_0(J_\tau)\equiv\Var(M_0)=\Var(\E[J_\tau\cond z_0])$ is the part of the variance that originates from the randomness of the initial state. We refer to it as the \emph{preparation} variance. It remains to express the second moment of the martingale increment through the dynamics. Divide $[0,\tau]$ into steps $0=t_0<t_1<\dots<t_n=\tau$ and write $M_\tau-M_0=\sum_{k=0}^{n-1}\Delta M_k$ with $\Delta M_k\equiv M_{t_{k+1}}-M_{t_k}$. Increments over different steps are uncorrelated: for $j<k$, $\Delta M_j$ is known at time $t_k$, and the martingale property gives $\E[\Delta M_k\cond\Fc_{t_k}]=0$, so that $\E[\Delta M_j\,\Delta M_k]=\E[\Delta M_j\,\E[\Delta M_k\cond\Fc_{t_k}]]=0$. Only the diagonal terms survive in the square of the sum, and refining the steps gives the martingale isometry
\begin{align}
	\E\big[(M_\tau-M_0)^2\big]&=\sum_k\E\big[(\Delta M_k)^2\big]\notag\\
	&=\int_0^\tau\E\big[(\dd{M_t})^2\big].
\label{eq:general.isometry}
\end{align}
The sum is the same for every choice of steps, so the integral, defined as its limit, is exact. Here, $\E[(\dd{M_t})^2]$ is the average of the conditional variance $\E[(\dd{M_t})^2\cond\Fc_{t^-}]$ of the next increment. This conditional variance depends on the state visited at time $t$, so its time integral along a trajectory is a random quantity, known as the predictable quadratic variation of $M$~\cite{Roldan.2023.AP}; the integral in Eq.~\eqref{eq:general.isometry} is its average over trajectories. Inserting Eq.~\eqref{eq:general.isometry} into Eq.~\eqref{eq:general.varstep} yields the exact variance identity
\begin{equation}
	\Var(J_\tau)=\Var_0(J_\tau)+\int_0^\tau\E\big[(\dd{M_t})^2\big].
\label{eq:general.var}
\end{equation}
Equation~\eqref{eq:general.var} expresses the law of total variance, with an explicit dynamical expression for its second term. Indeed, $M_\tau-M_0=J_\tau-\E[J_\tau\cond z_0]$, so the conditional second moment of the increment is $\Var(J_\tau\cond z_0)$. We call this term the \emph{dynamical} variance,
\begin{equation}
	\Vd\equiv\E[\Var(J_\tau\cond z_0)]=\int_0^\tau\E[(\dd{M_t})^2].
\label{eq:general.Vd}
\end{equation}
The first term in Eq.~\eqref{eq:general.var} measures how much the expected outcome varies across initial states, whereas the second accumulates the noise generated during the observation. A broad initial ensemble can thus produce a large total variance even when the noise generated during the observation, measured by $\Vd$, is small. Since the response bounds derived below involve only $\Vd$, using the total variance in their place can weaken them significantly. The identity holds at any finite time, for any initial ensemble, and for any Markov process for which $J_\tau$ has a finite variance.

The dynamical variance is a time integral of a local quantity. Since $M_t=M_0+N^H_t$, Eq.~\eqref{eq:general.Ncov} gives the conditional variance of its increment, $\E[(\dd{M_t})^2\cond\Fc_{t^-}]=\hGam_t[H_t,H_t](z_{t^-})\dd{t}$; it does not depend on $y$, because $\hLg_t^\dagger$ commutes with translations in $y$. Bilinearity splits it into three contributions,
\begin{equation}
	\hGam_t[H_t,H_t]
	=\underbrace{\hGam_t[y,y]}_{\text{direct}}
	+\underbrace{2\hGam_t[y,G_t]}_{\text{cross}}
	+\underbrace{\Gamma_t[G_t,G_t]}_{\text{propagated}},
\label{eq:general.threeterm}
\end{equation}
where $\Gamma_t$ is the local covariance rate of the state process, defined by Eq.~\eqref{eq:general.covrate} with $\Lg_t^\dagger$ in place of $\hLg_t^\dagger$. Hence, we obtain
\begin{equation}
	\Vd=\int_0^\tau\ev{\hGam_t[H_t,H_t]}_t\dd{t}.
\label{eq:general.Vdlocal}
\end{equation}
The direct term is the noise intensity of the measured increment, the propagated term is that of the expected future accumulation, and the cross term arises because both are driven by the same noise. The local variance rate $\Gamma_t[G_t,G_t]$ of the state process alone therefore misses two of the three contributions whenever the observable has a martingale part. For a residence observable, $K_t=0$, and only the propagated term remains.

The dynamical variance extends to a bilinear form of two observables. Let $J$ and $J'$ be arbitrary additive observables. For two observables, the local covariance rate $\hGam_t$ refers to the joint process $(z_t,J_t,J'_t)$, with $H_t=y+G_t(z)$ and $H'_t=y'+G'_t(z)$, where $y'$ denotes the value of $J'$. Polarizing Eqs.~\eqref{eq:general.Vd} and \eqref{eq:general.Vdlocal} then defines the stochastic metric
\begin{align}
	\Bform(J,J')&\equiv\E[\Cov(J_\tau,J'_\tau\cond z_0)]\notag\\
	&=\E[(M_\tau-M_0)(M'_\tau-M'_0)]\notag\\
	&=\int_0^\tau\ev{\hGam_t[H_t,H'_t]}_t\dd{t},
\label{eq:general.metricJ}
\end{align}
with $\Vd=\Bform(J,J)$. 

\subsection{Response identity}\label{sec:general.response}
The response of the mean observable follows from the backward equation, without differentiating path probabilities. Since $G_0(z)=\E[J_\tau\cond z_0=z]$, the mean observable is the average of the conditional future over the initial density,
\begin{equation}
	\E[J_\tau]=\inprod{G_0}{p_0}.
\label{eq:general.meanG}
\end{equation}
Consider an arbitrary weak perturbation of the initial density and of the generator, $p_0^\eps=p_0+\eps\delta p_0$ and $\Lg_t^\eps=\Lg_t+\eps\delta\Lg_t$, with fixed observable fields. Hereafter, $\delta X\equiv\partial_\eps X^\eps|_{\eps=0}$ denotes the first-order coefficient of a quantity $X$ along this perturbation, so that $X^\eps=X+\eps\delta X+o(\eps)$. The perturbation direction $(\delta p_0,\delta\Lg_t)$ is arbitrary. Both factors in Eq.~\eqref{eq:general.meanG} change: $p_0$ directly, and $G_0$ indirectly because it is determined by the generator through the backward equation~\eqref{eq:general.backward}. Varying that equation, including the change $\delta\Phi_t$ of the effective rate induced by $\delta\Lg_t$, gives
\begin{equation}
	\partial_t\delta G_t+\Lg_t^\dagger\delta G_t=-\big(\delta\Phi_t+\delta\Lg_t^\dagger G_t\big),
\label{eq:general.deltaG}
\end{equation}
where $\delta G_\tau=0$. Equation~\eqref{eq:general.deltaG} has the form of the backward equation~\eqref{eq:general.backward}, with the unperturbed generator and with $\Phi_t$ replaced by the source $s_t\equiv\delta\Phi_t+\delta\Lg_t^\dagger G_t$, a known function once the unperturbed $G_t$ is computed. Its solution is therefore the conditional future of the residence observable $\int_0^ts_u(z_u)\dd{u}$ under the unperturbed dynamics [cf.~Eq.~\eqref{eq:general.G}],
\begin{equation}
	\delta G_0(z)=\int_0^\tau\E[s_t(z_t)\cond z_0=z]\dd{t}.
\end{equation}
Averaging over $p_0$ with the tower property gives $\inprod{\delta G_0}{p_0}=\int_0^\tau\ev*{s_t}_t\dd{t}$. Adding the change of the initial density, we obtain
\begin{align}
	\delta\E[J_\tau]
	&=\inprod{G_0}{\delta p_0}+\int_0^\tau\ev{\delta\Phi_t+\delta\Lg_t^\dagger G_t}_t\dd{t}\notag\\
	&=\delta_0\E[J_\tau]+\int_0^\tau\ev{\delta\hLg_t^\dagger H_t}_t\dd{t},
\label{eq:general.response}
\end{align}
where $\delta_0\E[J_\tau]\equiv\inprod{G_0}{\delta p_0}$ is the response of the preparation; it equals $\Cov[G_0(z_0),\delta\ln p_0(z_0)]$ when $\delta p_0$ vanishes wherever $p_0$ does. In the second line, we used the fact that the joint generator acts on $H_t=y+G_t(z)$ term by term, $\hLg_t^\dagger y=\Phi_t$ and $\hLg_t^\dagger G_t=\Lg_t^\dagger G_t$, so that $\hLg_t^\dagger H_t=\Phi_t+\Lg_t^\dagger G_t$ for every generator of the perturbed family; varying the generator at fixed $H_t$ gives $\delta\hLg_t^\dagger H_t=\delta\Phi_t+\delta\Lg_t^\dagger G_t$. The integrand combines the direct change of the effective rate with the change of the expected future accumulation, and it is the variation of the joint generator applied to the same function $H_t$ whose local variance rate gives the variance. Equation~\eqref{eq:general.response} holds for any perturbation for which the first-order correction $\delta G_t$ exists and solves the linearized backward equation~\eqref{eq:general.deltaG}.

To turn Eq.~\eqref{eq:general.response} into a fluctuation-response relation, its dynamical part must be expressed in the metric~\eqref{eq:general.metricJ}. Since the perturbation itself is not an observable, we first extend the metric. Let $\Hdyn$ be the space of square-integrable functionals $\xi$ of the trajectory $z_{[0,\tau]}$ with $\E[\xi\cond z_0]=0$; for an observable $J$, $\xi_J\equiv M_\tau-M_0\in\Hdyn$. Each $\xi\in\Hdyn$ defines a martingale $X_t=\E[\xi\cond\Fc_t]$ with $X_0=0$, and the polarized isometry~\eqref{eq:general.isometry} gives
\begin{equation}
	\Bform(X,Y)\equiv\E[X_\tau Y_\tau]=\int_0^\tau\E[\dd{X_t}\dd{Y_t}].
\label{eq:general.metric}
\end{equation}
It induces the norm $\|\xi\|\equiv\E[\xi^2]^{1/2}$ on $\Hdyn$, so that $\|X_\tau\|^2=\Bform(X,X)$ and, for an observable, $\|\xi_J\|^2=\Vd$; closures and continuity below refer to this norm. Throughout this section, we identify an observable $J$ with its dynamical part $\xi_J=M_\tau-M_0$ and a martingale $X$ with $X_t-X_0$, and use the same symbol for the metric on all of them; for example, $\Bform(J,J')=\Bform(\xi_J,\xi_{J'})$ and $\Bform(J,X)=\E[\xi_JX_\tau]$. Equation~\eqref{eq:general.metricJ} is then the special case of Eq.~\eqref{eq:general.metric} for two observables. Let $\Cclass$ be a linear class of additive observables~\eqref{eq:general.additive} of a given process, and let $\Hdyn_{\Cclass}\subseteq\Hdyn$ be the closure of $\{\xi_J:J\in\Cclass\}$. The dynamical part of the response in Eq.~\eqref{eq:general.response},
\begin{equation}
	\Rc_J\equiv\int_0^\tau\ev{\delta\hLg_t^\dagger H_t}_t\dd{t},
\end{equation}
is linear in $J$. We call the perturbation admissible on $\Cclass$ if this response is bounded by the dynamical variance, i.e., if there is a constant $C>0$ such that
\begin{equation}
	\abs{\Rc_J}\le C\sqrt{\Vd}\quad\text{for all }J\in\Cclass.
\label{eq:general.admissible}
\end{equation}
Admissible perturbations are exactly those that can be represented by a martingale: a perturbation is admissible on $\Cclass$ if and only if there is a martingale $S_t$ with $S_0=0$ and $S_\tau\in\Hdyn_{\Cclass}$ such that
\begin{equation}
	\Rc_J=\int_0^\tau\E[\dd{M_t}\dd{S_t}]\quad\text{for all }J\in\Cclass,
\label{eq:general.Sdef}
\end{equation}
and this martingale is unique. If $S$ exists, combining the Cauchy-Schwarz inequality and Eq.~\eqref{eq:general.Vd} gives $|\Rc_J|\le[\Bform(S,S)\Vd]^{1/2}$, which is Eq.~\eqref{eq:general.admissible} with $C^2=\Bform(S,S)$. Conversely, every admissible perturbation is represented by a unique martingale $S$; the proof, based on the Riesz representation theorem, is given in Appendix~\ref{app:riesz}. Admissibility is therefore not an additional assumption of the present approach, but exactly the condition under which a response bound can hold on $\Cclass$. If it fails, $\Cclass$ contains observables with arbitrarily small dynamical variance and a response of order unity, so no inequality between the response and the dynamical fluctuations can hold. 

For admissible perturbations, the response takes the form of a correlation. By the polarized isometry, $\int_0^\tau\E[\dd{M_t}\dd{S_t}]=\E[(M_\tau-M_0)S_\tau]$, and since $M_\tau=J_\tau$ and $\E[M_0S_\tau]=\E[G_0(z_0)\,\E[S_\tau\cond\Fc_0]]=0$, Eq.~\eqref{eq:general.response} becomes
\begin{equation}
	\delta\E[J_\tau]=\delta_0\E[J_\tau]+\Bform(J,S).
\label{eq:general.responseS}
\end{equation}
We call $S$ the representing martingale of the perturbation. In the cases considered in this work, it is explicit. For a perturbation that leaves the path measure absolutely continuous, $S$ is the projection onto $\Hdyn_{\Cclass}$ of the derivative of the logarithm of the path probability with respect to the perturbation, Eq.~\eqref{eq:general.score} below; the unprojected score reproduces the integrand of Eq.~\eqref{eq:general.response} at every state, as a conditional covariance with $M$ [Eq.~\eqref{eq:general.tilt}]. For diffusion and jump processes, $S$ is given in terms of the injected source current (Table~\ref{tab:realizations}), which also covers changes of the noise intensity, for which the path probability cannot be differentiated. The dynamical response is thus the covariance of the Doob martingale of the observable with the representing martingale of the perturbation, which carries no information about the preparation. 

A natural measure of the strength of a perturbation is the largest squared response that it can produce per unit of dynamical fluctuation. We call this quantity the cost of the perturbation. By Eq.~\eqref{eq:general.responseS} and the Cauchy-Schwarz inequality, it is the squared norm of the representing martingale,
\begin{equation}
	\Icost\equiv\sup_{J\in\Cclass}\frac{\Rc_J^2}{\Vd}=\Bform(S,S)=\E[S_\tau^2],
\label{eq:general.cost}
\end{equation}
where the ratio is taken as $0$ if $\Rc_J=\Vd=0$ and as $+\infty$ if $\Vd=0$ but $\Rc_J\ne0$. The supremum is attained by any observable of the form $J_\tau=a(z_0)+cS_\tau$ with $c\ne0$, i.e., one whose Doob-martingale increment $M_t-M_0$ is proportional to $S_t$, provided that such an observable belongs to $\Cclass$; otherwise, it is approached by observables in $\Cclass$. The supremum is defined for any perturbation, and the perturbation is admissible if and only if it is finite; the second equality holds for admissible perturbations. As shown below, for a perturbation that leaves the path measure absolutely continuous, $\Icost$ is the part of the dynamical Fisher information that observables in $\Cclass$ can detect, and it therefore measures how distinguishable the perturbed dynamics is; for the friction perturbation of Sec.~\ref{sec:under}, it equals half the entropy production.

For perturbations that leave the path measure absolutely continuous, the representing martingale is obtained from the derivative of the logarithm of the path probability with respect to the perturbation, known in statistics as the score. Let $P^\eps$ be the path measure generated by $p_0^\eps$ and $\Lg_t^\eps$, differentiable in $\eps$ with $P^0=P$, and write
\begin{equation}
	\partial_\eps\ln\frac{\dd{P^\eps}}{\dd{P}}\bigg|_{\eps=0}=q_0(z_0)+S^{\mathrm{dyn}}_\tau,
\label{eq:general.score}
\end{equation}
where $q_0\equiv\delta\ln p_0=\delta p_0/p_0$ is the score of the preparation and the dynamical score $S^{\mathrm{dyn}}_t$ is a martingale with $S^{\mathrm{dyn}}_0=0$, both assumed to be square integrable. Differentiating $\E^\eps[J_\tau]=\E[J_\tau\dd{P^\eps}/\dd{P}]$ at $\eps=0$ under the expectation gives $\delta\E[J_\tau]=\E[J_\tau q_0(z_0)]+\E[J_\tau S^{\mathrm{dyn}}_\tau]$, because $\partial_\eps(\dd{P^\eps}/\dd{P})=\partial_\eps\ln(\dd{P^\eps}/\dd{P})$ at $\eps=0$. The first term is the preparation response, $\E[J_\tau q_0]=\E[G_0(z_0)q_0(z_0)]=\delta_0\E[J_\tau]$. For the second term, $J_\tau=G_0(z_0)+\xi_J$ and $\E[G_0(z_0)S^{\mathrm{dyn}}_\tau]=\E[G_0(z_0)\,\E[S^{\mathrm{dyn}}_\tau\cond z_0]]=0$ give
\begin{equation}
	\Rc_J=\E[\xi_JS^{\mathrm{dyn}}_\tau].
\end{equation}
The score $S^{\mathrm{dyn}}_\tau$ lies in $\Hdyn$ but need not lie in $\Hdyn_{\Cclass}$. Its component orthogonal to $\Hdyn_{\Cclass}$ is orthogonal to every $\xi_J$ with $J\in\Cclass$ and does not contribute, so that $\Rc_J=\E[\xi_J\Pi_{\Cclass}S^{\mathrm{dyn}}_\tau]$, where $\Pi_{\Cclass}$ denotes the orthogonal projection onto $\Hdyn_{\Cclass}$. By uniqueness, the representing martingale is $S_t=\E[\Pi_{\Cclass}S^{\mathrm{dyn}}_\tau\cond\Fc_t]$. The projection is trivial when $\Cclass$ contains the score, as for drift perturbations of a diffusion process with the full observable class, and nontrivial, for example, for jump observables with antisymmetric increments (Table~\ref{tab:realizations}). Girsanov's theorem moreover identifies the perturbation of the joint generator pointwise (Appendix~\ref{app:cov.gen}),
\begin{equation}
	\delta\hLg_t^\dagger H_t(z_{t^-})\dd{t}=\E[\dd{M_t}\dd{S^{\mathrm{dyn}}_t}\cond\Fc_{t^-}],
\label{eq:general.tilt}
\end{equation}
so that the integrand $\delta\hLg_t^\dagger H_t$ of the response identity~\eqref{eq:general.response} is, at every state, the conditional covariance of the increments of $M$ and $S^{\mathrm{dyn}}$ per unit time. Averaging over the state and integrating over time recovers Eq.~\eqref{eq:general.Sdef} with $S^{\mathrm{dyn}}$ in place of $S$. The cost satisfies $\Icost=\E[(\Pi_{\Cclass}S^{\mathrm{dyn}}_\tau)^2]\le I_{\mathrm{dyn}}\equiv\E[(S^{\mathrm{dyn}}_\tau)^2]$, because the component of the score orthogonal to $\Hdyn_{\Cclass}$ only adds to its norm; equality holds if and only if the score lies in $\Hdyn_{\Cclass}$. Here, $I_{\mathrm{dyn}}$ is the dynamical part of the path Fisher information $I\equiv\E[(q_0+S^{\mathrm{dyn}}_\tau)^2]=I_0+I_{\mathrm{dyn}}$, with $I_0\equiv\E[q_0^2]$; the cross term vanishes since $\E[q_0(z_0)S^{\mathrm{dyn}}_\tau]=\E[q_0(z_0)\,\E[S^{\mathrm{dyn}}_\tau\cond z_0]]=0$. Because every function of the initial state is uncorrelated with every element of $\Hdyn$, the variance, the response, and the Fisher information all split into a preparation term and a dynamical term. Specifically, the response can be written as
\begin{equation}
	\delta\E[J_\tau]=\Cov[G_0(z_0),q_0(z_0)]+\Bform(J,S),
\label{eq:general.channels}
\end{equation}
which recovers Eq.~\eqref{eq:general.responseS}.

\subsection{Response inequalities, variational forms, and matrix bounds}
For an admissible perturbation with representing martingale $S$, Eq.~\eqref{eq:general.responseS} shows that the response at fixed preparation, $\Rc_J=\Bform(J,S)$, is a value of the same metric as the dynamical variance, $\Vd=\Bform(J,J)$, and the cost, $\Icost=\Bform(S,S)$. Response inequalities thus follow from the geometry of this metric. In particular, the Cauchy-Schwarz inequality yields the fluctuation-response inequality and the variational form of the dynamical variance,
\begin{align}
	\Rc_J^2&\le\Vd\Icost,\label{eq:general.CSdual1}\\
	\Vd&=\sup_{X\in\Hdyn_{\Cclass}}\,\{2\Bform(J,X)-\Bform(X,X)\}.\label{eq:general.CSdual2}
\end{align}
Both relations follow from the identity
\begin{equation}
	2\Bform(J,X)-\Bform(X,X)=\Vd-\Bform(\xi_J-X,\xi_J-X),
\label{eq:general.square}
\end{equation}
obtained by expanding the last term and using $\Bform(\xi_J,\xi_J)=\Vd$. Since the last term is nonnegative, the left-hand side never exceeds $\Vd$, and it equals $\Vd$ if and only if $X=\xi_J$, which lies in $\Hdyn_{\Cclass}$ for $J\in\Cclass$; this proves the variational formula. Choosing $X=cS$ and maximizing the left-hand side of Eq.~\eqref{eq:general.square}, $2c\Rc_J-c^2\Icost$, over $c\in\mathbb{R}$ gives $\Rc_J^2/\Icost\le\Vd$, which is the first relation. Equality holds if and only if the Doob martingale of the observable is proportional to the representing martingale, $M_t-M_0=cS_t$ for all $t\in[0,\tau]$; whether such an observable exists in $\Cclass$ depends on the dynamics and on the class of observables. If $X$ is restricted to a closed subspace $\mathscr{W}$, such as the closed linear span of the representing martingales of perturbations accessible in an experiment, then $\Bform(J,X)=\Bform(\Pi_{\mathscr{W}}\xi_J,X)$ for $X\in\mathscr{W}$, where $\Pi_{\mathscr{W}}$ denotes the orthogonal projection onto $\mathscr{W}$. Equation~\eqref{eq:general.square} with $\Pi_{\mathscr{W}}\xi_J$ in place of $\xi_J$ then shows that the supremum equals $\Bform(\Pi_{\mathscr{W}}\xi_J,\Pi_{\mathscr{W}}\xi_J)$, which is smaller than $\Vd$ unless $\mathscr{W}$ contains $\xi_J$. In particular, if the accessible perturbations are complete, i.e., $\mathscr{W}=\Hdyn_{\Cclass}$, the dynamical variance follows from responses alone,
\begin{equation}
	\Vd=\sup_{S\in\Hdyn_{\Cclass}}\frac{\Bform(J,S)^2}{\Bform(S,S)}=\sup_{S\in\Hdyn_{\Cclass}}\frac{\Rc_J^2}{\Icost},
\label{eq:general.dualnorm}
\end{equation}
because each ratio is bounded from above by $\Vd$, with equality attained when $S_\tau\propto\xi_J$. Equation~\eqref{eq:general.dualnorm} is dual to Eq.~\eqref{eq:general.cost}: the cost of a perturbation is the largest squared response it produces per unit dynamical variance, whereas the dynamical variance of an observable is the largest squared response it shows per unit cost. A single perturbation probes $\xi_J$ only along the direction of its representing martingale and therefore yields only an inequality, whereas a complete set of perturbations yields the identity. For diffusion and jump processes, perturbations that inject probability current at a single time and state form such a set, and Eq.~\eqref{eq:general.dualnorm} becomes the local fluctuation-response identities~\eqref{eq:general.FRRdiff} and \eqref{eq:general.FRRjump} of Sec.~\ref{sec:general.realizations}.

The preparation term obeys its own inequality. Since $\delta_0\E[J_\tau]=\Cov[G_0(z_0),q_0(z_0)]$ with $\E[q_0]=0$, the Cauchy-Schwarz inequality gives $(\delta_0\E[J_\tau])^2\le\Var_0(J_\tau)I_0$. Combining it with $\Rc_J^2\le\Icost\Vd$ through $|\delta\E[J_\tau]|\le|\delta_0\E[J_\tau]|+|\Rc_J|$ and the inequality $(\sqrt{ab}+\sqrt{cd})^2\le(a+c)(b+d)$ yields 
\begin{equation}
	(\delta\E[J_\tau])^2\le\Var(J_\tau)(I_0+\Icost).
\end{equation}
For an absolutely continuous perturbation and a class $\Hdyn_{\Cclass}$ that contains its score, $\Icost=I_{\mathrm{dyn}}$, and this is the linear-response form of the information inequality, $(\delta\E[J_\tau])^2\le\Var(J_\tau)I$~\cite{Dechant.2020.PNAS}. The two separate inequalities are quantitatively sharper: their combination is tight only if $\Var_0(J_\tau)/I_0=\Vd/\Icost$, and at fixed preparation, $I_0=0$, it replaces $\Vd$ by the larger $\Var(J_\tau)$. Moreover, the dynamical inequality also holds for admissible perturbations without a path score, such as changes of friction or temperature, for which the information inequality is not defined.

The same structure extends to multiple observables and multiple perturbations at once. For observables $\{J^\alpha\}_\alpha$ and admissible perturbations with representing martingales $\{S^i\}_i$, we define
\begin{align}
	\Cm_{\alpha\beta}&\equiv\Cov(J_\tau^\alpha,J_\tau^\beta),\\
	\Cm_{0,\alpha\beta}&\equiv\Cov_0(J_\tau^\alpha,J_\tau^\beta)\equiv\Cov[G_0^\alpha(z_0),G_0^\beta(z_0)],\\
	\Rm_{\alpha i}&\equiv\Bform(J^\alpha,S^i),\\
	\Imat_{ij}&\equiv\Bform(S^i,S^j).
\end{align}
By the law of total covariance,
\begin{align}
	&\Cov(J_\tau^\alpha,J_\tau^\beta)\notag\\
	&=\Cov(\E[J_\tau^\alpha\cond z_0],\E[J_\tau^\beta\cond z_0])+\E[\Cov(J_\tau^\alpha,J_\tau^\beta\cond z_0)],
\end{align}
and with $\E[J_\tau^\alpha\cond z_0]=G_0^\alpha(z_0)$ and Eq.~\eqref{eq:general.metricJ}, this reads
\begin{equation}
	\Cm-\Cm_0=[\Bform(J^\alpha,J^\beta)],
\label{eq:general.covariance}
\end{equation}
which is the polarized form of the variance identity~\eqref{eq:general.var}. The three matrices $\Cm-\Cm_0$, $\Rm$, and $\Imat$ therefore consist of the values $\Bform(J^\alpha,J^\beta)$, $\Bform(J^\alpha,S^i)$, and $\Bform(S^i,S^j)$ of the metric on the vectors $\xi_{J^\alpha}$ and $S^i_\tau$ in $\Hdyn$. Arranged as blocks, they form the Gram matrix of these vectors, which is also the averaged conditional covariance matrix of $(J_\tau^\alpha,S_\tau^i)$ given $z_0$. Since a Gram matrix is positive semidefinite, we have
\begin{equation}
	\begin{pmatrix}\Cm-\Cm_0&\Rm\\\Rm^\top&\Imat\end{pmatrix}\succeq0.
\end{equation}
Taking the Schur complement of its lower-right block gives the matrix fluctuation-response inequality
\begin{equation}
	\Cm-\Cm_0\succeq\Rm\Imat^{-1}\Rm^\top.
\label{eq:general.matrix}
\end{equation}
For every combination $\sum_\alpha a_\alpha J^\alpha$ of the observables, this relation is the scalar bound $\Rc_J^2\le\Icost\,\Vd$ with the combination $\sum_ib_iS^i$ of the perturbations chosen optimally. It assumes $\Imat\succ0$. For a singular $\Imat$, the Moore-Penrose inverse applies, because a combination of perturbations with zero cost has zero response, so that the null space of $\Imat$ lies in that of $\Rm$. The Schur complement of the upper-left block gives instead 
\begin{equation}
	\Imat\succeq\Rm^\top(\Cm-\Cm_0)^{+}\Rm,
\end{equation}
in which $+$ denotes the pseudoinverse and the observable coefficients are optimized for every combination of perturbations; for a single perturbation, this is the multivariate bound of Sec.~\ref{sec:under}. The matrix bounds thus quantify how well a set of measured observables, taken together, resolves a set of perturbations.

\subsection{Diffusion and jump processes}\label{sec:general.realizations}
The martingales and the metric introduced so far take an explicit form once the noise is specified. By the martingale representation theorem, every element of $\Hdyn$ is a stochastic integral against the driving noise, i.e., the Wiener processes of a diffusion process or the compensated transition counts of a jump process. For the Doob martingale of an additive observable, the integrand is set by a function of time and the current state, or of time and the transition for a jump process. The metric then becomes a quadratic form in the amplitudes, weighted by the noise intensity. Table~\ref{tab:realizations} lists these expressions for the two classes of processes considered in this work. For processes with both diffusive and jump components, the two contributions add, because the two noises are uncorrelated.

\begin{table*}[t]
\caption{Explicit form of the general framework for diffusion and jump processes. The variance identity~\eqref{eq:general.var}, the response identity~\eqref{eq:general.responseS}, and the bounds of Sec.~\ref{sec:general} hold for any Markov process; the table expresses the quantities that enter them in terms of the drift and the diffusion matrix of a diffusion process and the transition rates of a jump process. For jump processes, the metric, the response, and the cost are given for observables with antisymmetric increments, $d_{mn}=-d_{nm}$, which include currents and residence observables; for general increments, they are given by Eqs.~\eqref{eq:general.jumpmet}, \eqref{eq:general.jumpresponse}, and \eqref{eq:general.jumpIcost}, respectively, with sums over ordered pairs $m\ne n$.}
\label{tab:realizations}
\setlength{\tabcolsep}{4pt}
\renewcommand{\arraystretch}{1.3}
\begin{ruledtabular}
\begin{tabular}{lll}
& Diffusion process (noisy coordinates $\bq$) & Jump process (transitions $n\to m$)\\
\colrule
Observable increment & $\dd{J_t}=\phi_t\dd{t}+\bw_t\circ\dd{\bq_t}$ & $\dd{J_t}=\phi_{z_{t^-}}\dd{t}+\sum_{m\ne n}d_{mn}\dd{\hat{N}_{mn}}$\\
Effective rate $\Phi_t$ & $\phi_t+\bw_t\cdot\bb_t+\D_t:\nabla_{\bq}\bw_t$ & $\phi_n+\sum_{m(\ne n)}w_{mn}d_{mn}$\\
Martingale increment & $\dd{M_t}=\bu_t^\top\sqrt{2\D_t}\,\dd{\bW_t}$ & $\dd{M_t}=\sum_{m\ne n}u_{mn}\dd{\widetilde{N}_{mn}}$\\
Martingale amplitude & $\bu_t=\bw_t+\nabla_{\bq}G_t$ & $u_{mn}=d_{mn}+G_t(m)-G_t(n)$\\
Metric $\Bform(J,J')$ & $2\int_0^\tau\ev{\bu_t^\top\D_t\bu'_t}_t\dd{t}$ & $\int_0^\tau\sum_{m<n}a_{mn}u_{mn}u'_{mn}\dd{t}$\\
Admissible perturbations & drift and diffusion in the noisy coordinates & rates of existing transitions\\
Source current & $\delta\bj_{t,\mathrm{src}}=\delta\bb_tp_t-\nabla_{\bq}\cdot(\delta\D_tp_t)$ & $\delta j_{mn}^{\mathrm{src}}=\delta w_{mn}p_n-\delta w_{nm}p_m$\\
Response $\Rc_J=\Bform(J,S)$ & $\int_0^\tau\inprod{\bu_t}{\delta\bj_{t,\mathrm{src}}}\dd{t}$ & $\int_0^\tau\sum_{m<n}u_{mn}\delta j_{mn}^{\mathrm{src}}\dd{t}$\\
Cost $\Icost=\Bform(S,S)$ & $(1/2)\int_0^\tau\ev{\bg_t^\top\D_t^{-1}\bg_t}_t\dd{t}$;\quad $\bg_t=\delta\bj_{t,\mathrm{src}}/p_t$ & $\int_0^\tau\sum_{m<n}(\delta j_{mn}^{\mathrm{src}})^2/a_{mn}\dd{t}$\\
\end{tabular}
\end{ruledtabular}
\end{table*}

\subsubsection{Diffusion processes}
We split the state as $\bz=(\br,\bq)$ into coordinates $\bq$ that are directly driven by noise and coordinates $\br$ that evolve deterministically given the trajectory, and write
\begin{align}
	\dd{\br_t}&=\ba_t(\bz_t)\dd{t},\notag\\
	\dd{\bq_t}&=\bb_t(\bz_t)\dd{t}+\sqrt{2\D_t(\bz_t)}\dd{\bW_t}.
\label{eq:general.diffusion}
\end{align}
Here, $\bW_t$ is a vector of independent standard Wiener processes, $\E[\dd{W_{t,i}}\dd{W_{t,j}}]=\delta_{ij}\dd{t}$, the drift $\bb_t$ collects the deterministic forces acting on the noisy coordinates, and the diffusion matrix $\D_t$, which is symmetric and positive definite, sets the strength and correlations of the noise. Both may depend on time and on the full state, and a state-dependent $\D_t$ is understood in the It\^o convention. In overdamped dynamics, all coordinates are noisy, $\bq=\bx$; in underdamped dynamics, the velocity is noisy, $\bq=\bv$, whereas the position follows $\dd{\bx_t}=\bv_t\dd{t}$, so that $\br=\bx$ and $\ba_t=\bv$. The backward generator reads
\begin{equation}
	\Lg_t^\dagger=\ba_t\cdot\nabla_{\br}+\bb_t\cdot\nabla_{\bq}+\D_t:\nabla_{\bq}\nabla_{\bq},
\end{equation}
where $\D_t:\nabla_{\bq}\nabla_{\bq}=\sum_{ij}\D_{t,ij}\partial_{q_i}\partial_{q_j}$.

We consider observables that accumulate both during residence and along the noisy displacements,
\begin{equation}
	\dd{J_t}=\phi_t(\bz_t)\dd{t}+\bw_t(\bz_t)\circ\dd{\bq_t},
\label{eq:general.Phi.diff}
\end{equation}
where the residence weight $\phi_t$ is an arbitrary scalar function, the current weight $\bw_t$ is an arbitrary vector field, and $\circ$ denotes the Stratonovich product. A current along the deterministic coordinates, such as $\int\vb*{\alpha}_t\cdot\dd{\bx_t}$ in underdamped dynamics, enters through the residence weight $\phi_t=\vb*{\alpha}_t\cdot\ba_t$. Converting the Stratonovich product to the It\^o form, $\bw_t\circ\dd{\bq_t}=\bw_t\cdot\dd{\bq_t}+\D_t:\nabla_{\bq}\bw_t\dd{t}$ with $\D_t:\nabla_{\bq}\bw_t=\sum_{ij}\D_{t,ij}\partial_{q_i}w_{t,j}$, separates the increment into the form~\eqref{eq:general.additive}, with the effective rate
\begin{equation}
	\Phi_t=\phi_t+\bw_t\cdot\bb_t+\D_t:\nabla_{\bq}\bw_t
\end{equation}
and the martingale part
\begin{equation}
	\dd{K_t}=\bw_t\cdot\sqrt{2\D_t}\dd{\bW_t}.
\end{equation}
Since $K_t$ is driven by the same Wiener process as $\bq_t$, the observable and the state fluctuate jointly, which gives rise to the cross term in Eq.~\eqref{eq:general.threeterm}.

To evaluate the dynamical variance, we combine the state and the observable into the joint process $Z_t=(\bz_t,J_t)$. By Eqs.~\eqref{eq:general.diffusion} and \eqref{eq:general.Phi.diff}, the increments of the noisy coordinates and of the observable value $y$ are $\dd{\bq_t}=\bb_t\dd{t}+\sqrt{2\D_t}\dd{\bW_t}$ and $\dd{y_t}=\Phi_t\dd{t}+\bw_t\cdot\sqrt{2\D_t}\dd{\bW_t}$. They are driven by the same Wiener process, with the conditional covariances $\E[\dd{\bq_t}\dd{\bq_t}^\top\cond\Fc_t]=2\D_t\dd{t}$, $\E[\dd{\bq_t}\dd{y_t}\cond\Fc_t]=2\D_t\bw_t\dd{t}$, and $\E[(\dd{y_t})^2\cond\Fc_t]=2\bw_t^\top\D_t\bw_t\dd{t}$. It\^o's formula therefore gives the joint generator
\begin{align}
	\hLg_t^\dagger&=\Lg_t^\dagger+\Phi_t\partial_y+2(\D_t\bw_t)\cdot\nabla_{\bq}\partial_y+(\bw_t^\top\D_t\bw_t)\partial_y^2.
\label{eq:general.jointdiff}
\end{align}
Only the second-order part of the generator contributes to the local covariance rate~\eqref{eq:general.covrate}, because first-order terms satisfy the product rule and cancel. Equation~\eqref{eq:general.jointdiff} thus gives
\begin{equation}
	\hGam_t[f,g]=2(\nabla_{\bq}f+\bw_t\partial_yf)^\top\D_t(\nabla_{\bq}g+\bw_t\partial_yg),
\label{eq:general.covratediff}
\end{equation}
in which the deterministic coordinates do not appear. For $H_t=y+G_t$, we have $\partial_yH_t=1$ and $\nabla_{\bq}H_t=\nabla_{\bq}G_t$, so that the three terms of Eq.~\eqref{eq:general.threeterm} are $\hGam_t[y,y]=2\bw_t^\top\D_t\bw_t$, $2\hGam_t[y,G_t]=4\bw_t^\top\D_t\nabla_{\bq}G_t$, and $\Gamma_t[G_t,G_t]=2(\nabla_{\bq}G_t)^\top\D_t\nabla_{\bq}G_t$. These three terms combine into a perfect square. To see its structure, note that It\^o's formula gives $\dd{G_t(\bz_t)}=(\partial_t+\Lg_t^\dagger)G_t\dd{t}+(\nabla_{\bq}G_t)^\top\sqrt{2\D_t}\dd{\bW_t}$, where the noise multiplies only the gradient with respect to the noisy coordinates, so that the martingale part of $G_t(\bz_t)$ is $\dd{N^G_t}=(\nabla_{\bq}G_t)^\top\sqrt{2\D_t}\dd{\bW_t}$. Adding $\dd{K_t}=\bw_t^\top\sqrt{2\D_t}\dd{\bW_t}$ according to Eq.~\eqref{eq:general.dM} shows that the increment of the Doob martingale is linear in the noise increment,
\begin{equation}
	\dd{M_t}=\bu_t^\top\sqrt{2\D_t}\dd{\bW_t},
\label{eq:general.diffdM}
\end{equation}
where we call its coefficient $\bu_t$ the martingale amplitude:
\begin{equation}
	\bu_t\equiv\bw_t+\nabla_{\bq}G_t.
\label{eq:general.diffamp}
\end{equation}
Accordingly, the three terms of Eq.~\eqref{eq:general.threeterm} sum to $\hGam_t[H_t,H_t]=2\bu_t^\top\D_t\bu_t$. The amplitude combines the direct weight $\bw_t$ of the observable with the sensitivity $\nabla_{\bq}G_t$ of its expected future to a kick of the noisy coordinates. The metric follows from $\E[\dd{M_t}\dd{M_t'}\cond\Fc_t]=2\bu_t^\top\D_t\bu_t'\dd{t}$ and reads
\begin{equation}
	\Bform(J,J')=2\int_0^\tau\ev{\bu_t^\top\D_t\bu'_t}_t\dd{t}.
\label{eq:general.diffmet}
\end{equation}
We note that only the gradient with respect to the noisy coordinates enters, because the noise acts on them alone. The deterministic coordinates nonetheless enter through $G_t$: a noise kick to $\bq_t$ changes their future evolution and hence the expected future accumulation. In underdamped dynamics, for example, a position-dependent observable has no direct velocity weight, $\bw_t=0$, but generally a nonzero amplitude $\nabla_{\bv}G_t$, because its expected future depends on the present velocity.

We now perturb the noise-driven part of the dynamics in the direction $(\delta\bb_t,\delta\D_t)$, i.e., $\bb_t\to\bb_t+\eps\,\delta\bb_t$ and $\D_t\to\D_t+\eps\,\delta\D_t$ in the notation of Sec.~\ref{sec:general.response}, while keeping the deterministic dynamics of $\br$ fixed. The forward generator, $\Lg_tp=-\nabla_{\br}\cdot(\ba_tp)-\nabla_{\bq}\cdot(\bb_tp)+\sum_{ij}\partial_{q_i}\partial_{q_j}(\D_{t,ij}p)$, is linear in $\bb_t$ and $\D_t$, so that its first variation follows by replacing them with $\delta\bb_t$ and $\delta\D_t$ and dropping the term with $\ba_t$. The perturbation therefore enters the forward equation as the divergence of a source current,
\begin{equation}
\begin{gathered}
	\delta\Lg_tp_t=-\nabla_{\bq}\cdot\delta\bj_{t,\mathrm{src}},\\
	\delta\bj_{t,\mathrm{src}}=\delta\bb_tp_t-\nabla_{\bq}\cdot(\delta\D_tp_t).
\end{gathered}
\label{eq:general.source}
\end{equation}
  The source current is the first variation of the probability current at fixed density, i.e., the current that the perturbation injects directly; the actual injected current is $\eps\delta\bj_{t,\mathrm{src}}$. The full first variation of the probability current splits as $\delta\bj_t=\delta\bj_{t,\mathrm{src}}+\bb_t\delta p_t-\nabla_{\bq}\cdot(\D_t\delta p_t)$, where the second part is generated by the unperturbed dynamics from the density variation $\delta p_t$. For a spatially uniform $\delta\D_t$, $\delta\bj_{t,\mathrm{src}}=\delta\bb_tp_t-\delta\D_t\nabla_{\bq}p_t$, which is the form relevant for the thermal perturbations of underdamped dynamics in Sec.~\ref{sec:under}.

The dynamical response follows from Eq.~\eqref{eq:general.response}. At fixed observable fields, the effective rate changes by $\delta\Phi_t=\bw_t\cdot\delta\bb_t+\delta\D_t:\nabla_{\bq}\bw_t$, and the generator acting on $G_t$ changes by $\delta\Lg_t^\dagger G_t=\delta\bb_t\cdot\nabla_{\bq}G_t+\delta\D_t:\nabla_{\bq}\nabla_{\bq}G_t$. Their sum depends on the observable only through the martingale amplitude, $\delta\Phi_t+\delta\Lg_t^\dagger G_t=\delta\bb_t\cdot\bu_t+\delta\D_t:\nabla_{\bq}\bu_t$. Integrating the second term by parts in $\bq$ and writing $\bg_t\equiv\delta\bj_{t,\mathrm{src}}/p_t$ for the injected current per unit density, we obtain
\begin{equation}
	\Rc_J=\int_0^\tau\inprod{\bu_t}{\delta\bj_{t,\mathrm{src}}}\dd{t}=\int_0^\tau\ev{\bu_t\cdot\bg_t}_t\dd{t}.
\label{eq:general.diffresponse}
\end{equation}
The martingale amplitude that sets the fluctuations in Eq.~\eqref{eq:general.diffmet} is thus also the susceptibility of the mean to an injected current. The representing martingale of the perturbation follows by matching this response to the metric. A martingale $\dd{S_t}=\bh_t^\top\sqrt{2\D_t}\dd{\bW_t}$ has $\Bform(J,S)=2\int_0^\tau\ev*{\bu_t^\top\D_t\bh_t}_t\dd{t}$, which equals $\Rc_J$ for every observable if $2\D_t\bh_t=\bg_t$. Hence,
\begin{equation}
	\dd{S_t}=\bh_t^\top\sqrt{2\D_t}\dd{\bW_t},\quad \bh_t=(2\D_t)^{-1}\bg_t.
\label{eq:general.diffS}
\end{equation}
Since the amplitude $\bh_t$ is a function of time and state, $S$ lies in $\Hdyn_{\Cclass}$ for the full class of observables~\eqref{eq:general.Phi.diff}. The cost of the perturbation reads
\begin{equation}
	\Icost=\Bform(S,S)=\frac{1}{2}\int_0^\tau\ev{\bg_t^\top\D_t^{-1}\bg_t}_t\dd{t}.
\label{eq:general.diffI}
\end{equation}
The cost measures the injected current in the inverse noise metric: a source is expensive when it pushes probability in directions in which the noise is weak.

Drift and diffusion perturbations differ in their relation to the path measure. For a pure drift perturbation, $\delta\D_t=0$ and $\bg_t=\delta\bb_t$, the path measures remain absolutely continuous, and $S$ is the dynamical score, i.e., the dynamical part of the derivative of the log-likelihood ratio~\eqref{eq:general.score}. On the other hand, a perturbation of $\D_t$ generally destroys absolute continuity. A uniform change of the diffusion matrix, for instance, alters the quadratic variation of almost every path, so that the perturbed and unperturbed path measures are mutually singular and no score exists; thermal responses of this kind have therefore been treated by dedicated methods~\cite{Baiesi.2014.EPJB,Klinger.2025.PRL}. The perturbation is nevertheless admissible whenever the cost $\Icost$ is finite, because the Cauchy-Schwarz inequality applied to Eq.~\eqref{eq:general.diffresponse} gives $\Rc_J^2\le\Icost\,\Vd$. On the full class of observables, its representing martingale~\eqref{eq:general.diffS} coincides with the dynamical score of an auxiliary drift perturbation in the direction $\bg_t$ at fixed $\D_t$. This auxiliary perturbation injects the same source current and therefore has the same first-order effect on the mean of every observable, although it does not reproduce the perturbed path measure \cite{Klinger.2025.PRL}.

Perturbations of the deterministic coordinates, by contrast, are generally not admissible, as the following example of a nonzero response for an observable with zero dynamical variance shows. Consider underdamped dynamics, $\br=\bx$ and $\bq=\bv$ with $\ba_t=\bv$, and the residence observable with weights $\phi_t(\bx,\bv)=f(\bx)-(\tau-t)\bv\cdot\nabla_{\bx}f(\bx)$ and $\bw_t=0$,
\begin{equation}
	J_\tau=\int_0^\tau\qty[f(\bx_t)-(\tau-t)\bv_t\cdot\nabla_{\bx}f(\bx_t)]\dd{t}.
\end{equation}
Using $\dd{f(\bx_t)}=\bv_t\cdot\nabla_{\bx}f(\bx_t)\dd{t}$ and an integration by parts gives
\begin{equation}
	\int_t^\tau(\tau-s)\bv_s\cdot\nabla_{\bx}f(\bx_s)\dd{s}=-(\tau-t)f(\bx_t)+\int_t^\tau f(\bx_s)\dd{s}.
\end{equation}
Hence, $J_\tau-J_t=(\tau-t)f(\bx_t)$, so that $G_t(\bx,\bv)=(\tau-t)f(\bx)$ and $M_t=J_t+G_t(\bz_t)=\tau f(\bx_0)$ at all times. The amplitude $\bu_t=\nabla_{\bv}G_t$ vanishes, and $\Vd=0$: the value of the observable is fixed by the initial position. Now, we perturb the kinematics in a direction $\delta\ba_t$, i.e., $\dd{\bx_t}=(\bv_t+\eps\delta\ba_t)\dd{t}$, at fixed initial density and observable fields. Then, $\delta\Phi_t=0$ and $\delta\Lg_t^\dagger=\delta\ba_t\cdot\nabla_{\bx}$, and Eq.~\eqref{eq:general.response} gives
\begin{equation}
	\Rc_J=\int_0^\tau\ev{\delta\ba_t\cdot\nabla_{\bx}G_t}_t\dd{t}=\int_0^\tau(\tau-t)\ev{\delta\ba_t\cdot\nabla_{\bx}f}_t\dd{t},
\end{equation}
which is generically nonzero; for $f(\bx)=\vb*{k}\cdot\bx$ and a constant $\delta\ba$, $\Rc_J=\tau^2(\vb*{k}\cdot\delta\ba)/2$. Since $\Vd=0$ whereas $\Rc_J\ne0$, no bound of the form $\Rc_J^2\le C^2\Vd$ can hold for such a perturbation.

\subsubsection{Jump processes}
For a discrete-state Markov jump process, the state changes only through transitions $n\to m$, which occur at the rates $w_{mn}(t)\ge0$. Let $\hat{N}_{mn}(t)$ count the transitions $n\to m$ up to time $t$. Given the past, such a transition occurs in $[t,t+\dd{t})$ with probability $w_{mn}(t)\mathds{1}_{\{z_{t^-}=n\}}\dd{t}$, so that the compensated counting processes
\begin{equation}
	\dd{\widetilde{N}_{mn}(t)}\equiv\dd{\hat{N}_{mn}(t)}-w_{mn}(t)\mathds{1}_{\{z_{t^-}=n\}}\dd{t}
\label{eq:general.counting}
\end{equation}
are martingales. Here, the indicator $\mathds{1}_{\{A\}}$ equals $1$ if the condition $A$ holds and $0$ otherwise. These martingales play the role of the Wiener increments in diffusion processes. Since two distinct transitions never occur at the same time, the compensated counting processes of distinct transitions are orthogonal, and the variance rate of each equals its rate, 
\begin{equation}
	\E[\dd{\widetilde{N}_{mn}}\dd{\widetilde{N}_{m'n'}}\cond\Fc_{t^-}]=\delta_{mm'}\delta_{nn'}w_{mn}(t)\mathds{1}_{\{z_{t^-}=n\}}\dd{t}.
\end{equation}
The forward and backward generators read
\begin{align}
	(\Lg_tp)_m&=\sum_{n(\ne m)}(w_{mn}p_n-w_{nm}p_m),\\
	(\Lg_t^\dagger f)(n)&=\sum_{m(\ne n)}w_{mn}[f(m)-f(n)],
\end{align}
and we denote the net probability current and the traffic of the edge between $n$ and $m$ by $j_{mn}\equiv w_{mn}p_n-w_{nm}p_m$ and $a_{mn}\equiv w_{mn}p_n+w_{nm}p_m$. Unlike diffusion processes, jump processes have no deterministic coordinates: every transition with a nonzero rate is a source of noise.

We consider observables that accumulate both during residence and at transitions,
\begin{equation}
	\dd{J_t}=\phi_{z_{t^-}}(t)\dd{t}+\sum_{m\ne n}d_{mn}(t)\dd{\hat{N}_{mn}(t)},
\label{eq:general.Jjump}
\end{equation}
where the residence weights $\{\phi_n\}$ and the increments $\{d_{mn}\}$ are arbitrary. For example, $\phi_n=\delta_{nk}$ gives the occupation time of state $k$; $d_{mn}=-d_{nm}=1$ on a single edge and zero otherwise gives the net number of transitions along that edge; $d_{mn}=f(m)-f(n)$ gives the increment $f(z_\tau)-f(z_0)$; and, under local detailed balance, $d_{mn}=\ln(w_{mn}/w_{nm})$ gives the entropy flow into the environment. Splitting each counting increment into its conditional mean and its compensated part, $\dd{\hat{N}_{mn}}=w_{mn}\mathds{1}_{\{z_{t^-}=n\}}\dd{t}+\dd{\widetilde{N}_{mn}}$, brings Eq.~\eqref{eq:general.Jjump} into the form~\eqref{eq:general.additive}, with the effective rate and the martingale part
\begin{gather}
	\Phi_t(n)=\phi_n+\sum_{m(\ne n)}w_{mn}d_{mn},\\
	\dd{K_t}=\sum_{m\ne n}d_{mn}\dd{\widetilde{N}_{mn}}.
\end{gather}
Since $K_t$ is driven by the same transitions as the state, the observable and the state jump together, which gives rise to the cross term in Eq.~\eqref{eq:general.threeterm}.

To evaluate the dynamical variance, we again use the joint process $Z_t=(z_t,J_t)$. Between transitions, the observable value $y$ grows at the rate $\phi_n$; at a transition $n\to m$, the joint state jumps from $(n,y)$ to $(m,y+d_{mn})$. The joint generator therefore reads
\begin{align}
	(\hLg_t^\dagger H)(n,y)
	&=\phi_n\partial_y H(n,y)\notag\\
	&\quad+\sum_{m(\ne n)}w_{mn}[H(m,y+d_{mn})-H(n,y)].
\label{eq:general.jointjump}
\end{align}
The first-order term $\phi_n\partial_y$ obeys the product rule and drops out of the local covariance rate~\eqref{eq:general.covrate}. Each jump term instead contributes the product of the changes of $f$ and $g$ across the jump,
\begin{equation}
	\hGam_t[f,g](n,y)=\sum_{m(\ne n)}w_{mn}\Delta_{mn}f\Delta_{mn}g,
\label{eq:general.covratejump}
\end{equation}
where $\Delta_{mn}f\equiv f(m,y+d_{mn})-f(n,y)$. For $H_t=y+G_t$, the change across a jump, $\Delta_{mn}H_t=d_{mn}+G_t(m)-G_t(n)$, does not depend on $y$, so that the three terms of Eq.~\eqref{eq:general.threeterm} are functions of the present state $n$ alone: $\hGam_t[y,y](n)=\sum_{m(\ne n)}w_{mn}d_{mn}^2$, $2\hGam_t[y,G_t](n)=2\sum_{m(\ne n)}w_{mn}d_{mn}[G_t(m)-G_t(n)]$, and $\Gamma_t[G_t,G_t](n)=\sum_{m(\ne n)}w_{mn}[G_t(m)-G_t(n)]^2$. As for diffusion processes, they combine into a perfect square. At a transition $n\to m$, $G_t(z_t)$ jumps by $G_t(m)-G_t(n)$, so that its martingale part is $\dd{N^G_t}=\sum_{m\ne n}[G_t(m)-G_t(n)]\dd{\widetilde{N}_{mn}}$. Adding $\dd{K_t}$ according to Eq.~\eqref{eq:general.dM} shows that the increment of the Doob martingale is linear in the compensated counting increments, 
\begin{equation}
	\dd{M_t}=\sum_{m\ne n}u_{mn}\dd{\widetilde{N}_{mn}},
\label{eq:general.jumpdM}
\end{equation}
with the amplitude $u_{mn}$:
\begin{equation}
	u_{mn}\equiv d_{mn}+G_t(m)-G_t(n).
\label{eq:general.jumpamp}
\end{equation}
Accordingly, the three terms sum to $\hGam_t[H_t,H_t](n)=\sum_{m(\ne n)}w_{mn}u_{mn}^2$. The amplitude combines the direct increment $d_{mn}$ of the observable with the change $G_t(m)-G_t(n)$ of its expected future accumulation caused by the jump. A residence observable, $d_{mn}=0$, thus generally has a nonzero amplitude, because a transition changes the expected future occupation. Since the compensated counting processes are orthogonal, $\E[\dd{M_t}\dd{M_t'}\cond\Fc_{t^-}]=\sum_{m\ne n}w_{mn}\mathds{1}_{\{z_{t^-}=n\}}u_{mn}u'_{mn}\dd{t}$, and the metric reads
\begin{equation}
	\Bform(J,J')=\int_0^\tau\sum_{m\ne n}w_{mn}p_n u_{mn}u'_{mn}\dd{t}.
\label{eq:general.jumpmet}
\end{equation}

For antisymmetric increments, $d_{mn}=-d_{nm}$, which include currents, residence observables, and their combinations, the amplitude is antisymmetric, $u_{mn}=-u_{nm}$. The two directions of an edge then contribute $u_{mn}(\dd{\widetilde{N}_{mn}}-\dd{\widetilde{N}_{nm}})$ to $\dd{M_t}$, and they combine into the traffic,
\begin{equation}
	\Bform(J,J')=\int_0^\tau\sum_{m<n}a_{mn}u_{mn}u'_{mn}\dd{t}.
\end{equation}
Note the relative factor of two between the coordinate forms of the metric, $2\D_t$ for diffusions and $a_{mn}$ for jumps. Both are the noise intensities of the corresponding state increments: $\E[\dd{\bq_t}\dd{\bq_t}^\top\cond\Fc_t]=2\D_t\dd{t}$, whereas the net number of transitions along an edge, $\dd{\hat{N}_{mn}}-\dd{\hat{N}_{nm}}$, has the conditional variance $(w_{mn}\mathds{1}_{\{z_{t^-}=n\}}+w_{nm}\mathds{1}_{\{z_{t^-}=m\}})\dd{t}$, whose average is $a_{mn}\dd{t}$. Both therefore represent the same metric~\eqref{eq:general.metric}.

We now perturb the rates of the existing transitions in the direction $\delta w_{mn}(t)$, i.e., $w_{mn}\to w_{mn}+\eps\,\delta w_{mn}$ for $w_{mn}>0$, at fixed observable fields; the opening of new transitions is discussed below. The forward generator is linear in the rates, so that its first variation follows by replacing $w_{mn}$ with $\delta w_{mn}$. The perturbation therefore enters the master equation as the net inflow of a source current along the edges,
\begin{gather}
	(\delta\Lg_tp_t)_m=\sum_{n(\ne m)}\delta j_{mn}^{\mathrm{src}},
\label{eq:general.jumpsource}
\end{gather}
where $\delta j_{mn}^{\mathrm{src}}\equiv\delta w_{mn}p_n-\delta w_{nm}p_m$.
As for diffusions, the source current is the first variation of the edge current $j_{mn}$ at fixed density, i.e., the current that the perturbation injects directly. The remaining part of the first variation, $w_{mn}\delta p_n-w_{nm}\delta p_m$, is generated by the unperturbed dynamics from the density variation $\delta p_t$ and is accounted for through $G_t$. 

The dynamical response follows from Eq.~\eqref{eq:general.response}. At fixed observable fields, the first variations of the effective rate and of the generator acting on $G_t$ are $\delta\Phi_t(n)=\sum_{m(\ne n)}\delta w_{mn}d_{mn}$ and $(\delta\Lg_t^\dagger G_t)(n)=\sum_{m(\ne n)}\delta w_{mn}[G_t(m)-G_t(n)]$. Their sum depends on the observable only through the martingale amplitude, $\delta\Phi_t(n)+(\delta\Lg_t^\dagger G_t)(n)=\sum_{m(\ne n)}\delta w_{mn}u_{mn}$, and averaging over $p_t$ gives
\begin{equation}
	\Rc_J=\int_0^\tau\sum_{m\ne n}p_n\delta w_{mn}u_{mn}\dd{t}.
\label{eq:general.jumpresponse}
\end{equation}
For antisymmetric increments, the two directions of each edge combine into the source current, and the response reads
\begin{equation}
	\Rc_J=\int_0^\tau\sum_{m<n}u_{mn}\delta j_{mn}^{\mathrm{src}}\dd{t}.
\label{eq:general.jumpresponse.asym}
\end{equation}
This is the counterpart of Eq.~\eqref{eq:general.diffresponse}: the martingale amplitude is the susceptibility of the mean to a current injected along an edge, and the edge source $g_{mn}\equiv\delta j_{mn}^{\mathrm{src}}$ plays the role of $\bg_t$.

The representing martingale follows by matching this response to the metric. A martingale of the form $\dd{S_t}=\sum_{m\ne n}h_{mn}\dd{\widetilde{N}_{mn}}$ has
\begin{equation}
	\Bform(J,S)=\int_0^\tau\sum_{m\ne n}w_{mn}p_nu_{mn}h_{mn}\dd{t}.
\end{equation}
On the full class of additive observables, whose martingale amplitudes need not be antisymmetric, this equals $\Rc_J$ for every observable if $h_{mn}=\delta w_{mn}/w_{mn}=\delta\ln w_{mn}$. The representing martingale is then the dynamical score of the rate perturbation,
\begin{equation}
	\dd{S_t}=\sum_{m\ne n}\delta\ln w_{mn}\dd{\widetilde{N}_{mn}},
\label{eq:general.jumpscore}
\end{equation}
and its cost is the dynamical Fisher information,
\begin{equation}
	\Icost=I_{\mathrm{dyn}}=\int_0^\tau\sum_{m\ne n}\frac{p_n(\delta w_{mn})^2}{w_{mn}}\dd{t}.
\label{eq:general.jumpIcost}
\end{equation}
On the class of observables with antisymmetric increments, the amplitudes $u_{mn}$ are antisymmetric. Since $S_\tau$ must lie in $\Hdyn_{\Cclass}$, whose elements all have antisymmetric amplitudes, the amplitude $h_{mn}$ of the representing martingale is antisymmetric as well. Matching then requires only $\int_0^\tau\sum_{m<n}a_{mn}u_{mn}h_{mn}\dd{t}=\int_0^\tau\sum_{m<n}u_{mn}\delta j_{mn}^{\mathrm{src}}\dd{t}$ for all antisymmetric $u_{mn}$, so that
\begin{gather}
	h_{mn}=\frac{\delta j_{mn}^{\mathrm{src}}}{a_{mn}},\quad
	\Icost=\int_0^\tau\sum_{m<n}\frac{(\delta j_{mn}^{\mathrm{src}})^2}{a_{mn}}\dd{t}.
\label{eq:general.jumpS}
\end{gather}
This amplitude is the orthogonal projection of the score onto the antisymmetric class, as described in Sec.~\ref{sec:general.response}, and hence $\Icost\le I_{\mathrm{dyn}}$. Equality holds when the score is itself antisymmetric, $\delta\ln w_{mn}=-\delta\ln w_{nm}$, as for a perturbation that only biases the ratio of the two rates of each edge. 

We now discuss the admissibility of perturbations. A change of the rates of existing transitions, $w_{mn}>0$, leaves the path measure absolutely continuous. Its cost $\Icost$ is finite whenever the integrand $\sum_{m\ne n}p_n(\delta w_{mn})^2/w_{mn}$ in Eq.~\eqref{eq:general.jumpIcost} is integrable, and the perturbation is then admissible on every class of observables. Opening a transition that is absent in the reference dynamics, $w_{mn}=0\to\eps\,\delta w_{mn}$ with $\delta w_{mn}>0$, is defined only for $\eps\ge0$, so that the response is a right derivative at $\eps=0$. It makes the perturbed path measure not absolutely continuous with respect to the reference one, so that no score exists. Such a perturbation is not admissible on the full class of observables. To see this, consider the observable that counts only this transition, $J_\tau=\hat{N}_{mn}(\tau)$, i.e., $d_{mn}=1$ and all other weights zero. Under the reference dynamics, the transition never occurs, so that $J_\tau=0$ along every trajectory; accordingly, $\Phi_t=0$, $G_t=0$, and $\Vd=\int_0^\tau w_{mn}p_n\dd{t}=0$. Under the perturbed dynamics, the transition occurs at the rate $\eps\,\delta w_{mn}$ whenever the system is in state $n$, so that the mean count changes by $\eps\int_0^\tau p_n\delta w_{mn}\dd{t}$ to first order; equivalently, Eq.~\eqref{eq:general.response} with $G_t=0$ and $\delta\Phi_t(n)=\delta w_{mn}$ gives $\Rc_J=\int_0^\tau p_n\,\delta w_{mn}\dd{t}$. This response is positive whenever the system visits $n$ while the transition is open, whereas $\Vd=0$, so that no bound of the form $\Rc_J^2\le C^2\Vd$ can hold. 

However, on the antisymmetric class, the cost~\eqref{eq:general.jumpS} depends only on the traffic $a_{mn}$ and the source current $\delta j_{mn}^{\mathrm{src}}$ of the edge. If the reverse transition carries positive reference traffic, $a_{mn}=w_{nm}p_m>0$, opening $n\to m$ gives $\delta j_{mn}^{\mathrm{src}}=\delta w_{mn}p_n$ and the cost $\int_0^\tau(\delta w_{mn}p_n)^2/(w_{nm}p_m)\dd{t}$, which is finite, e.g., if $\delta w_{mn}$ is bounded and the reverse traffic $a_{mn}$ stays positive on the closed interval $[0,\tau]$. The perturbation is then admissible although no score exists, because an observable with antisymmetric increments that counts the opened transition must also count the reverse one, whose fluctuations bound the response. This is the jump counterpart of a change of the diffusion matrix in diffusion processes. Since the response on this class of observables depends only on the source current, the same response can be produced by the auxiliary change $\delta w_{nm}=-\delta w_{mn}p_n/p_m$ of the existing reverse rate, which leaves the path measure absolutely continuous. Its score, $\dd{S_t}=\delta\ln w_{nm}\dd{\widetilde{N}_{nm}}$, has the antisymmetric amplitude $h_{mn}=-\delta\ln w_{nm}=\delta w_{mn}p_n/(w_{nm}p_m)$, in agreement with Eq.~\eqref{eq:general.jumpS}, and is the representing martingale of the opening. If neither direction exists, $a_{mn}=0$. The net count along the edge, with $d_{mn}=-d_{nm}=1$, then vanishes identically under the reference dynamics, so that $\Phi_t=0$, $G_t=0$, and $\Vd=0$, whereas Eq.~\eqref{eq:general.response} gives $\Rc_J=\int_0^\tau p_n\delta w_{mn}\dd{t}$, which is positive whenever the system visits $n$ while the transition is open. Such an edge carries no noise and is the counterpart of the deterministic coordinates of a diffusion.

\subsubsection{Local fluctuation-response relations}
Since the response is linear in the direction of the perturbation, it can be resolved in time and state. Let the perturbation change a local parameter $\vb*{\theta}_t(\bz)$ with $k$ components, such as a force field or a temperature profile. We write $\delta\E[J_\tau]=\int_0^\tau\int\vb*{R}_\theta(t,\bz)\cdot\delta\vb*{\theta}_t(\bz)\dd{\bz}\dd{t}$, so that the local response $\vb*{R}_\theta(t,\bz)\equiv\delta\E[J_\tau]/\delta\vb*{\theta}_t(\bz)$ is the functional derivative of the mean, i.e., its response to a change of $\vb*{\theta}$ localized at time $t$ and state $\bz$. The local response depends on the choice of parameter only through the source current that a unit change of $\vb*{\theta}$ injects. If $\delta\bj_{t,\mathrm{src}}=\mathsf{N}_\theta\,\delta\vb*{\theta}_t$, where $\mathsf{N}_\theta(t,\bz)$ is a $d_q\times k$ matrix and $d_q$ is the number of noisy coordinates, Eq.~\eqref{eq:general.diffresponse} gives $\vb*{R}_\theta=\mathsf{N}_\theta^\top\bu_t$. For a scalar parameter $\theta_t$, $\mathsf{N}_\theta$ is a vector and $R_\theta=\mathsf{N}_\theta\cdot\bu_t$ is a scalar. For jump processes, an edge parameter with $\delta j_{mn}^{\mathrm{src}}=N_{\theta,mn}\delta\theta_{mn}$ gives $R_{\theta,mn}=N_{\theta,mn}u_{mn}$. The source current itself gives $\vb*{R}_{\mathrm{src}}=\bu_t$ and $R_{\mathrm{src},mn}=u_{mn}$, whereas a drift, $\mathsf{N}_\theta=p_t\mathds{I}$ with the $d_q\times d_q$ identity matrix $\mathds{I}$, gives $p_t\bu_t$. Responses to local forces or drifts, which are the ones usually measured, therefore carry a density factor that must be removed before they are compared with the martingale amplitude. 

When the parameter is the source current itself, the explicit forms of the metric, Eq.~\eqref{eq:general.diffmet} and the traffic form of Eq.~\eqref{eq:general.jumpmet}, become exact local fluctuation-response relations,
\begin{align}
	\Vd&=2\int_0^\tau\ev{\vb*{R}_{\mathrm{src}}^\top\D_t\vb*{R}_{\mathrm{src}}}_t\dd{t}
\label{eq:general.FRRdiff}
\end{align}
for diffusion processes and
\begin{align}
	\Vd&=\int_0^\tau\sum_{m<n}a_{mn}R_{\mathrm{src},mn}^2\dd{t}
\label{eq:general.FRRjump}
\end{align}
for jump processes. For any other parameter, the same relations hold with $\bu_t$ reconstructed from $\vb*{R}_\theta=\mathsf N_\theta^\top\bu_t$, which is possible wherever $\mathsf N_\theta$ has rank $d_q$. Elsewhere, some components of $\bu_t$ do not affect the response to $\vb*{\theta}$ and must be obtained from another perturbation. Equations~\eqref{eq:general.FRRdiff} and \eqref{eq:general.FRRjump} are the explicit form of the dual-norm identity~\eqref{eq:general.dualnorm}: local perturbations are complete, and the supremum is attained by any source proportional to the amplitude, $\bg_t\propto\D_t\vb*{R}_{\mathrm{src}}$ or $g_{mn}\propto a_{mn}R_{\mathrm{src},mn}$, so that no optimization is needed once the local responses are known.

For later use, we write the response and the cost of a source in a common form. For diffusion processes, with the injected current per unit density $\bg_t$, we write $\Rc[\bg]\equiv\int_0^\tau\ev*{\bu_t\cdot\bg_t}_t\dd{t}$, which equals $\Rc_J$ by Eq.~\eqref{eq:general.diffresponse}, with the cost $\Ac[\bg]\equiv(1/2)\int_0^\tau\ev*{\bg_t^\top\D_t^{-1}\bg_t}_t\dd{t}$, which equals $\Icost$ by Eq.~\eqref{eq:general.diffI}. For jump processes and observables with antisymmetric increments, with the edge source $g_{mn}$, we write $\Rc[g]\equiv\int_0^\tau\sum_{m<n}u_{mn}g_{mn}\dd{t}$, which equals $\Rc_J$ by Eq.~\eqref{eq:general.jumpresponse.asym}, and $\Ac[g]\equiv\int_0^\tau\sum_{m<n}g_{mn}^2/a_{mn}\dd{t}$, which equals $\Icost$ by Eq.~\eqref{eq:general.jumpS}. In both realizations, the fluctuation-response inequality and the variational formula in Eqs.~\eqref{eq:general.CSdual1} and \eqref{eq:general.CSdual2} read
\begin{equation}
\begin{aligned}
	\Rc[g]^2&\le\Vd\Ac[g],\\
	\Vd&=\max_g\{2\Rc[g]-\Ac[g]\},
\end{aligned}
\label{eq:general.sourcebounds}
\end{equation}
where $g$ stands for $\bg$ or $g_{mn}$. The maxima are attained at $\bg_t=2\D_t\bu_t$ and $g_{mn}=a_{mn}u_{mn}$. These relations hold for arbitrary time- and state-dependent sources; restricting to the sources available in an experiment can only lower the variational estimate.

\subsection{Long-time diffusivity}\label{sec:general.longtime}
To study the long-time limit, we consider autonomous dynamics, with a time-independent generator and time-independent observable fields, in the steady state $p_{\st}$, $\Lg p_{\st}=0$. Hereafter, the subscript ``$\st$'' denotes evaluation at the stationary state, and $\jr_{\st}=\ev*{\Phi}_{\st}$ is the mean rate of the observable. With $\mathcal{P}_t\equiv e^{t\Lg^\dagger}$, the solution of the backward equation~\eqref{eq:general.backward} is $G_t=\int_0^{\tau-t}\mathcal{P}_s\Phi\dd{s}$, the expected accumulation over the remaining time $\tau-t$. Far from the terminal time, this accumulation grows at the mean rate, up to a time-independent function of the current state. To make this precise, we assume that the dynamics is mixing, i.e., that $\mathcal{P}_sf\to\ev*{f}_{\st}$ as $s\to\infty$, fast enough that $\int_0^\infty\|\mathcal{P}_s(\Phi-\jr_{\st})\|_{\st}\dd{s}<\infty$, where $\|f\|_{\st}^2\equiv\ev*{f^2}_{\st}$. Then, $G_t(z)=(\tau-t)\jr_{\st}+\Psi(z)+o(1)$ as $\tau-t\to\infty$, with
\begin{equation}
	\Psi(z)\equiv\int_0^\infty \mathcal{P}_s(\Phi-\jr_{\st})(z)\dd{s}.
\label{eq:general.Psi}
\end{equation}
The function $\Psi(z)$ is the expected accumulation from the initial state $z$ in excess of the mean rate. Since $\Lg^\dagger \mathcal{P}_s=\partial_s\mathcal{P}_s$ and $\mathcal{P}_s(\Phi-\jr_{\st})\to\ev*{\Phi-\jr_{\st}}_{\st}=0$ as $s\to\infty$, applying $\Lg^\dagger$ to Eq.~\eqref{eq:general.Psi} gives the Poisson equation \cite{Cattiaux.2012}
\begin{equation}
	\Lg^\dagger\Psi=-(\Phi-\jr_{\st}),\quad \ev{\Psi}_{\st}=0,
\label{eq:general.poisson}
\end{equation}
where the normalization follows from the stationarity of $p_{\st}$, $\ev*{\mathcal{P}_sf}_{\st}=\ev*{f}_{\st}$. Equation~\eqref{eq:general.poisson} is the stationary counterpart of the backward equation~\eqref{eq:general.backward}.

The function $\Psi$ removes the drift of the observable at the stationary rate. With $H_\Psi(z,y)\equiv y+\Psi(z)$, the properties $\hLg^\dagger y=\Phi$ and $\hLg^\dagger\Psi=\Lg^\dagger\Psi$ of the joint generator give $\hLg^\dagger H_\Psi=\Phi+\Lg^\dagger\Psi=\jr_{\st}$. The process $M^\Psi_t\equiv J_t-\jr_{\st}t+\Psi(z_t)=H_\Psi(Z_t)-\jr_{\st}t$ therefore has no drift, and Eq.~\eqref{eq:general.Nf} shows that $M^\Psi_t-M^\Psi_0$ is a martingale, with the increment $\dd{M^\Psi_t}=\dd{K_t}+\dd{N^\Psi_t}$, as in Eq.~\eqref{eq:general.dM} with $\Psi$ in place of $G_t$. Evaluating $M^\Psi$ at $t=\tau$ and $t=0$ gives
\begin{equation}
	J_\tau-\jr_{\st}\tau=(M^\Psi_\tau-M^\Psi_0)+\Psi(z_0)-\Psi(z_\tau),
\label{eq:general.Psidecomp}
\end{equation}
i.e., a martingale integral plus a boundary term.

Although the two parts contribute differently to the variance, both have zero mean in the steady state. By the isometry~\eqref{eq:general.isometry} and Eq.~\eqref{eq:general.Ncov}, the martingale part has the variance $\E[(M^\Psi_\tau-M^\Psi_0)^2]=\tau\ev*{\hGam[H_\Psi,H_\Psi]}_{\st}$, which grows linearly in time; note that the local covariance rate does not depend on $y$, as shown above. If $\ev*{\Psi^2}_{\st}$ is finite, the boundary term has a variance bounded by $4\ev*{\Psi^2}_{\st}$ at all times, $\E[(\Psi(z_0)-\Psi(z_\tau))^2]\le 4\ev*{\Psi^2}_{\st}$. The Cauchy-Schwarz inequality bounds the covariance of the two parts by $2[\tau\ev*{\hGam[H_\Psi,H_\Psi]}_{\st}\ev*{\Psi^2}_{\st}]^{1/2}=O(\sqrt\tau)$. Therefore, only the variance of the martingale part survives in the long-time limit. Defining the diffusivity of the observable as
\begin{equation}
	D_J\equiv\lim_{\tau\to\infty}\frac{\Var(J_\tau)}{2\tau},
\label{eq:general.DJ}
\end{equation}
we obtain
\begin{equation}
	D_J=\frac{1}{2}\ev{\hGam[H_\Psi,H_\Psi]}_{\st}.
\label{eq:general.DJgeneral}
\end{equation}
The diffusivity is thus half the stationary average of the local variance rate of the corrected observable, with the same direct, cross, and propagated terms as Eq.~\eqref{eq:general.threeterm} and $\Psi$ in place of $G_t$. In the two realizations, Eqs.~\eqref{eq:general.covratediff} and \eqref{eq:general.covratejump} with $\partial_yH_\Psi=1$ give the corrected amplitudes $\bw+\nabla_{\bq}\Psi$ and $d_{mn}+\Psi(m)-\Psi(n)$, and we obtain
\begin{align}
	D_J&=\ev{(\bw+\nabla_{\bq}\Psi)^\top\D(\bw+\nabla_{\bq}\Psi)}_{\st}
\label{eq:general.DJdiff}
\end{align}
for diffusion processes and
\begin{align}
	D_J&=\frac{1}{2}\sum_{m\ne n}w_{mn}p_n^{\st}[d_{mn}+\Psi(m)-\Psi(n)]^2
\label{eq:general.DJjump}
\end{align}
for jump processes.
For observables with antisymmetric increments, $d_{mn}=-d_{nm}$, the diffusivity for jump processes becomes $D_J=(1/2)\sum_{m<n}a_{mn}^{\st}[d_{mn}+\Psi(m)-\Psi(n)]^2$.

\subsection{Frequency-domain identities}\label{sec:general.frequency}
The finite-time martingale framework admits a direct frequency-domain representation without any additional stochastic construction: a Fourier-filtered observable remains within the class~\eqref{eq:general.additive}, with explicitly time-dependent weights. For a real frequency $\omega$, we define
\begin{equation}
	J_\tau[\omega]=\int_0^\tau e^{-i\omega t}\dd{J_t},
\label{eq:general.filter}
\end{equation}
which combines the experimentally accessible cosine and sine quadratures; for complex variables, the variance is defined as $\Var(X)\equiv\E[|X-\E[X]|^2]$. The filtered observable has the effective rate $e^{-i\omega t}\Phi_t$ and the martingale part $\int e^{-i\omega t}\dd{K_t}$, so its conditional future $G_t[\omega]$ solves the backward equation with the source $e^{-i\omega t}\Phi_t$. Its Doob martingale is $M_t[\omega]=\E[J_\tau[\omega]\cond\Fc_t]=J_t[\omega]+G_t[\omega](z_t)$. For complex martingales, we extend the metric~\eqref{eq:general.metric} to the Hermitian form
\begin{align}
	\Bform(X,Y)&\equiv\E[(X_\tau-X_0)(Y_\tau-Y_0)^*]\notag\\
	&=\int_0^\tau\E[\dd{X_t}\dd{Y_t}^*],
\label{eq:general.hermitian}
\end{align}
which is linear in its first argument, conjugate-linear in its second, and reduces to Eq.~\eqref{eq:general.metric} for real martingales; the second equality holds because increments over disjoint time intervals are uncorrelated, as in Eq.~\eqref{eq:general.isometry}. Splitting $J_\tau[\omega]-\E[J_\tau[\omega]]=(M_\tau[\omega]-M_0[\omega])+(M_0[\omega]-\E[M_0[\omega]])$ and using $\E[(M_\tau[\omega]-M_0[\omega])(M_0[\omega]-\E[M_0[\omega]])^*]=0$, which follows from the martingale property, gives the exact finite-time identity
\begin{equation}
	\Var(J_\tau[\omega])-\Var_0(J_\tau[\omega])=\Bform(J[\omega],J[\omega]).
\end{equation}
The filter multiplies the weights of the observable by $e^{-i\omega t}$, so that $\bw_t$ and $d_{mn}$ are replaced by $e^{-i\omega t}\bw_t$ and $e^{-i\omega t}d_{mn}$, and the martingale amplitude and the metric become
\begin{equation}
\begin{gathered}
	\bu_t[\omega]=e^{-i\omega t}\bw_t+\nabla_{\bq}G_t[\omega],\\
	\Bform(J[\omega],J'[\omega])=2\int_0^\tau\ev{\bu_t[\omega]^\top\D_t\bu'_t[\omega]^*}_t\dd{t}
\end{gathered}
\end{equation}
for diffusion processes and 
\begin{equation}
\begin{gathered}
	u_{mn}[\omega]=e^{-i\omega t}d_{mn}+G_t[\omega](m)-G_t[\omega](n),\\
	\Bform(J[\omega],J'[\omega])=\int_0^\tau\sum_{m\ne n}w_{mn}p_nu_{mn}[\omega]u'_{mn}[\omega]^*\dd{t}
\end{gathered}
\end{equation}
for jump processes.

Resolving the response in time retains information that would otherwise be lost to phase cancellation. For an admissible perturbation with representing martingale $S$, let $r_t[\omega]\dd{t}\equiv\E[\dd{M_t}[\omega]\dd{S_t}]$ be the time-resolved dynamical response, $c_t\dd{t}\equiv\E[(\dd{S_t})^2]$ its cost rate, and $\Rc_{J[\omega]}\equiv\int_0^\tau r_t[\omega]\dd{t}$, which reduces to $\Rc_J$ at $\omega=0$. Since the conditional covariances of the increments form a positive semidefinite Hermitian form at every state, the Cauchy-Schwarz inequality applied first in state space and then in time gives
\begin{align}
	&\Var(J_\tau[\omega])-\Var_0(J_\tau[\omega])\notag\\
	&\quad\ge\int_0^\tau\frac{|r_t[\omega]|^2}{c_t}\dd{t}
	\ge\frac{\qty(\int_0^\tau|r_t[\omega]|\dd{t})^2}{\Icost}
	\ge\frac{|\Rc_{J[\omega]}|^2}{\Icost}.
\label{eq:general.frequencyhierarchy}
\end{align}
Saturation of the three inequalities requires, respectively, that the representing amplitude be aligned with the martingale amplitude of the observable, that $|r_t[\omega]|/c_t$ be constant in time, and that the response have a common phase. For a source $\bg$ in a diffusion process, $r_t[\omega]=\ev*{\bu_t[\omega]\cdot\bg_t}_t$ and $c_t=(1/2)\ev*{\bg_t^\top\D_t^{-1}\bg_t}_t$; for an edge source $g_{mn}$ in a jump process and an observable with antisymmetric increments, $r_t[\omega]=\sum_{m<n}u_{mn}[\omega]g_{mn}$ and $c_t=\sum_{m<n}g_{mn}^2/a_{mn}$. The last inequality is generally not tight, because it loses information through phase cancellation. 

In the steady state, the covariance of the instantaneous current, $C_{\dot{J}}(s)\equiv\Cov(\dot{J}_{t+s},\dot{J}_t)$, depends only on the time difference $s$. The spectrum of the current is its Fourier transform,
\begin{equation}
	\mathcal{S}_{\dot{J}}(\omega)\equiv\int_{-\infty}^{\infty}e^{-i\omega s}C_{\dot{J}}(s)\dd{s}.
\label{eq:general.spectrumdef}
\end{equation}
By the Wiener-Khinchin theorem, it equals the long-time variance rate of the filtered observable, $\mathcal{S}_{\dot{J}}(\omega)=\lim_{\tau\to\infty}\Var(J_\tau[\omega])/\tau$, and it reduces to $2D_J$ at $\omega=0$ by Eq.~\eqref{eq:general.DJ}. As for the diffusivity, the spectrum is obtained from a frequency-dependent counterpart of $\Psi$, which carries the phase of the filter. Under the mixing condition of Sec.~\ref{sec:general.longtime}, the conditional future of the filtered observable, $G_t[\omega]=\int_t^\tau e^{-i\omega s}\mathcal{P}_{s-t}\Phi\dd{s}$, behaves far from the terminal time as
\begin{equation}
	G_t[\omega](z)=e^{-i\omega t}\qty[\jr_{\st}\int_0^{\tau-t}e^{-i\omega s}\dd{s}+\Psi_\omega(z)]+o(1),
\label{eq:general.Gomega}
\end{equation}
with $\Psi_\omega(z)\equiv\int_0^\infty e^{-i\omega s}\mathcal{P}_s(\Phi-\jr_{\st})(z)\dd{s}$, the phase-weighted excess accumulation expected from the state $z$. Since $\partial_s(e^{-i\omega s}\mathcal{P}_s)=e^{-i\omega s}(\Lg^\dagger-i\omega)\mathcal{P}_s$, we obtain a frequency-domain counterpart of Eq.~\eqref{eq:general.poisson},
\begin{equation}
	(\Lg^\dagger-i\omega)\Psi_\omega=-(\Phi-\jr_{\st}),\quad
	\ev{\Psi_\omega}_{\st}=0,
\label{eq:general.resolvent}
\end{equation}
which reduces to the Poisson equation~\eqref{eq:general.poisson} at $\omega=0$.

The function $\Psi_\omega$ removes the drift of the filtered observable. The increment of $e^{-i\omega t}\Psi_\omega(z_t)$ is $e^{-i\omega t}[(\Lg^\dagger-i\omega)\Psi_\omega\dd{t}+\dd{N^{\Psi_\omega}_t}]$, and that of $J_t[\omega]$ is $e^{-i\omega t}(\Phi\dd{t}+\dd{K_t})$. Adding the two and using Eq.~\eqref{eq:general.resolvent} gives $e^{-i\omega t}(\jr_{\st}\dd{t}+\dd{K_t}+\dd{N^{\Psi_\omega}_t})$, so that
\begin{align}
	&J_\tau[\omega]-\jr_{\st}\int_0^\tau e^{-i\omega t}\dd{t}\notag\\
	&=\int_0^\tau e^{-i\omega t}\dd{N^{H_\omega}_t}+\Psi_\omega(z_0)-e^{-i\omega\tau}\Psi_\omega(z_\tau),
\label{eq:general.omegadecomp}
\end{align}
where $N^{H_\omega}=K+N^{\Psi_\omega}$ is the martingale of Eq.~\eqref{eq:general.Nf} for $H_\omega(z,y)\equiv y+\Psi_\omega(z)$, with $y$ the value of the unfiltered observable $J_t$. The first term is a martingale integral; the second is a boundary term, whose variance is bounded by $4\ev*{|\Psi_\omega|^2}_{\st}$. As for $D_J$, only the martingale part survives in the variance per unit time, and the Hermitian isometry~\eqref{eq:general.hermitian} gives
\begin{equation}
	\mathcal{S}_{\dot{J}}(\omega)=\ev{\hGam[H_\omega,H_\omega^*]}_{\st},
\label{eq:general.spectrum}
\end{equation}
where $\hGam$ is extended bilinearly to complex functions, so that $\hGam[H_\omega,H_\omega^*]\dd{t}=\E[|\dd{N^{H_\omega}_t}|^2\cond\Fc_{t^-}]$. In the two realizations of Markov processes, with the corrected amplitudes $\bu_\omega\equiv\bw+\nabla_{\bq}\Psi_\omega$ and $u_{\omega,mn}\equiv d_{mn}+\Psi_\omega(m)-\Psi_\omega(n)$, the spectrum reads
\begin{equation}
	\mathcal{S}_{\dot{J}}(\omega)=2\ev{\bu_\omega^\dagger\D\bu_\omega}_{\st}
\label{eq:general.spectrumcoord.diff}
\end{equation}
for diffusion processes and
\begin{equation}
	\mathcal{S}_{\dot{J}}(\omega)=\sum_{m\ne n}w_{mn}p_n^{\st}|u_{\omega,mn}|^2
\label{eq:general.spectrumcoord.jump}
\end{equation}
for jump processes.
Here, the dagger denotes the conjugate transpose. For observables with antisymmetric increments, the spectrum becomes $\sum_{m<n}a_{mn}^{\st}|u_{\omega,mn}|^2$. At $\omega=0$, $\Psi_0=\Psi$, and Eq.~\eqref{eq:general.DJgeneral} gives $\mathcal{S}_{\dot{J}}(0)=2D_J$. As $\omega\to\infty$, $\Psi_\omega\to0$, and the spectrum approaches the direct term $\ev*{\hGam[y,y]}_{\st}$, i.e., $2\ev*{\bw^\top\D\bw}_{\st}$ or $\sum_{m\ne n}w_{mn}p_n^{\st}d_{mn}^2$. This high-frequency limit is the white-noise floor of a diffusion process or the shot-noise floor of a jump process. It arises from the delta-correlated part of the covariance of the instantaneous current, and it vanishes for state observables.

The time-resolved bound in the hierarchy~\eqref{eq:general.frequencyhierarchy} recovers the known stationary bound. Consider a stationary diffusion process with an admissible perturbation characterized by a time-independent source $\bg$. By Eq.~\eqref{eq:general.Gomega}, the martingale amplitude of the filtered observable approaches $e^{-i\omega t}\bu_\omega$ far from the terminal time, so that the time-resolved response $r_t[\omega]$ approaches $e^{-i\omega t}\chi_{\bg}(\omega)$, with the stationary susceptibility $\chi_{\bg}(\omega)\equiv\ev*{\bu_\omega\cdot\bg}_{\st}$, whereas the cost rate is $c_{\st}=(1/2)\ev*{\bg^\top\D^{-1}\bg}_{\st}$. Note that the susceptibility is the Fourier transform of the response of the mean observable to an impulsive source, $\chi_{\bg}(\omega)=\int_0^\infty e^{-i\omega t}\mathcal{K}_{\bg}(t)\dd{t}$, where $\mathcal{K}_{\bg}(t)$ is the response at time $t$ to a pulse at time $0$ (see Appendix~\ref{app:susceptibility}). We divide the first inequality in Eq.~\eqref{eq:general.frequencyhierarchy} by $\tau$ and let $\tau\to\infty$. The left-hand side tends to $\mathcal{S}_{\dot{J}}(\omega)$, because $\Var_0(J_\tau[\omega])$ remains bounded, and the time average of $|r_t[\omega]|^2/c_t$ approaches $|\chi_{\bg}(\omega)|^2/c_{\st}$. Hence, we obtain
\begin{equation}
	|\chi_{\bg}(\omega)|^2\le\mathcal{S}_{\dot{J}}(\omega)\,c_{\st}.
\label{eq:general.spectralbound}
\end{equation}
This is the finite-frequency response inequality obtained in Ref.~\cite{Dechant.2026.PRL}. For jump processes and observables with antisymmetric increments, the same argument with a time-independent edge source $g_{mn}$ gives Eq.~\eqref{eq:general.spectralbound} with $\chi_g(\omega)=\sum_{m<n}u_{\omega,mn}g_{mn}$ and $c_{\st}=\sum_{m<n}g_{mn}^2/a_{mn}^{\st}$.

The susceptibility also resolves the spectrum in state space. Let $\vb*{R}_\theta(\omega,\bz)$ be the one-sided Fourier transform of the response of $\ev*{\dot{J}_t}$ to an impulsive change of a local parameter $\vb*{\theta}$ at the state $\bz$. Since $\chi_{\bg}(\omega)$ is linear in the source, a source concentrated at $\bz$, $\bg p_{\st}=\mathsf{N}_\theta\,\delta\vb*{\theta}\,\delta(\cdot-\bz)$, gives $\vb*{R}_\theta(\omega,\bz)=\mathsf{N}_\theta^\top\bu_\omega(\bz)$, the frequency-domain counterpart of $\vb*{R}_\theta=\mathsf{N}_\theta^\top\bu_t$; likewise, $R_{\theta,mn}(\omega)=N_{\theta,mn}u_{\omega,mn}$ for an edge parameter. Wherever $\mathsf{N}_\theta$ is invertible, or $N_{\theta,mn}\ne0$, Eqs.~\eqref{eq:general.spectrumcoord.diff} and \eqref{eq:general.spectrumcoord.jump} then become
\begin{align}
	\mathcal{S}_{\dot{J}}(\omega)&=\int\vb*{R}_\theta^\dagger\mathsf{A}_\theta^{-1}\vb*{R}_\theta\dd{\bz},~
	\mathsf{A}_\theta\equiv\frac{1}{2p_{\st}}\mathsf{N}_\theta^\top\D^{-1}\mathsf{N}_\theta,
\label{eq:general.spectrumlocal.diff}\\
	\mathcal{S}_{\dot{J}}(\omega)&=\sum_{m<n}a_{mn}^{\st}\left|\frac{R_{\theta,mn}(\omega)}{N_{\theta,mn}}\right|^2,
\label{eq:general.spectrumlocal.jump}
\end{align}
where the second form applies to observables with antisymmetric increments. For the force, mobility, and temperature in overdamped dynamics (Sec.~\ref{sec:over}) and the kinetic and entropic parameters of jump processes (Sec.~\ref{sec:jump}), these are the fluctuation-response identities of Ref.~\cite{Kwon.2026.arxiv}, up to the sign convention of the Fourier transform, and polarization gives their matrix form. The diffusion form also covers underdamped dynamics. With the corresponding sources, Eq.~\eqref{eq:general.spectralbound} gives the response inequalities of Refs.~\cite{Kwon.2026.arxiv,Zheng.2026a.arxiv}. For instance, a uniform mobility rescaling in overdamped dynamics injects $\bg=\bnu_{\st}$; since $\nabla\cdot\bj_{\st}=0$, the susceptibility of a current equals its mean rate at every frequency, $\chi_{\bg}(\omega)=\jr_{\st}$, so that $\mathcal{S}_{\dot{J}}(\omega)\ge2\jr_{\st}^2/\sigma_{\st}$ with $\sigma_{\st}=\ev*{\bnu_{\st}^\top\D^{-1}\bnu_{\st}}_{\st}$.

\section{Underdamped Langevin dynamics}\label{sec:under}
\subsection{Dynamics, observables, and variance identity}
We now specialize to an underdamped particle of unit mass in $d$ dimensions, described by the Langevin equation
\begin{equation}
\begin{aligned}
	\dd{\bx_t}&=\bv_t\dd{t},\\
	\dd{\bv_t}&=[\bF_t(\bx_t)-\gamma\bv_t]\dd{t}+\sqrt{2\D}\dd{\bW_t},
\end{aligned}
\label{eq:und.dynamics}
\end{equation}
where $\gamma>0$ is the friction coefficient, $\D$ is a constant positive-definite diffusion matrix, and $\bF_t$ is a mechanical force that may be nonconservative and time dependent but does not depend on the velocity. We restrict ourselves to velocity-independent forces for simplicity; velocity-dependent forces can be included once their parity under time reversal is specified \cite{Lee.2019.PRE}. The position may live on a torus or in an unbounded domain, provided that the density decays sufficiently fast at infinity. This is the diffusion process of Sec.~\ref{sec:general.realizations} with the deterministic coordinates $\br=\bx$ and the noisy coordinates $\bq=\bv$, so that $\ba_t=\bv$ and $\bb_t=\bF_t-\gamma\bv$: the noise acts directly only on the velocity, and the position responds to it through the time integral of the velocity. In the thermal setting, $\D=\gamma\T$ with a constant positive-definite temperature matrix $\T$. Hereafter, Boltzmann's constant is set to unity.

With $\bz=(\bx,\bv)$, the forward and backward generators read
\begin{align}
	\Lg_t p&=-\nabla_{\bx}\cdot(\bv p)-\nabla_{\bv}\cdot[(\bF_t-\gamma\bv)p]+\D:\nabla_{\bv}\nabla_{\bv}p,\notag\\
	\Lg_t^\dagger a&=\bv\cdot\nabla_{\bx}a+(\bF_t-\gamma\bv)\cdot\nabla_{\bv}a+\D:\nabla_{\bv}\nabla_{\bv}a,
\label{eq:und.generator}
\end{align}
respectively, and the probability currents in position and velocity space are $\bj_t^{\bx}\equiv\bv p_t$ and $\bj_t^{\bv}\equiv(\bF_t-\gamma\bv)p_t-\D\nabla_{\bv}p_t$. Friction and noise then combine into the irreversible part of the velocity current, 
\begin{equation}
	\bj_t^{\irr}\equiv-\gamma\bv p_t-\D\nabla_{\bv}p_t,\quad\bnu_t^{\irr}\equiv\bj_t^{\irr}/p_t,
\label{eq:und.irrcurr}
\end{equation}
whose square, weighted by the inverse diffusion matrix, gives the total entropy production \cite{Spinney.2012.PRE,Seifert.2012.RPP},
\begin{equation}
	\Sigma_\tau=\int_0^\tau\ev{(\bnu_t^{\irr})^\top\D^{-1}\bnu_t^{\irr}}_t\dd{t}.
\label{eq:und.entropy}
\end{equation}

We consider observables that accumulate both during residence and along the velocity,
\begin{equation}
	J_\tau=\int_0^\tau\phi_t(\bz_t)\dd{t}+\int_0^\tau\bw_t(\bz_t)\circ\dd{\bv_t}.
\label{eq:und.J}
\end{equation}
Their effective rate is the general expression of Sec.~\ref{sec:general.realizations} with $\bb_t=\bF_t-\gamma\bv$, namely $\Phi_t=\phi_t+\bw_t\cdot(\bF_t-\gamma\bv)+\D:\nabla_{\bv}\bw_t$. A position current $\int\vb*{\alpha}_t(\bz_t)\cdot\dd{\bx_t}$ enters through the residence weight, $\phi_t=\vb*{\alpha}_t\cdot\bv$ and $\bw_t=0$, because the position increment $\dd{\bx_t}=\bv_t\dd{t}$ carries no noise. A velocity-space line integral, by contrast, requires $\bw_t\ne 0$; for example, $\bw_t=\bv$ gives the kinetic-energy increment $\int_0^\tau\bv_t\circ\dd{\bv_t}=(\|\bv_\tau\|^2-\|\bv_0\|^2)/2$.

Because the noise acts only on the velocity, only the velocity gradient of the conditional future enters the martingale amplitude~\eqref{eq:general.diffamp},
\begin{align}
	\dd{M_t}=\bu_t^\top\sqrt{2\D}\dd{\bW_t},\quad\bu_t&=\bw_t+\nabla_{\bv}G_t,
\label{eq:und.kernel}
\end{align}
and the dynamical variance~\eqref{eq:general.diffmet} becomes
\begin{equation}
	\Vd=2\int_0^\tau\ev{\bu_t^\top\D\bu_t}_t\dd{t}.
\label{eq:und.variance}
\end{equation}
There is no separate contribution from position noise. A position-dependent observable nevertheless generally has a nonzero amplitude, because its expected future value depends on the present velocity.

\subsection{Fluctuation-response relations}
We specify a perturbation of the underdamped dynamics by its direction $(\delta\bb_t,\delta\D)$, i.e., $\bb_t\to\bb_t+\eps\delta\bb_t$ and $\D\to\D+\eps\delta\D$, where $\bb_t=\bF_t-\gamma\bv$ is the velocity drift. This covers changes of the force, $\delta\bb_t=\delta\bF_t$; of the friction, $\delta\bb_t=-\delta\gamma\bv$ together with $\delta\D=\delta\gamma\T$ at fixed temperature; and of the temperature, $\delta\D=\gamma\delta\T$. For a constant $\delta\D$, the general source current~\eqref{eq:general.source} reduces to a current in velocity space,
\begin{equation}
\begin{gathered}
	\delta\bj_{t,\mathrm{src}}^{\bv}=\delta\bb_tp_t-\delta\D\nabla_{\bv}p_t,\\
	\quad \delta\Lg_t p_t=-\nabla_{\bv}\cdot\delta\bj_{t,\mathrm{src}}^{\bv},
\end{gathered}
\label{eq:und.source}
\end{equation}
and Eq.~\eqref{eq:general.diffresponse} gives the dynamical response at fixed initial density,
\begin{equation}
	\delta\E[J_\tau]=\int_0^\tau\inprod{\bu_t}{\delta\bj_{t,\mathrm{src}}^{\bv}}\dd{t}.
\label{eq:und.response}
\end{equation}
The martingale amplitude that generates the fluctuations in Eq.~\eqref{eq:und.variance} is thus also the susceptibility of the mean observable to a current injected in velocity space: the local response $\vb*{R}_{\mathrm{src}}(t,\bz)\equiv\delta\E[J_\tau]/\delta\bj_{t,\mathrm{src}}^{\bv}(\bz)$ equals $\bu_t(\bz)$. Combining this with Eqs.~\eqref{eq:general.var} and \eqref{eq:und.variance} expresses the preparation-subtracted variance entirely through local responses,
\begin{equation}
	\Var(J_\tau)-\Var_0(J_\tau)=2\int_0^\tau\ev{\vb*{R}_{\mathrm{src}}^\top\D\vb*{R}_{\mathrm{src}}}_t\dd{t},
\label{eq:und.FRR}
\end{equation}
which is the underdamped form of Eq.~\eqref{eq:general.FRRdiff}. The left-hand side follows from the statistics of $J_\tau$ together with its conditional mean given the initial state. The right-hand side involves responses to sources of velocity current localized in phase space, which are difficult to realize directly; Eq.~\eqref{eq:und.FRR} is therefore mainly a structural identity, whereas the bounds derived from it below involve only responses to global perturbations, such as a change of friction. No current can be injected along the position, because the kinematics $\dd{\bx_t}=\bv_t\dd{t}$ carries no noise.

Changes of friction and temperature alter $\D$, which can be read off from a single velocity path: the matrix of quadratic covariations over $[0,t]$, $\sum_k\Delta\bv_k\Delta\bv_k^\top$, converges to $2\D t$ along almost every path as the time step goes to zero. The perturbed and unperturbed path measures are therefore mutually singular, and no score exists. Equation~\eqref{eq:und.response} nevertheless holds, because it follows from the forward equation and the effective rate rather than from the path measure. As shown in Sec.~\ref{sec:general.realizations}, such a perturbation is admissible whenever its cost is finite, and its representing martingale is the dynamical score of the auxiliary drift perturbation $\bg_t=\delta\bj_{t,\mathrm{src}}^{\bv}/p_t$, which injects the same source current. Friction and temperature responses can therefore be treated in the same way as force responses.

For the source $\bg_t=\delta\bj_{t,\mathrm{src}}^{\bv}/p_t$, the response and the cost of Sec.~\ref{sec:general.realizations} read
\begin{gather}
	\Rc[\bg]=\int_0^\tau\ev{\bu_t\cdot\bg_t}_t\dd{t},\\
	\Ac[\bg]=\frac{1}{2}\int_0^\tau\ev{\bg_t^\top\D^{-1}\bg_t}_t\dd{t},
\label{eq:und.cost}
\end{gather}
and the fluctuation-response inequality~\eqref{eq:general.CSdual1} and the variational formula~\eqref{eq:general.CSdual2} become
\begin{gather}
	\Rc[\bg]^2\le[\Var(J_\tau)-\Var_0(J_\tau)]\Ac[\bg],\label{eq:und.bounds}\\
	\Var(J_\tau)-\Var_0(J_\tau)=\max_{\bg}\{2\Rc[\bg]-\Ac[\bg]\}.
\end{gather}
Equality in Eq.~\eqref{eq:und.bounds} holds if and only if $\bg_t\propto\D\bu_t$, and the maximum is attained when $\bg_t=2\D\bu_t$. These relations hold for arbitrary time- and state-dependent sources; restricting to the perturbations available in an experiment can only lower the variational estimate.

\subsection{Friction-response thermodynamic uncertainty relation}
For a suitable perturbation, the fluctuation-response inequality becomes a thermodynamic uncertainty relation. In the thermal setting, $\D=\gamma\T$, a change of friction at fixed temperature injects precisely the irreversible probability current. Under $\gamma\to(1+\eps)\gamma$, the fluctuation-dissipation relation requires $\D\to(1+\eps)\D$, so that the perturbation has the direction $\delta\bb_t=-\gamma\bv$ and $\delta\D=\D$, and the two terms in Eq.~\eqref{eq:und.source} combine into
\begin{equation}
	\delta\bj_{t,\mathrm{src}}^{\bv}=-\gamma\bv p_t-\D\nabla_{\bv}p_t=\bnu_t^{\irr}p_t.
\label{eq:und.frictionsource}
\end{equation}
Here, the mechanical force, the temperature, the observation interval, the initial density, and the observable fields are held fixed, so that $\delta\E[J_\tau]=\gamma\partial_\gamma\E[J_\tau]$. The response and the cost of this perturbation are therefore
\begin{equation}
	\Rgamma\equiv\gamma\partial_\gamma\E[J_\tau]=\Rc[\bnu^{\irr}],\quad
	\Ac[\bnu^{\irr}]=\frac{\Sigma_\tau}{2}.
\label{eq:und.frictionresponse}
\end{equation}
The cost is half the entropy production, because the injected current is exactly the irreversible current, whose squared norm in the inverse noise metric is $\Sigma_\tau$ [Eq.~\eqref{eq:und.entropy}]. Inserting Eq.~\eqref{eq:und.frictionresponse} into Eq.~\eqref{eq:und.bounds} yields the friction-response thermodynamic uncertainty relation
\begin{equation}
	\frac{\Var(J_\tau)-\Var_0(J_\tau)}{\Rgamma^2}\ge\frac{2}{\Sigma_\tau}
\label{eq:und.TUR}
\end{equation}
for $\Rgamma\ne0$; the equivalent product form, $[\Var(J_\tau)-\Var_0(J_\tau)]\Sigma_\tau\ge2\Rgamma^2$, also covers a vanishing response. The bound holds at arbitrary finite times, for any initial density, and for arbitrary time-dependent driving. It is tight: within the observable class~\eqref{eq:und.J}, an observable that saturates it exists at every finite time (Sec.~\ref{sec:under.saturation}). It sharpens the full-variance friction-response inequality of Ref.~\cite{Kwon.2022.NJP} by removing the preparation variance $\Var_0(J_\tau)$, and it extends that inequality to the observable class~\eqref{eq:und.J}. In the overdamped limit, it reduces to the mobility-response bound~\eqref{eq:od.mobility} for observables with a well-defined limit and, for pure position currents, recovers the TUR obtained in Ref.~\cite{Koyuk.2020.PRL} for time-dependent driving.

If the initial density itself depends on the friction, its change contributes a separate response, which must not be attributed to dissipation during the observation. For an initial density $p_0^\gamma$ that depends on $\gamma$, we distinguish the total derivative $\gamma\dd{\E[J_\tau]}/\dd{\gamma}$ from the response $\Rgamma$ at fixed preparation and define
\begin{gather}
	q_0\equiv\gamma\partial_\gamma\ln p_0^\gamma,\quad I_0\equiv\ev{q_0^2}_0,\\
	B_\gamma\equiv\gamma\int G_0\partial_\gamma p_0^\gamma\dd{\bz}=\Cov_0(G_0,q_0),
\label{eq:und.preparation}
\end{gather}
where the last equality uses $\ev*{q_0}_0=0$, which follows from normalization. For fixed observable fields, the total response splits as $\gamma\dd{\E[J_\tau]}/\dd{\gamma}=\Rgamma+B_\gamma$, which is Eq.~\eqref{eq:general.channels} for this perturbation. The bound~\eqref{eq:und.TUR} applies to $\Rgamma$, and the Cauchy-Schwarz inequality gives $B_\gamma^2\le\Var_0(J_\tau)I_0$, so that
\begin{equation}
	\qty|\gamma\frac{\dd{\E[J_\tau]}}{\dd{\gamma}}|\le\sqrt{\frac{\Vd\Sigma_\tau}{2}}+\sqrt{\Var_0(J_\tau)I_0}.
\label{eq:und.prepbound}
\end{equation}
A further application of the Cauchy-Schwarz inequality, $(\sqrt{ab}+\sqrt{cd})^2\le(a+c)(b+d)$, yields
\begin{equation}
	\qty(\gamma\frac{\dd{\E[J_\tau]}}{\dd{\gamma}})^2\le\Var(J_\tau)\left(\frac{\Sigma_\tau}{2}+I_0\right).
\label{eq:und.fullbound}
\end{equation}
Both preparation terms are relevant at finite times. For autonomous mixing dynamics, $B_\gamma$ and $I_0$ remain bounded as $\tau\to\infty$, whereas the response of an observable and the entropy production grow linearly in time, so that the preparation terms become negligible at long times. The term $I_0$ is nonzero, for example, when the system is prepared in the stationary state of each friction value and this state depends on the friction, as in nonequilibrium steady states; in equilibrium, the thermal Gibbs density does not depend on $\gamma$, and $I_0=0$.

Operationally, the bound~\eqref{eq:und.TUR} requires the preparation to be controlled separately from the response measurement: the same initial ensemble is used for the reference and perturbed runs, the bath temperature is held fixed, and the friction is changed only during the observation interval. Recording the initial state together with $J_\tau$ allows one to estimate $G_0(\bz_0)=\E[J_\tau\cond\bz_0]$ and hence $\Var_0(J_\tau)$. In practice, the initial state may be recorded only partially, for example only some of its degrees of freedom, such as the position but not the velocity, or only with finite resolution. Let $\by_0$ denote the recorded information, which is a function of $\bz_0$. By the tower property, the conditional mean given this information is an average of the full one, $\E[J_\tau\cond\by_0]=\E[G_0(\bz_0)\cond\by_0]$. Applying the law of total variance to $G_0(\bz_0)$ then gives
\begin{equation}
	\Var_0(J_\tau)=\Var(\E[J_\tau\cond\by_0])+\E[\Var(G_0(\bz_0)\cond\by_0)],
\end{equation}
so that subtracting $\Var(\E[J_\tau\cond\by_0])$ instead of $\Var_0(J_\tau)$ leaves
\begin{align}
	&\Var(J_\tau)-\Var(\E[J_\tau\cond\by_0])\notag\\
	&\quad=\Vd+\E[\Var(G_0(\bz_0)\cond\by_0)]\ge\frac{2\Rgamma^2}{\Sigma_\tau}.
\end{align}
The bound therefore remains valid and is weaker than Eq.~\eqref{eq:und.TUR} only by the part of the preparation variance that the recorded information does not resolve. It interpolates between the full-variance bound, obtained when nothing is recorded, and Eq.~\eqref{eq:und.TUR}, obtained when the full initial state is recorded; each additional recorded quantity can only tighten it. Without any conditional information, only the full-variance inequality~\eqref{eq:und.fullbound} is available, and it gives a weaker estimate of the dissipation: the subtraction uses information about the preparation that cannot be inferred from the distribution of $J_\tau$ alone.

\subsubsection{Finite-time saturation and exact dissipation inference}\label{sec:under.saturation}
We show that the friction-response TUR~\eqref{eq:und.TUR} can always be saturated at finite times, allowing dissipation to be inferred exactly. Equality in Eq.~\eqref{eq:und.TUR} requires $\bu_t\propto\D^{-1}\bnu_t^{\irr}$, with a proportionality constant independent of time and state. Within the full observable class of Eq.~\eqref{eq:und.J}, this condition can be fulfilled at every time. To construct the corresponding observable, we introduce
\begin{equation}
	\Lambda_t(\bz)\equiv -\ln\qty[p_t(\bz)\exp\qty(\frac{1}{2}\bv^\top\T^{-1}\bv)].
\label{eq:und.Lambda}
\end{equation}
The fluctuation-dissipation relation $\D=\gamma\T$ implies $\D\nabla_{\bv}\Lambda_t=\bnu_t^{\irr}$, so the required martingale amplitude is a velocity gradient, which can be realized consistently with the terminal condition on $G_t$.

A saturating observable is obtained by separating the terminal gradient from the remaining future contribution:
\begin{equation}
\begin{gathered}
	\bw^{\mathrm{opt}}=\nabla_{\bv}\Lambda_\tau,\\
	G_t^{\mathrm{opt}}=\Lambda_t-\Lambda_\tau,\\
	\Phi_t^{\mathrm{opt}}=-\partial_tG_t^{\mathrm{opt}}-\Lg_t^\dagger G_t^{\mathrm{opt}},\\
	\phi_t^{\mathrm{opt}}=\Phi_t^{\mathrm{opt}}-\bw^{\mathrm{opt}}\cdot(\bF_t-\gamma\bv)-\D:\nabla_{\bv}\bw^{\mathrm{opt}}.
\end{gathered}
\label{eq:und.optimal}
\end{equation}
The backward equation and the terminal condition hold by construction, so that $\bu_t^{\mathrm{opt}}=\nabla_{\bv}\Lambda_t=\D^{-1}\bnu_t^{\irr}$. The velocity weight $\bw^{\mathrm{opt}}$ is time independent, because it is fixed by the terminal density, whereas the residence weight $\phi_t^{\mathrm{opt}}$ depends explicitly on time through $p_t$ and the driving. In a steady state under autonomous driving, $\Lambda_t$ does not depend on time, so that $G_t^{\mathrm{opt}}=0$ and both weights are time independent. All quantities in Eq.~\eqref{eq:und.optimal} are computed from the reference process and kept fixed when the response is measured.

For this observable, Eqs.~\eqref{eq:und.variance} and \eqref{eq:und.frictionresponse} give $\Vd^{\mathrm{opt}}=2\Sigma_\tau$ and $\Rgamma^{\mathrm{opt}}=\Sigma_\tau$. The friction-response TUR therefore becomes an exact variational principle for the entropy production,
\begin{equation}
	\Sigma_\tau=\sup_{\phi,\bw}\frac{2(\gamma\partial_\gamma\E[J_\tau])^2}{\Var(J_\tau)-\Var_0(J_\tau)}.
\label{eq:und.entropyvariational}
\end{equation}
The supremum is attained by the observable~\eqref{eq:und.optimal}, provided that the required regularity and integrability conditions are satisfied. This construction establishes the attainability of the bound; in practice, a saturating observable can be obtained by optimization using trajectory data. Restricting the class of observables may yield an underestimate of $\Sigma_\tau$.

Saturation within the full observable class does not by itself require subtracting the preparation variance, because the full-variance bound is also saturated by a suitable observable with an explicitly time-dependent velocity weight. If the velocity weight may depend explicitly on time, the choice $G_t\equiv0$, $\bw_t=\nabla_{\bv}\Lambda_t$, and $\phi_t=-\bw_t\cdot(\bF_t-\gamma\bv)-\D:\nabla_{\bv}\bw_t$, i.e., $\Phi_t=0$, yields the same amplitude $\bu_t=\D^{-1}\bnu_t^{\irr}$. This observable is a pure martingale, $J_\tau=M_\tau-M_0$, with $\Var_0(J_\tau)=0$, and it saturates the full-variance bound as well. Nevertheless, the subtraction is important in two respects. First, it tightens the bound by $\Var_0(J_\tau)$ for every given observable. Second, it permits saturation with a time-independent velocity weight, as in Eq.~\eqref{eq:und.optimal}. For such a weight, saturation fixes $\nabla_{\bv}G_0=\nabla_{\bv}(\Lambda_0-\Lambda_\tau)=\nabla_{\bv}\ln p_\tau-\nabla_{\bv}\ln p_0$, which is nonzero unless the velocity dependence of $\ln p_t$ is the same at $t=0$ and $t=\tau$; the preparation variance is then nonzero, and only the subtracted bound is saturated.

\begin{figure}[t]
\centering\includegraphics[width=1.0\linewidth]{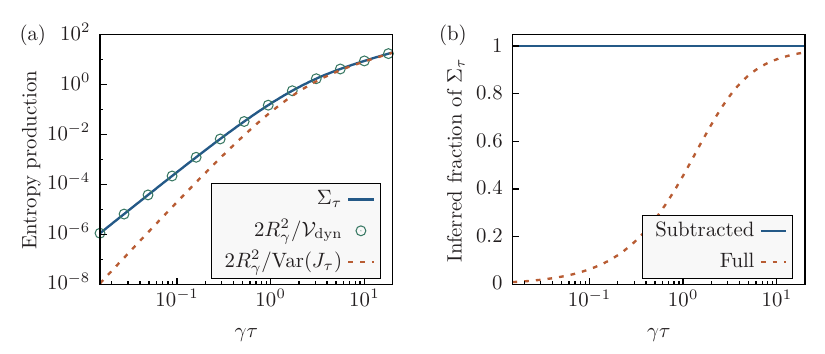}
\caption{Finite-time saturation of the friction-response TUR for a driven rotor relaxing from equilibrium, with $\gamma=T=f=1$. The initial density is uniform in the angle and Gaussian in the velocity, $v_0\sim\mathscr{N}(0,1)$, and $\bar{v}_\tau=1-e^{-\tau}$ is the mean velocity. (a) For the saturating observable of Eq.~\eqref{eq:und.rotorJ}, the preparation-subtracted estimate $2R_\gamma^2/\Vd$ equals $\Sigma_\tau$ at every time (circles: exact evaluation), whereas the full-variance estimate $2R_\gamma^2/\Var(J_\tau)=2\Sigma_\tau^2/(2\Sigma_\tau+\bar{v}_\tau^2)$ (dashed line) falls below it. (b) Fraction of $\Sigma_\tau$ recovered by each estimate. The full-variance fraction vanishes as $2\tau/3$ at short times and only approaches unity at long times.}
\label{fig:inference}
\end{figure}

We demonstrate the saturation of the friction-response TUR using an example with a nonzero preparation contribution. Consider a rotor with angle $\alpha_t$ and angular velocity $v_t$, 
\begin{equation}
\begin{gathered}
	\dd{\alpha_t}=v_t\dd{t},\\
	\dd{v_t}=(f-\gamma v_t)\dd{t}+\sqrt{2\gamma T}\dd{W_t},
\end{gathered}
\end{equation}
which starts in equilibrium and relaxes after a constant force $f$ is switched on at $t=0$. We set $\gamma=T=f=1$ and take a uniformly distributed initial angle and $v_0\sim\mathscr{N}(0,1)$. The angle remains uniformly distributed, and the velocity distribution remains Gaussian; its mean obeys $\dot{\bar{v}}_t=1-\bar{v}_t$ with $\bar{v}_0=0$ and its variance $s_t$ obeys $\dot{s}_t=2-2s_t$ with $s_0=1$. Hence, $p_t(v)$ has unit variance and mean $\bar{v}_t=1-e^{-t}$. Since $\partial_vp_t=-(v-\bar{v}_t)p_t$, the irreversible current is $j_t^{\irr}=-vp_t-\partial_vp_t=-\bar{v}_tp_t$, so that $\nu_t^{\irr}=-\bar{v}_t$ is uniform in phase space and
\begin{equation}
	\Sigma_\tau=\int_0^\tau\bar{v}_t^2\dd{t}=\tau-2(1-e^{-\tau})+\tfrac12(1-e^{-2\tau}).
\label{eq:und.rotorSigma}
\end{equation}
The optimal observable follows from Eq.~\eqref{eq:und.optimal}. With $\Lambda_t=-\ln p_t-v^2/2=-\bar{v}_tv+\bar{v}_t^2/2$ up to a constant, it has $w=-\bar{v}_\tau$, $G_t=-(\bar{v}_t-\bar{v}_\tau)v+(\bar{v}_t^2-\bar{v}_\tau^2)/2$, and $\phi_t=(1-2\bar{v}_t)v+\bar{v}_t^2$. The observable explicitly reads
\begin{equation}
	J_\tau=\int_0^\tau[(1-2\bar{v}_t)v_t+\bar{v}_t^2]\dd{t}-\int_0^\tau\bar{v}_\tau\dd{v_t}.
\label{eq:und.rotorJ}
\end{equation}
The martingale amplitude is $u_t=w+\partial_vG_t=-\bar{v}_t=\nu_t^{\irr}/D$, as required for saturation, so that $\Vd=2\Sigma_\tau$ and $\Rgamma=\Sigma_\tau$. The preparation variance is nonvanishing because the conditional mean depends on the initial velocity, $G_0(v_0)=\bar{v}_\tau v_0-\bar{v}_\tau^2/2$, so that $\Var_0(J_\tau)=\bar{v}_\tau^2$ and $\Var(J_\tau)=2\Sigma_\tau+\bar{v}_\tau^2$. The subtracted bound~\eqref{eq:und.TUR} therefore returns the entropy production exactly, 
\begin{equation}
	\frac{2\Rgamma^2}{\Var(J_\tau)-\Var_0(J_\tau)}=\Sigma_\tau,
\end{equation}
whereas the full-variance bound gives only a lower bound on it,
\begin{equation}
	\frac{2\Rgamma^2}{\Var(J_\tau^{\mathrm{opt}})}=\frac{2\Sigma_\tau^2}{2\Sigma_\tau+\bar{v}_\tau^2}<\Sigma_\tau.
\end{equation}
Figure~\ref{fig:inference} compares the two estimates. At short times $\tau\ll 1$, $\bar{v}_\tau\simeq\tau$ and $\Sigma_\tau\simeq\tau^3/3$, so that the full-variance estimate for this observable recovers only the fraction $2\tau/3$ of the entropy production. The pure-martingale observable with the time-dependent weight $w_t=-\bar{v}_t$ saturates even the full-variance bound, but the observable~\eqref{eq:und.rotorJ}, with its time-independent weight, saturates only the subtracted one. At long times, $\Var_0(J_\tau)$ remains bounded while $\Sigma_\tau$ grows linearly, and the full-variance estimate recovers an asymptotically unit fraction of the entropy production, although its shortfall $\Sigma_\tau\bar{v}_\tau^2/(2\Sigma_\tau+\bar{v}_\tau^2)$ tends to $1/2$.

\subsubsection{Short-time asymptotics}\label{sec:under.short}
We highlight the role of the preparation variance in the friction-response TUR by examining the short-time regime. Consider autonomous dynamics and the observable $J_\tau=\int_0^\tau\phi(\bz_t)\dd{t}$ without a direct velocity weight (i.e., $\bw=0$). Expanding the solution $G_t=\int_0^{\tau-t}e^{s\Lg^\dagger}\Phi\dd{s}$ in terms of the backward generator gives, for a smooth force and weight, to any finite order $N$,
\begin{equation}
	G_t=\sum_{n=0}^{N-1}\frac{(\tau-t)^{n+1}}{(n+1)!}(\Lg^\dagger)^n\phi+O((\tau-t)^{N+1}),
\label{eq:und.shortG}
\end{equation}
where we used $\Phi=\phi$, which holds for $\bw=0$. Let $k\ge1$ be the smallest index for which
\begin{equation}
	\btheta=\nabla_{\bv}(\Lg^\dagger)^{k-1}\phi\ne0.
\label{eq:und.k}
\end{equation}
A velocity-dependent weight has $k=1$. If $\nabla_{\bv}\phi=0$, then $\phi=\phi(\bx)$ and $\nabla_{\bv}\Lg^\dagger\phi=\nabla_{\bx}\phi$, so that a nonconstant position-dependent weight has $k=2$. For the dynamics considered here, these are the only two cases.

The first nonvanishing gradient $\btheta$ controls both the dynamical variance and the friction response. With $\sigma_0\equiv\ev*{(\bnu_0^{\irr})^\top\D^{-1}\bnu_0^{\irr}}_0$, we find
\begin{align}
	\Var_0(J_\tau)&=\tau^2\Var_{p_0}(\phi)+O(\tau^3),\label{eq:und.shortexp1}\\
	\Vd&=\frac{2\tau^{2k+1}}{(k!)^2(2k+1)}\ev{\btheta^\top\D\btheta}_0+O(\tau^{2k+2}),\\
	\Rgamma&=\frac{\tau^{k+1}}{(k+1)!}\ev{\bnu_0^{\irr}\cdot\btheta}_0+O(\tau^{k+2}),\\
	\Sigma_\tau&=\tau\sigma_0+O(\tau^2).\label{eq:und.shortexp2}
\end{align}
Hereafter, the subscript $0$ denotes evaluation at $t=0$. For $\Var_{p_0}(\phi)>0$, the preparation variance is of lower order in $\tau$ than the dynamical variance: particles prepared in different states accumulate different values of the observable immediately, before the thermal noise has had time to act.

Subtracting the preparation variance therefore turns a divergent short-time uncertainty factor into a finite one. To quantify how tight the two bounds are, we define the quality factors
\begin{equation}
	\Qc_{\mathrm{sub}}\equiv\frac{\Vd\Sigma_\tau}{2\Rgamma^2},\quad
	\Qc_{\mathrm{full}}\equiv\frac{\Var(J_\tau)\Sigma_\tau}{2\Rgamma^2},
\label{eq:und.Qshort}
\end{equation}
which satisfy $\Qc_{\mathrm{full}}\ge\Qc_{\mathrm{sub}}\ge1$ by Eq.~\eqref{eq:und.TUR} and $\Var_0(J_\tau)\ge0$, with $\Qc_{\mathrm{sub}}=1$ when the subtracted bound is saturated. Their short-time behavior follows from the orders in Eqs.~\eqref{eq:und.shortexp1}--\eqref{eq:und.shortexp2}. In $\Qc_{\mathrm{sub}}$, the dynamical variance, of order $\tau^{2k+1}$, and the entropy production, of order $\tau$, together match the squared response, of order $\tau^{2k+2}$, so that the powers of $\tau$ cancel. In $\Qc_{\mathrm{full}}$, by contrast, the variance is dominated by the preparation variance, of order $\tau^2$, which exceeds the dynamical variance by the factor $\tau^{-(2k-1)}$. If $\ev*{\bnu_0^{\irr}\cdot\btheta}_0\ne0$ and $\Var_{p_0}(\phi)>0$, we obtain
\begin{align}
	\lim_{\tau\to0}\Qc_{\mathrm{sub}}
	&=\frac{(k+1)^2}{2k+1}
	\frac{\ev{\btheta^\top\D\btheta}_0\sigma_0}
	{\ev{\bnu_0^{\irr}\cdot\btheta}_0^2}
	\ge\frac{(k+1)^2}{2k+1},\label{eq:und.Qsub}\\
	\Qc_{\mathrm{full}}&\simeq
	\frac{[(k+1)!]^2\Var_{p_0}(\phi)\sigma_0}
	{2\ev{\bnu_0^{\irr}\cdot\btheta}_0^2}\,\tau^{-(2k-1)},
\label{eq:und.Qfull}
\end{align}
where the inequality follows from the Cauchy-Schwarz inequality $\ev*{\bnu_0^{\irr}\cdot\btheta}_0^2\le\ev*{\btheta^\top\D\btheta}_0\sigma_0$. The two quality factors thus behave in qualitatively different ways. The quality factor of the full-variance bound diverges as $\tau\to0$, as $\tau^{-1}$ for $k=1$ and as $\tau^{-3}$ for $k=2$: the full-variance bound becomes arbitrarily loose, and the entropy production it infers, $2\Rgamma^2/\Var(J_\tau)=\Sigma_\tau/\Qc_{\mathrm{full}}$, vanishes as $\tau^{2k}$, much faster than $\Sigma_\tau$ itself. The quality factor of the subtracted bound, in contrast, remains finite, so that the subtracted bound recovers a fixed fraction of the entropy production arbitrarily close to $\tau=0$. This fraction is at most $(2k+1)/(k+1)^2$, i.e., $3/4$ for $k=1$ and $5/9$ for $k=2$. Observables without a direct velocity weight therefore cannot saturate the friction-response TUR at short times, in contrast to the full observable class used in Sec.~\ref{sec:under.saturation}, and in contrast to overdamped steady states, where suitable currents saturate the conventional TUR at short times~\cite{Manikandan.2020.PRL,Vu.2020.PRE,Otsubo.2020.PRE}. If the leading response vanishes by symmetry, or if the preparation is singular or deterministic, the next nonzero orders must be examined.

Both cases, $k=1$ and $k=2$, can be illustrated using an exactly solvable Gaussian example. Consider a particle moving on a line with $\gamma=T=f=1$, governed by $\dd{x_t}=v_t\dd{t}$ and $\dd{v_t}=(1-v_t)\dd{t}+\sqrt2\dd{W_t}$. The initial position and velocity are independently sampled from $x_0\sim\mathscr{N}(0,1)$ and the stationary velocity distribution $v_0\sim\mathscr{N}(1,1)$. The conditional mean velocity evolves as $\E[v_s\cond v_0]=1+(v_0-1)e^{-s}$, giving rise to two functions that recur throughout the analysis:
\begin{equation}
	h_s\equiv1-e^{-s},\quad \tilde{h}_s\equiv\int_0^sh_r\dd{r}=s-h_s,
\label{eq:und.hfunctions}
\end{equation}
which behave as $h_s\simeq s$ and $\tilde{h}_s\simeq s^2/2$ at short times. The position spreads and becomes correlated with the velocity: $(x_t,v_t)$ is jointly Gaussian with $\Var(x_t)=1+2\tilde{h}_t$, $\Var(v_t)=1$, and $\Cov(x_t,v_t)=h_t$. For a Gaussian density with covariance matrix $C_t$, $\nu_t^{\irr}=-v-\partial_v\ln p_t$ has mean $-1$ and variance $(C_t^{-1})_{vv}-1$, so that the entropy production is
\begin{equation}
	\Sigma_\tau=\int_0^\tau\sigma_t\dd{t},\quad
	\sigma_t=(C_t^{-1})_{vv}=\frac{1+2\tilde{h}_t}{1+2\tilde{h}_t-h_t^2}.
\label{eq:und.gaussSigma}
\end{equation}
For $k=1$, consider the displacement $J_\tau=\int_0^\tau v_t\dd{t}$, i.e., $\phi=v$. The conditional future is $G_t=(\tau-t)+(v-1)h_{\tau-t}$, so that $u_t=h_{\tau-t}$ and $G_0=\tau+(v_0-1)h_\tau$. Since $u_t$ does not depend on the state and $\ev*{\nu_t^{\irr}}_t=-1$, Eqs.~\eqref{eq:und.variance} and \eqref{eq:und.response} give
\begin{equation}
	\Vd=2\int_0^\tau h_s^2\dd{s},~ \Var_0(J_\tau)=h_\tau^2,~
	\Rgamma=-\int_0^\tau h_s\dd{s}.
\label{eq:und.gaussk1}
\end{equation}
For $k=2$, consider the time-integrated position $J_\tau=\int_0^\tau x_t\dd{t}$, i.e., $\phi=x$. Now, $G_t=(\tau-t)x+(\tau-t)^2/2+(v-1)\tilde{h}_{\tau-t}$, so that $u_t=\tilde{h}_{\tau-t}$ and $G_0=\tau x_0+\tau^2/2+(v_0-1)\tilde{h}_\tau$. Hence, $h$ is replaced by $\tilde{h}$ in $\Vd$ and $\Rgamma$, and $\Var_0(J_\tau)=\tau^2+\tilde{h}_\tau^2$.

\begin{figure}[t]
 \centering\includegraphics[width=1.0\linewidth]{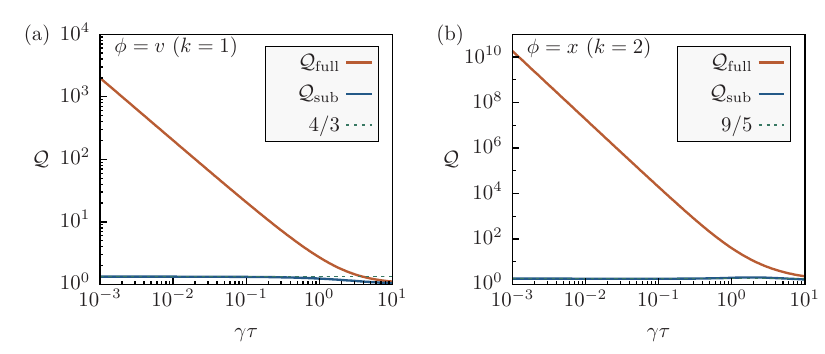}
 \caption{Finite-time uncertainty factors and their short-time limits, with and without preparation subtraction, for the exactly solvable Gaussian example with $\gamma=T=f=1$ and independent $x_0\sim\mathscr{N}(0,1)$ and $v_0\sim\mathscr{N}(1,1)$. (a) Velocity observable $\phi=v$ ($k=1$): $\Qc_{\mathrm{sub}}\to4/3$, whereas $\Qc_{\mathrm{full}}\propto\tau^{-1}$. (b) Position observable $\phi=x$ ($k=2$): $\Qc_{\mathrm{sub}}\to9/5$, whereas $\Qc_{\mathrm{full}}\propto\tau^{-3}$. Dotted lines mark the short-time limits. All curves are exact evaluations of Gaussian covariance integrals.}
 \label{fig:short}
\end{figure}

For both observables, $\vartheta=1$, $\ev*{\nu_0^{\irr}\vartheta}_0=-1$, and $\sigma_0=1$, so that the Cauchy-Schwarz inequality in Eq.~\eqref{eq:und.Qsub} is an equality. The subtracted quality factor therefore tends to its lower bound, $4/3$ for $k=1$ and $9/5$ for $k=2$, whereas Eq.~\eqref{eq:und.Qfull} with $\Var_{p_0}(\phi)=1$ gives $\Qc_{\mathrm{full}}\simeq2/\tau$ and $18/\tau^3$, respectively. Figure~\ref{fig:short} confirms these finite limits of $\Qc_{\mathrm{sub}}$ and the divergences of $\Qc_{\mathrm{full}}$.

\subsubsection{Multivariate friction-response relation}
Several correlated observables can resolve the dissipation better than any single one~\cite{Dechant.2019.JPA,Vu.2020.PRE}. In underdamped dynamics, the blocks of the Gram matrix behind Eq.~\eqref{eq:general.matrix} are explicit: for observables $J_\tau^\alpha$ with amplitudes $\bu_t^\alpha$ and sources $\bg_t^i$,
\begin{align}
	\Cm_{\alpha\beta}-\Cm_{0,\alpha\beta}&=2\int_0^\tau\ev{(\bu_t^\alpha)^\top\D\bu_t^\beta}_t\dd{t},\\
	\Rm_{\alpha i}&=\int_0^\tau\ev{\bu_t^\alpha\cdot\bg_t^i}_t\dd{t},\\
	\Imat_{ij}&=\frac{1}{2}\int_0^\tau\ev{(\bg_t^i)^\top\D^{-1}\bg_t^j}_t\dd{t},
\end{align}
so that the matrix inequality $\Cm-\Cm_0\succeq\Rm\Imat^{+}\Rm^\top$ applies without change.

For the friction direction alone, $\Rm$ reduces to the vector of friction responses $R_\gamma^\alpha=\gamma\partial_\gamma\E[J_\tau^\alpha]$ and $\Imat$ to $\Sigma_\tau/2$. The matrix inequality then reads $\Cm-\Cm_0\succeq2\bR_\gamma\bR_\gamma^\top/\Sigma_\tau$, which is equivalent to the multivariate friction-response TUR
\begin{equation}
	\Sigma_\tau\ge2\bR_\gamma^\top(\Cm-\Cm_0)^+\bR_\gamma.
\label{eq:und.multiTUR}
\end{equation}
It is the scalar bound~\eqref{eq:und.TUR} applied to the best linear combination of the measured observables. For any observable of the form $J_\tau^{\vb*{\ell}}=\sum_\alpha \ell_\alpha J_\tau^\alpha$ with $\vb*{\ell}=(\ell_\alpha)^\top$, the preparation-subtracted variance is $\vb*{\ell}^\top(\Cm-\Cm_0)\vb*{\ell}$ and the friction response is $\vb*{\ell}\cdot\bR_\gamma$, so that Eq.~\eqref{eq:und.TUR} gives $\Sigma_\tau\ge 2(\vb*{\ell}\cdot\bR_\gamma)^2/[\vb*{\ell}^\top(\Cm-\Cm_0)\vb*{\ell}]$. Maximizing this ratio over $\vb*{\ell}$ yields Eq.~\eqref{eq:und.multiTUR}, with the optimal coefficients $\vb*{\ell}\propto(\Cm-\Cm_0)^+\bR_\gamma$. A combination with zero preparation-subtracted variance has zero friction response by Eq.~\eqref{eq:und.TUR}, so that the pseudoinverse discards no information.

Three properties make the multivariate bound useful in practice. First, it can only improve as observables are added \cite{Dechant.2021.PRX}, and it is never weaker than the best single-observable bound, $\max_\alpha 2(R_\gamma^\alpha)^2/(\Cm-\Cm_0)_{\alpha\alpha}$. Second, the improvement comes from correlations of the dynamical fluctuations. For two observables, we write $V_\alpha\equiv(\Cm-\Cm_0)_{\alpha\alpha}=\Var(J_\tau^\alpha)-\Var_0(J_\tau^\alpha)$ for their preparation-subtracted variances and $\rho\equiv(\Cm-\Cm_0)_{\alpha\beta}/\sqrt{V_\alpha V_\beta}$ for the correlation coefficient of their dynamical fluctuations, which differs in general from that of $J_\tau^\alpha$ and $J_\tau^\beta$ themselves. For $|\rho|<1$, Eq.~\eqref{eq:und.multiTUR} becomes
\begin{equation}
	\Sigma_\tau\ge\frac{2}{1-\rho^2}\qty[\frac{(R_\gamma^\alpha)^2}{V_\alpha}+\frac{(R_\gamma^\beta)^2}{V_\beta}-\frac{2\rho R_\gamma^\alpha R_\gamma^\beta}{\sqrt{V_\alpha V_\beta}}].
\label{eq:und.twoobs}
\end{equation}
For uncorrelated fluctuations, $\rho=0$, the single-observable bounds simply add. For $\rho R_\gamma^\alpha R_\gamma^\beta<0$, the bound exceeds their sum, because the optimal combination cancels part of the common noise while adding the responses. Third, equality holds if and only if the span of the amplitudes $\bu_t^\alpha$ contains $\D^{-1}\bnu_t^{\irr}$, the saturating amplitude of Sec.~\ref{sec:under.saturation}. For observables built from a basis of weights $\phi_t$ and $\bw_t$, Eq.~\eqref{eq:und.multiTUR} is therefore the solution of the variational principle~\eqref{eq:und.entropyvariational} restricted to their span.

\subsection{Coherence bound for underdamped limit cycles}\label{sec:under.cycle}
Noisy self-sustained oscillators must dissipate to remain coherent~\cite{Cao.2015.NP,Barato.2017.PRE,Oberreiter.2022.PRE}. We now apply the friction-response TUR to the phase of such an oscillator. In the weak-noise limit, the resulting bound relates the entropy produced per oscillation to the number of coherent oscillations and to the squared friction sensitivity of the deterministic frequency. This response factor distinguishes the bound from the dissipation-coherence bound obtained in Refs.~\cite{Santolin.2025.PRL,Nagayama.2025.arxiv,Kolchinsky.2026.PRE} for overdamped limit cycles. Below, we define the phase, derive its fluctuations and the friction sensitivity of the frequency, evaluate each term of the friction-response TUR over one period, and discuss three examples demonstrating the relevance of the dissipation-coherence bound.

\subsubsection{Dynamics and asymptotic phase}
We consider Eq.~\eqref{eq:und.dynamics} with an autonomous force and weak thermal noise,
\begin{equation}
\begin{aligned}
	\dd{\bx_t}&=\bv_t\dd{t},\\
	\dd{\bv_t}&=[\bF(\bx_t)-\gamma\bv_t]\dd{t}+\sqrt{2\eps\D_\gamma}\dd{\bW_t},
\end{aligned}
\label{eq:cycle.dynamics}
\end{equation}
where $\D_\gamma=\gamma\T$ and $0<\eps\ll1$, i.e., $\D=\eps\D_\gamma$. At a reference friction coefficient $\gamma$, the deterministic dynamics is $\dot{\bz}=\vb*{\beta}_\gamma(\bz)$ with the phase-space drift
\begin{equation}
	\vb*{\beta}_\gamma(\bz)=\begin{pmatrix}\bv\\ \bF(\bx)-\gamma\bv\end{pmatrix}.
\label{eq:cycle.drift}
\end{equation}
We assume that, for friction coefficients near the reference value, this dynamics admits a smooth family of stable hyperbolic limit cycles with period $\tau_p(\gamma)$ and frequency $\omega_\gamma=2\pi/\tau_p(\gamma)>0$. Such cycles require a nonconservative force, because friction alone makes the mechanical energy of a conservative system decrease monotonically. Stability means that small transverse displacements relax back to the cycle. We denote the periodic orbit at the reference friction by $\bz_*(t)=(\bx_*(t),\bv_*(t))^\top$, so that $\dot{\bz}_*(t)=\vb*{\beta}_\gamma(\bz_*(t))$ and $\bz_*(t+\tau_p)=\bz_*(t)$.

For $\eps>0$, we assume that a smooth stationary density $p_\st^\eps$ exists at the reference friction and concentrates on this cycle as $\eps\to0$. More precisely, for the functions used below, stationary averages converge to time averages along the cycle,
\begin{equation}
	\ev{g}_\st\equiv\int g(\bz)p_\st^\eps(\bz)\dd{\bz}\xrightarrow{\eps\to0}\overline{g}\equiv\frac{1}{\tau_p}\int_0^{\tau_p}g(\bz_*(t))\dd{t}.
\label{eq:cycle.stationary}
\end{equation}
These assumptions exclude switching between competing attractors on the time scale of phase coherence.

The asymptotic phase $\Theta_\gamma(\bz)$, defined modulo $2\pi$, assigns to each point $\bz$ near the cycle the phase of the point on the cycle that the deterministic trajectory from $\bz$ approaches~\cite{Kuramoto.1984,Nakao.2016.CP}. Let $\bz^\gamma(s;\bz)$ denote the deterministic trajectory that starts from $\bz$ at time $0$. With the phase origin fixed by $\Theta_\gamma(\bz_*(0))=0$, the asymptotic phase is characterized by
\begin{equation}
	\lim_{s\to\infty}\left\lVert\bz^\gamma(s;\bz)-\bz_*\qty(s+\frac{\Theta_\gamma(\bz)}{\omega_\gamma})\right\rVert=0,
\label{eq:cycle.phasedef}
\end{equation}
where $\lVert\cdot\rVert$ denotes the Euclidean norm. Because the trajectory from $\bz^\gamma(t;\bz)$ is the same trajectory shifted in time, $\bz^\gamma(s;\bz^\gamma(t;\bz))=\bz^\gamma(s+t;\bz)$, Eq.~\eqref{eq:cycle.phasedef} implies that the phase advances uniformly along every deterministic trajectory,
\begin{equation}
	\Theta_\gamma(\bz^\gamma(t;\bz))=\Theta_\gamma(\bz)+\omega_\gamma t\pmod{2\pi}.
\label{eq:cycle.flow}
\end{equation}
Differentiating this identity with respect to $t$ at $t=0$ gives the phase equation
\begin{equation}
	\vb*{\beta}_\gamma\cdot\nabla\Theta_\gamma=\bv\cdot\nabla_{\bx}\Theta_\gamma+(\bF-\gamma\bv)\cdot\nabla_{\bv}\Theta_\gamma=\omega_\gamma,
\label{eq:cycle.phase}
\end{equation}
which holds wherever the asymptotic phase is smooth, not only on the cycle.

The phase response along the cycle is described by the gradient
\begin{equation}
	\bZ(t)\equiv\nabla\Theta_\gamma(\bz_*(t))=\begin{pmatrix}\bZ_{\bx}(t)\\ \bZ_{\bv}(t)\end{pmatrix},
\label{eq:cycle.Z}
\end{equation}
which is $\tau_p$-periodic and satisfies the normalization $\bZ(t)\cdot\dot{\bz}_*(t)=\omega_\gamma$ by Eq.~\eqref{eq:cycle.phase}. Its velocity component $\bZ_{\bv}$ measures how much a small velocity kick advances or delays the subsequent oscillation. Let $\M(t)\equiv\nabla\vb*{\beta}_\gamma(\bz_*(t))$ be the Jacobian of the drift along the cycle. Taking the gradient of Eq.~\eqref{eq:cycle.phase}, evaluating it on the cycle, and using $\dot{\bZ}=(\nabla\nabla\Theta_\gamma)\dot{\bz}_*$ yields the adjoint equation~\cite{Kuramoto.1984,Nakao.2016.CP}
\begin{equation}
	\dot{\bZ}(t)=-\M(t)^\top\bZ(t),
\label{eq:cycle.adjoint}
\end{equation}
or, in components, $\dot{\bZ}_{\bx}=-(\nabla_{\bx}\bF)^\top\bZ_{\bv}$ and $\dot{\bZ}_{\bv}=-\bZ_{\bx}+\gamma\bZ_{\bv}$, with the force Jacobian evaluated along the cycle.

\subsubsection{Phase diffusion and the number of coherent oscillations}
Along a noisy trajectory, we follow the unwrapped phase $\Theta_t\equiv\operatorname{unwrap}[\Theta_\gamma(\bz_t)]$, the continuous lift of $\Theta_\gamma(\bz_t)$ that counts completed turns instead of reducing the phase modulo $2\pi$. The phase inherits noise through its velocity gradient, and It\^o's formula gives
\begin{align}
	\dd{\Theta_t}&=[\omega_\gamma+\eps c(\bz_t)]\dd{t}+\nabla_{\bv}\Theta_\gamma(\bz_t)^\top\sqrt{2\eps\D_\gamma}\dd{\bW_t},\notag\\
	c(\bz)&\equiv\D_\gamma:\nabla_{\bv}\nabla_{\bv}\Theta_\gamma(\bz).
\label{eq:cycle.phaseIto}
\end{align}
By Eq.~\eqref{eq:cycle.phase}, the deterministic part of the drift is exactly $\omega_\gamma$, and $\eps c$ is the It\^o correction. Both this correction and the noise amplitude depend on the transverse coordinates. In the weak-noise limit, transverse deviations remain small after relaxation, and we evaluate both coefficients at the point on the cycle with the same phase. With the $2\pi$-periodic functions
\begin{equation}
	c_*(\theta)\equiv c\qty(\bz_*(\theta/\omega_\gamma)),\quad
	q_*(\theta)\equiv\bZ_{\bv}^\top\D_\gamma\bZ_{\bv}\big|_{t=\theta/\omega_\gamma},
\label{eq:cycle.coefficients}
\end{equation}
this gives the reduced phase process $\theta_t$~\cite{Yoshimura.2008.PRL,Teramae.2009.PRL}, which obeys a closed It\^o equation with a scalar Wiener process $B_t$ and has the backward generator $\Lg_\theta^\dagger$,
\begin{equation}
\begin{gathered}
	\dd{\theta_t}=[\omega_\gamma+\eps c_*(\theta_t)]\dd{t}+\sqrt{2\eps q_*(\theta_t)}\dd{B_t},\\
	\Lg_\theta^\dagger=\omega_\gamma\partial_\theta+\eps\qty[c_*(\theta)\partial_\theta+q_*(\theta)\partial_\theta^2].
\end{gathered}
\label{eq:cycle.reduced}
\end{equation}
We introduce the phase-diffusion integral and the cycle averages
\begin{equation}
\begin{gathered}
	\mathcal{D}_\Theta\equiv\int_0^{\tau_p}\bZ_{\bv}(t)^\top\D_\gamma\bZ_{\bv}(t)\dd{t},\\
	\bar{q}\equiv\frac{\mathcal{D}_\Theta}{\tau_p},\quad
	\bar{c}\equiv\frac{1}{\tau_p}\int_0^{\tau_p}c(\bz_*(t))\dd{t},
\end{gathered}
\label{eq:cycle.D}
\end{equation}
and assume $\mathcal{D}_\Theta>0$. To leading order, $\eps\bar{q}$ is the diffusion coefficient of the phase~\cite{Kuramoto.1984,Teramae.2009.PRL}, so that the phase variance accumulated over one period is $2\eps\mathcal{D}_\Theta$.

The decay of oscillatory correlations follows from the eigenvalues of $\Lg_\theta^\dagger$ in the Fourier modes $e^{in\theta}$, $n\in\mathbb{Z}$. At zero noise, $\Lg_\theta^\dagger e^{in\theta}=in\omega_\gamma e^{in\theta}$. For each fixed $n$, we substitute the eigenfunction $e^{in\theta}+\eps\rho_n(\theta)+o(\eps)$ and the eigenvalue $\lambda_n=in\omega_\gamma+\eps\lambda_n^{(1)}+o(\eps)$ into the eigenvalue equation, multiply the terms of order $\eps$ by $e^{-in\theta}$, and integrate over the circle. The correction $\rho_n$ drops out, because $\int_0^{2\pi}e^{-in\theta}(\partial_\theta-in)\rho_n\dd{\theta}=0$ by integration by parts and periodicity. Since $(2\pi)^{-1}\int_0^{2\pi}c_*\dd{\theta}=\bar{c}$ and likewise for $q_*$, we obtain
\begin{align}
	\lambda_n&=in\omega_\gamma+\frac{\eps}{2\pi}\int_0^{2\pi}e^{-in\theta}\qty[c_*\partial_\theta+q_*\partial_\theta^2]e^{in\theta}\dd{\theta}+o(\eps)\notag\\
	&=in\omega_\gamma+\eps(in\bar{c}-n^2\bar{q})+o(\eps).
\label{eq:cycle.eigen}
\end{align}
The decay rate grows as $n^2$, so that the slowest-decaying nonzero modes form the fundamental pair $n=\pm1$, with
\begin{equation}
\begin{gathered}
	\lambda_{\pm1}=-r_\eps\pm i\Omega_\eps,\\
	r_\eps=\eps\bar{q}+o(\eps),\quad \Omega_\eps=\omega_\gamma+\eps\bar{c}+o(\eps).
\end{gathered}
\label{eq:cycle.rates}
\end{equation}

For a real observable $A$ whose fundamental harmonic along the cycle, $a_1\equiv(2\pi)^{-1}\int_0^{2\pi}e^{-i\theta}A(\bz_*(\theta/\omega_\gamma))\dd{\theta}$, does not vanish, the stationary connected correlation $C_A(t)\equiv\ev*{A(\bz_t)A(\bz_0)}_\st-\ev{A}_\st^2$ is then approximated by
\begin{equation}
	C_A(t)\simeq\Re\qty[C_1e^{(-r_\eps+i\Omega_\eps)t}]
\label{eq:cycle.correlation}
\end{equation}
when the fundamental pair dominates. Here, $C_1=2|a_1|^2+o(1)$ is an amplitude set by the fundamental harmonic of $A$. This approximation describes phase coherence at times when transverse transients and higher phase harmonics are negligible. Under this assumption, the envelope decays by a factor $e$ over the coherence time $t_{\mathrm{coh}}=1/r_\eps$, while the oscillation period is $\tau_{\mathrm{osc}}=2\pi/\Omega_\eps$. The number of coherent oscillations~\cite{Barato.2017.PRE,Oberreiter.2022.PRE} is therefore
\begin{equation}
	\Nc_\eps\equiv\frac{t_{\mathrm{coh}}}{\tau_{\mathrm{osc}}}=\frac{\Omega_\eps}{2\pi r_\eps}=\frac{1}{\eps\mathcal{D}_\Theta}+o(\eps^{-1}).
\label{eq:cycle.N}
\end{equation}
For overdamped oscillators, the corresponding weak-noise quality factor, which is proportional to $\Nc_\eps$, was determined in Ref.~\cite{Remlein.2022.PRE}.

\subsubsection{Friction sensitivity of the frequency}
Both the limit cycle and its frequency depend on the friction coefficient. Choosing the phase origin smoothly in $\gamma$, we define $\Theta_\gamma'(\bz)\equiv\partial_\gamma\Theta_\gamma(\bz)$ at fixed $\bz$ and fixed force field, and $\omega_\gamma'\equiv\dd{\omega_\gamma}/\dd{\gamma}$. Since $\partial_\gamma\vb*{\beta}_\gamma=(\vb*0,-\bv)$, differentiating Eq.~\eqref{eq:cycle.phase} with respect to $\gamma$ gives the exact deterministic identity
\begin{equation}
	\omega_\gamma'=\vb*{\beta}_\gamma\cdot\nabla\Theta_\gamma'-\bv\cdot\nabla_{\bv}\Theta_\gamma.
\label{eq:cycle.dphase}
\end{equation}
On the reference cycle, $\vb*{\beta}_\gamma\cdot\nabla\Theta_\gamma'$ is the total time derivative of the periodic function $\Theta_\gamma'(\bz_*(t))$, so that its integral over one period vanishes. Averaging Eq.~\eqref{eq:cycle.dphase} over the period, multiplying by $\gamma/\omega_\gamma$, and using $\omega_\gamma\tau_p=2\pi$, we obtain the dimensionless friction sensitivity of the frequency,
\begin{equation}
	\varpi_\gamma\equiv\gamma\partial_\gamma\ln\omega_\gamma
	=\frac{1}{2\pi}\int_0^{\tau_p}\bZ_{\bv}(t)^\top[-\gamma\bv_*(t)]\dd{t}.
\label{eq:cycle.sensitivity}
\end{equation}
The integrand is the rate at which the friction force along the orbit shifts the phase. Equation~\eqref{eq:cycle.sensitivity} accounts for changes of both the orbit and the speed along it.

\subsubsection{Dissipation-coherence bound}
We apply the friction-response TUR~\eqref{eq:und.TUR} to the phase increment over one reference period,
\begin{align}
	J_{\tau_p}&=\Theta_{\tau_p}-\Theta_0\notag\\
	&=\int_0^{\tau_p}\qty(\nabla_{\bx}\Theta_\gamma\cdot\bv_t\dd{t}+\nabla_{\bv}\Theta_\gamma\circ\dd{\bv_t}),
\label{eq:cycle.observable}
\end{align}
where the phase gradients are evaluated at $\bz_t$. This observable has the form~\eqref{eq:und.J} with $\phi=\nabla_{\bx}\Theta_\gamma\cdot\bv$ and $\bw=\nabla_{\bv}\Theta_\gamma$. In the friction response, these fields and the observation time $\tau_p$ are held at their reference values, together with the temperature, the noise strength $\eps$, and the stationary reference preparation $p_\st^\eps$. All averages below are taken for the reference process started in $p_\st^\eps$. We evaluate the three terms of the TUR in turn.

\emph{Dynamical variance}---Integrating Eq.~\eqref{eq:cycle.phaseIto} from $t$ to $\tau_p$ and conditioning on $\bz_t=\bz$ gives the conditional future
\begin{equation}
\begin{gathered}
	G_t(\bz)=\omega_\gamma(\tau_p-t)+\eps\Xi_t(\bz),\\
	\Xi_t(\bz)\equiv\E\qty[\int_t^{\tau_p}c(\bz_s)\dd{s}\,\Big|\,\bz_t=\bz],
\end{gathered}
\label{eq:cycle.G}
\end{equation}
so that the martingale amplitude~\eqref{eq:und.kernel} is $\bu_t=\nabla_{\bv}\Theta_\gamma+\eps\nabla_{\bv}\Xi_t$. Although not indicated explicitly, $\Xi_t$ depends on $\eps$ through the reference dynamics. We assume that the stationary moments of $\nabla_{\bv}\Xi_t$ and $\nabla_{\bv}\nabla_{\bv}\Xi_t$ that appear below remain bounded uniformly in $t\in[0,\tau_p]$ as $\eps\to0$, and that all boundary terms in the integrations by parts vanish. With $\D=\eps\D_\gamma$, Eq.~\eqref{eq:und.variance} then gives
\begin{align}
	\Vd&=2\eps\int_0^{\tau_p}\ev{\bu_t^\top\D_\gamma\bu_t}_\st\dd{t}\notag\\
	&=2\eps\tau_p\ev{\nabla_{\bv}\Theta_\gamma^\top\D_\gamma\nabla_{\bv}\Theta_\gamma}_\st+O(\eps^2)\notag\\
	&=2\eps\mathcal{D}_\Theta+o(\eps),
\label{eq:cycle.variance}
\end{align}
where stationarity is used in the second line and Eq.~\eqref{eq:cycle.stationary} in the last. Because $G_0=\omega_\gamma\tau_p+\eps\Xi_0$, the preparation variance $\Var_0(J_{\tau_p})=\eps^2\Var[\Xi_0(\bz_0)]$ is of higher order whenever $\Xi_0$ has a bounded variance, so that the full variance has the same leading term.

\emph{Friction response}---For the stationary reference process, the irreversible local mean velocity is $\bnu^{\irr}=-\gamma\bv-\eps\D_\gamma\nabla_{\bv}\ln p_\st^\eps$. Inserting it and the amplitude $\bu_t$ into $\Rgamma=\int_0^{\tau_p}\ev*{\bu_t\cdot\bnu^{\irr}}_\st\dd{t}$ [Eq.~\eqref{eq:und.frictionresponse}], we integrate the terms that contain $\nabla_{\bv}\ln p_\st^\eps$ by parts. For example, $-\ev*{(\nabla_{\bv}\Theta_\gamma)^\top\D_\gamma\nabla_{\bv}\ln p_\st^\eps}_\st=\ev*{\D_\gamma:\nabla_{\bv}\nabla_{\bv}\Theta_\gamma}_\st=\ev*{c}_\st$, and the term with $\Xi_t$ is treated in the same way. This yields
\begin{align}
	\Rgamma&=\tau_p\ev{-\gamma\bv\cdot\nabla_{\bv}\Theta_\gamma}_\st+\eps\tau_p\ev{c}_\st\notag\\
	&\quad-\eps\gamma\int_0^{\tau_p}\ev{\bv\cdot\nabla_{\bv}\Xi_t}_\st\dd{t}\notag\\
	&\quad+\eps^2\int_0^{\tau_p}\ev{\D_\gamma:\nabla_{\bv}\nabla_{\bv}\Xi_t}_\st\dd{t}.
\label{eq:cycle.responseexp}
\end{align}
Under the stated boundedness assumptions, the last three terms vanish as $\eps\to0$, and Eqs.~\eqref{eq:cycle.stationary} and \eqref{eq:cycle.sensitivity} give
\begin{align}
	\Rgamma&=\int_0^{\tau_p}\bZ_{\bv}(t)^\top[-\gamma\bv_*(t)]\dd{t}+o(1)\notag\\
	&=2\pi\varpi_\gamma+o(1).
\label{eq:cycle.response}
\end{align}
The density-gradient part of the irreversible current thus contributes only at order $\eps$, through the It\^o correction, and the leading response is the deterministic cycle integral of Eq.~\eqref{eq:cycle.sensitivity}.

\emph{Entropy production}---With $\D=\eps\D_\gamma$, the stationary entropy production rate follows from Eq.~\eqref{eq:und.entropy}. Expanding the square of $\bnu^{\irr}$ and using $\ev*{\bv\cdot\nabla_{\bv}\ln p_\st^\eps}_\st=-d$, which follows from integration by parts, we obtain
\begin{equation}
\begin{gathered}
	\sigma_\eps=\frac{\gamma^2}{\eps}\ev{\bv^\top\D_\gamma^{-1}\bv}_\st-2\gamma d+\eps I_{\bv},\\
	I_{\bv}\equiv\ev{(\nabla_{\bv}\ln p_\st^\eps)^\top\D_\gamma\nabla_{\bv}\ln p_\st^\eps}_\st,
\end{gathered}
\label{eq:cycle.EPexp}
\end{equation}
where $I_{\bv}$ is the velocity-space Fisher information weighted by $\D_\gamma$. The stationary entropy balance implies $\eps I_{\bv}=\gamma d$. Hence,
\begin{equation}
	\sigma_\eps=\frac{\gamma^2}{\eps}\ev{\bv^\top\D_\gamma^{-1}\bv}_\st-\gamma d.
\label{eq:cycle.EP}
\end{equation}
The leading term is the friction force along the orbit, measured in the inverse noise metric. We therefore define the dissipation integral
\begin{equation}
	\mathcal{E}\equiv\int_0^{\tau_p}[-\gamma\bv_*(t)]^\top\D_\gamma^{-1}[-\gamma\bv_*(t)]\dd{t},
\label{eq:cycle.E}
\end{equation}
so that Eq.~\eqref{eq:cycle.stationary} gives 
\begin{equation}
	\Sigma_{\tau_p}=\tau_p\sigma_\eps=\frac{\mathcal{E}}{\eps}+o(\eps^{-1}).
\label{eq:cycle.entprod}
\end{equation}
Since $2\pi/\Omega_\eps=\tau_p+O(\eps)$, the entropy produced per noisy oscillation has the same leading term,
\begin{equation}
	\Sigma_{\mathrm{cyc}}^\eps\equiv\frac{2\pi}{\Omega_\eps}\sigma_\eps=\frac{\mathcal{E}}{\eps}+o(\eps^{-1}).
\label{eq:cycle.Scyc}
\end{equation}

\emph{Dissipation-coherence bound}---Substituting Eqs.~\eqref{eq:cycle.variance}, \eqref{eq:cycle.response}, and \eqref{eq:cycle.entprod} into the TUR~\eqref{eq:und.TUR} gives
\begin{equation}
	[2\eps\mathcal{D}_\Theta+o(\eps)]\qty[\frac{\mathcal{E}}{\eps}+o(\eps^{-1})]\ge2[2\pi\varpi_\gamma+o(1)]^2.
\label{eq:cycle.insertion}
\end{equation}
Taking $\eps\to0$ with the reference oscillator fixed yields $\mathcal{E}\mathcal{D}_\Theta\ge4\pi^2\varpi_\gamma^2$. Together with Eqs.~\eqref{eq:cycle.N} and \eqref{eq:cycle.Scyc}, this establishes the dissipation-coherence bound
\begin{equation}
	\lim_{\eps\to0}\frac{\Sigma_{\mathrm{cyc}}^\eps}{\Nc_\eps}=\mathcal{E}\mathcal{D}_\Theta\ge4\pi^2(\gamma\partial_\gamma\ln\omega_\gamma)^2.
\label{eq:cycle.bound}
\end{equation}
The entropy cost per coherent oscillation is thus bounded from below by the squared friction sensitivity of the deterministic frequency: at a given sensitivity, an oscillator that remains coherent for more cycles must dissipate proportionally more per cycle. The limit is taken at fixed friction and under the phase-reduction assumptions stated above, so that Eq.~\eqref{eq:cycle.bound} is a weak-noise statement. Once the cycle integrals are identified, the bound also follows from the Cauchy-Schwarz inequality applied directly to Eq.~\eqref{eq:cycle.sensitivity} with the metric $\D_\gamma$, $(2\pi\varpi_\gamma)^2\le\mathcal{D}_\Theta\mathcal{E}$. This makes the equality condition transparent: $\bZ_{\bv}$ must be proportional to $\D_\gamma^{-1}\bv_*$ along the whole orbit. In the overdamped limit, friction only rescales the frequency, $\varpi_\gamma=-1$, and Eq.~\eqref{eq:cycle.bound} reduces to the dissipation-coherence bound $\Sigma_{\mathrm{cyc}}/\Nc\ge4\pi^2$, conjectured for Markov jump processes in Ref.~\cite{Oberreiter.2022.PRE} and established for overdamped limit cycles in Refs.~\cite{Santolin.2025.PRL,Nagayama.2025.arxiv,Kolchinsky.2026.PRE}. In the following, we show that a driven rotor attains the bound, that a large friction sensitivity can raise it arbitrarily above $4\pi^2$, and that a weak friction sensitivity permits costs below $4\pi^2$, which we demonstrate by evaluating the cycle integrals explicitly.

\subsubsection{Saturation by a driven rotor}
A uniformly driven rotor saturates both the finite-time TUR and the dissipation-coherence bound. Let $\alpha$ be an angle defined modulo $2\pi$, and consider
\begin{equation}
	\dd{\alpha_t}=v_t\dd{t},\quad
	\dd{v_t}=(f-\gamma v_t)\dd{t}+\sqrt{2\eps\gamma T}\dd{W_t},
\label{eq:cycle.rotor}
\end{equation}
with $f,\gamma,T,\eps>0$. The deterministic cycle has the constant velocity $v_*=f/\gamma$ and the frequency $\omega_\gamma=f/\gamma$. Up to an additive constant, the asymptotic phase is $\Theta_\gamma=\alpha+v/\gamma$: it satisfies $\vb*{\beta}_\gamma\cdot\nabla\Theta_\gamma=v+(f-\gamma v)/\gamma=f/\gamma$, so that $Z_v=1/\gamma$ and $\varpi_\gamma=-1$. The stationary density is uniform in $\alpha$ and Gaussian in $v$, with mean $f/\gamma$ and variance $\eps T$. The irreversible local mean velocity is therefore $\nu^{\irr}=-\gamma v+\gamma(v-f/\gamma)=-f$, and the It\^o correction vanishes, $c=0$. Under the reference dynamics, the unwrapped phase increment over any observation time $\tau$ is exactly
\begin{equation}
	J_\tau=\frac{f\tau}{\gamma}+\sqrt{\frac{2\eps T}{\gamma}}\,W_\tau.
\label{eq:cycle.rotorJ}
\end{equation}
Its conditional mean does not depend on the initial state, so that $\Var_0(J_\tau)=0$. Because $c=0$, we have $\Xi_t=0$ and $u_t=1/\gamma$, and the response follows from $\ev*{-\gamma v\,\partial_v\Theta_\gamma}_\st=-f/\gamma$. With the initial density, the reference phase field, and $\tau$ held fixed, the three terms of the TUR are
\begin{equation}
	\Vd=\frac{2\eps T\tau}{\gamma},\quad
	\Rgamma=-\frac{f\tau}{\gamma},\quad
	\Sigma_\tau=\frac{f^2\tau}{\eps\gamma T}.
\label{eq:cycle.rotorterms}
\end{equation}
They satisfy $\Vd\Sigma_\tau=2\Rgamma^2$ at every $\tau$, as required by the equality condition: the source $\nu^{\irr}=-f$ is a constant multiple of $\D u_t=\eps T$. The phase mode has $r_\eps=\eps T/\gamma$ and $\Omega_\eps=f/\gamma$, and hence
\begin{equation}
	\Nc_\eps=\frac{f}{2\pi\eps T},\quad
	\Sigma_{\mathrm{cyc}}^\eps=\frac{2\pi f}{\eps T}=4\pi^2\Nc_\eps.
\label{eq:cycle.rotorN}
\end{equation}
With $\varpi_\gamma^2=1$, both the finite-time TUR and the coherence bound are saturated at every positive noise strength. 

\subsubsection{Enhanced friction sensitivity}
The friction sensitivity can exceed unity in magnitude, so that the bound may exceed the overdamped value $4\pi^2$. We demonstrate this with a two-dimensional oscillator, $\bx,\bv\in\mathbb{R}^2$, whose frequency and phase response can be evaluated analytically. In dimensionless units, we take Eq.~\eqref{eq:cycle.dynamics} with the force
\begin{equation}
	\bF(\bx)=(1-\|\bx\|^4)\bx+\|\bx\|^2\mathsf{J}\bx,
	\quad \mathsf{J}=\begin{pmatrix}0&-1\\1&0\end{pmatrix}.
\label{eq:cycle.force}
\end{equation}
The first term is the conservative force generated by the radial potential $U(r)=r^6/6-r^2/2$, with $r=\|\bx\|$, and the second exerts a nonconservative torque. The force field does not depend on $\gamma$ and is held fixed in the friction response.

The deterministic dynamics is conveniently expressed in polar coordinates. For $r>0$, we introduce the polar angle $\alpha$ and the orthonormal basis $\vb*{e}_r\equiv(\cos\alpha,\sin\alpha)^\top$ and $\vb*{e}_\alpha\equiv(-\sin\alpha,\cos\alpha)^\top=\mathsf{J}\vb*{e}_r$. Since $\dot{\vb*{e}}_r=\dot{\alpha}\vb*{e}_\alpha$ and $\dot{\vb*{e}}_\alpha=-\dot{\alpha}\vb*{e}_r$, differentiating $\bx=r\vb*{e}_r$ twice gives
\begin{equation}
\begin{aligned}
	\bv&=\dot{r}\vb*{e}_r+r\dot{\alpha}\vb*{e}_\alpha,\\
	\dot{\bv}&=(\ddot{r}-r\dot{\alpha}^2)\vb*{e}_r+(2\dot{r}\dot{\alpha}+r\ddot{\alpha})\vb*{e}_\alpha.
\end{aligned}
\label{eq:cycle.polarkin}
\end{equation}
In the same basis, $\bF=r(1-r^4)\vb*{e}_r+r^3\vb*{e}_\alpha$ and $-\gamma\bv=-\gamma \dot{r}\vb*{e}_r-\gamma r\dot{\alpha}\vb*{e}_\alpha$. Equating the radial and tangential components of $\dot{\bv}=\bF-\gamma\bv$ and dividing the latter by $r$, we obtain
\begin{equation}
\begin{aligned}
	\ddot{r}&=r(\dot{\alpha}^2+1-r^4)-\gamma \dot{r},\\
	\ddot{\alpha}&=r^2-\gamma \dot{\alpha}-2\dot{r}\dot{\alpha}/r.
\end{aligned}
\label{eq:cycle.polar}
\end{equation}

For each fixed $\gamma>1$, we seek a circular orbit of radius $R>0$ traversed at a constant angular velocity, $r_t=R$ and $\alpha_t=\omega_\gamma t+\alpha_0$, so that $\dot{r}=\ddot{r}=\ddot{\alpha}=0$ and $\dot{\alpha}=\omega_\gamma$. Equation~\eqref{eq:cycle.polar} then requires $R(\omega_\gamma^2+1-R^4)=0$ and $R^2=\gamma\omega_\gamma$, i.e., $R^4=1+\omega_\gamma^2$ and $R^2=\gamma\omega_\gamma$. Their unique positive solution and the resulting friction sensitivity are
\begin{equation}
	\omega_\gamma=\frac{1}{\sqrt{\gamma^2-1}},~R=\left(\frac{\gamma^2}{\gamma^2-1}\right)^{1/4},~\varpi_\gamma=-\frac{\gamma^2}{\gamma^2-1}.
\label{eq:cycle.enhanced}
\end{equation}
Thus, $|\varpi_\gamma|>1$ for every $\gamma>1$, and $|\varpi_\gamma|\to\infty$ as $\gamma\downarrow1$. Increasing the friction both slows the rotation directly and shrinks the orbit, $R^2=\gamma/\sqrt{\gamma^2-1}$, which weakens the torque. 

\begin{figure}[t]
\centering\includegraphics[width=1.0\linewidth]{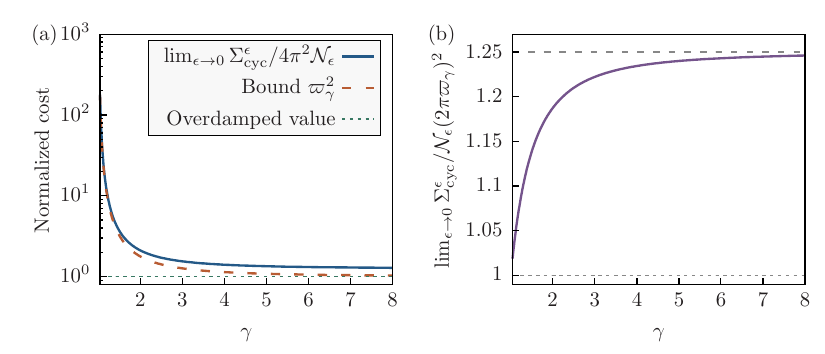}
\caption{Weak-noise dissipation-coherence bound for the enhanced-sensitivity oscillator, with the limit $\eps\to0$ taken at fixed $\gamma>1$. (a) Normalized cost per coherent oscillation, $\lim_{\eps\to0}\Sigma_{\mathrm{cyc}}^\eps/(4\pi^2\Nc_\eps)=\varpi_\gamma^2[1+(\gamma^2-1)/(4\gamma^2)]$, compared with the friction-response bound $\varpi_\gamma^2$ and with the overdamped value $1$ (dotted). (b) Ratio of the cost to the friction-response bound, which lies between $1$ and $5/4$.}
\label{fig:cycle}
\end{figure}

The circular orbit is transversely stable. Linearizing Eq.~\eqref{eq:cycle.polar} in $\vb*{\zeta}\equiv(r-R,\dot{r},\dot{\alpha}-\omega_\gamma)^\top$ gives $\dot{\vb*{\zeta}}=\M_\perp\vb*{\zeta}+O(\|\vb*{\zeta}\|^2)$ with
\begin{equation}
	\M_\perp=\begin{pmatrix}
	0&1&0\\-4R^4&-\gamma&2R\omega_\gamma\\
	2R&-2\omega_\gamma/R&-\gamma
	\end{pmatrix}.
\label{eq:cycle.stability}
\end{equation}
Using $R^2=\gamma\omega_\gamma$ and $R^4=1+\omega_\gamma^2$, its characteristic polynomial reduces to
\begin{equation}
	\det(\lambda-\M_\perp)=\lambda^3+2\gamma\lambda^2+(\gamma^2+4+8\omega_\gamma^2)\lambda+4\gamma.
\label{eq:cycle.charpoly}
\end{equation}
For a cubic polynomial $\lambda^3+a_1\lambda^2+a_2\lambda+a_3$, the Routh-Hurwitz conditions for all roots to have negative real parts are $a_1,a_2,a_3>0$ and $a_1a_2>a_3$. Since all coefficients are positive and $a_1a_2-a_3=2\gamma(\gamma^2+2+8\omega_\gamma^2)>0$, all three transverse eigenvalues have negative real parts, and the circular orbit is a stable limit cycle.

The phase response follows from the same linearization. By rotational symmetry, $\Theta_\gamma-\alpha$ depends only on $(r,\dot{r},\dot{\alpha})$, and we fix the phase origin so that it vanishes on the cycle. To first order in $\vb*{\zeta}$, $\Theta_\gamma=\alpha+l_1(r-R)+l_2\dot{r}+l_3(\dot{\alpha}-\omega_\gamma)$, and Eq.~\eqref{eq:cycle.phase} with $\dot{\alpha}=\omega_\gamma+\zeta_3$ requires $\M_\perp^\top(l_1,l_2,l_3)^\top=-(0,0,1)^\top$. Solving this linear system gives
\begin{equation}
	l_3=\frac{\gamma}{\gamma^2-1},~
	l_2=\frac{l_3}{2R^3},~
	l_1=\gamma l_2+\frac{2\omega_\gamma l_3}{R}.
\label{eq:cycle.lcoeff}
\end{equation}
Since $\dot{r}=\bv\cdot\vb*{e}_r$ and $\dot{\alpha}=\bv\cdot\vb*{e}_\alpha/r$, the velocity component of the phase response is $\bZ_{\bv}=l_2\vb*{e}_r+(l_3/R)\vb*{e}_\alpha$. 

For an isotropic temperature $\T=T\mathds{I}$, the cycle integrals are $\mathcal{D}_\Theta=\tau_p\gamma T[l_2^2+(l_3/R)^2]$ and $\mathcal{E}=\tau_p\gamma R^2\omega_\gamma^2/T$, so that
\begin{equation}
	\frac{\mathcal{E}\mathcal{D}_\Theta}{4\pi^2}
	=\varpi_\gamma^2\left(1+\frac{1}{4R^4}\right)
	=\varpi_\gamma^2\left(1+\frac{\gamma^2-1}{4\gamma^2}\right).
\label{eq:cycle.tightness}
\end{equation}
For every fixed $\gamma>1$, the entropy cost per coherent oscillation thus lies between one and $5/4$ times the friction-response bound (Fig.~\ref{fig:cycle}), while the bound $4\pi^2\varpi_\gamma^2$ itself exceeds the overdamped value $4\pi^2$ and diverges as $\gamma\downarrow1$. The weak-noise limit is taken first, at fixed $\gamma>1$. Because the radial component $l_2$ of the phase response is nonzero, $\bZ_{\bv}$ is not parallel to $\D_\gamma^{-1}\bv_*\propto\vb*{e}_\alpha$, and the bound is not saturated.

\subsubsection{Friction-insensitive frequency}
The friction sensitivity can also vanish. Since $\Sigma_{\mathrm{cyc}}^\eps/\Nc_\eps\ge4\pi^2\varpi_\gamma^2$ in the weak-noise limit, an underdamped oscillator can fall below the overdamped value $4\pi^2$ only if $|\varpi_\gamma|<1$. This occurs, for example, when the frequency is set by a conservative restoring force. Consider the force
\begin{equation}
	\bF(\bx)=-k\bx+h(\|\bx\|)\mathsf{J}\bx,
\label{eq:cycle.isoforce}
\end{equation}
with $k>0$, which combines a harmonic restoring force with a rotational force whose angular acceleration $h(r)$ is positive and decreases to zero at large radius, for example, $h(r)=h_0e^{-r^2/\ell^2}$. The polar equations~\eqref{eq:cycle.polar} become $\ddot{r}=r(\dot{\alpha}^2-k)-\gamma\dot{r}$ and $\ddot{\alpha}=h(r)-\gamma\dot{\alpha}-2\dot{r}\dot{\alpha}/r$. A circular orbit requires the radial balance $\omega_\gamma^2=k$ and the angular balance $\gamma\omega_\gamma=h(R)$, so that
\begin{equation}
	\omega_\gamma=\sqrt{k},\quad h(R)=\gamma\sqrt{k},\quad\varpi_\gamma=0
\end{equation}
for every such $h$ with $h(0)>\gamma\sqrt{k}$, which guarantees a unique radius $R$. The frequency is thus fixed by the restoring force, and friction changes only the radius. Since $h>0$, no circular orbit rotates in the opposite direction. Linearizing around the orbit in the variables of Eq.~\eqref{eq:cycle.stability} gives
\begin{equation}
	\M_\perp=\begin{pmatrix}0&1&0\\0&-\gamma&2R\sqrt{k}\\h'(R)&-2\sqrt{k}/R&-\gamma\end{pmatrix},
\end{equation}
with the characteristic polynomial $\lambda^3+2\gamma\lambda^2+(\gamma^2+4k)\lambda-2R\sqrt{k}\,h'(R)$. For $h'(R)<0$, all coefficients are positive, and the Routh-Hurwitz conditions reduce to $\gamma(\gamma^2+4k)>-R\sqrt{k}\,h'(R)$. The phase-response condition $\M_\perp^\top(l_1,l_2,l_3)^\top=-(0,0,1)^\top$ gives $l_3=0$, $l_2=-1/(2R\sqrt{k})$, and $l_1=\gamma l_2$, independently of $h$. The velocity component of the phase response, $\bZ_{\bv}=l_2\vb*{e}_r$, is therefore radial and orthogonal to the friction force $-\gamma\bv_*\propto\vb*{e}_\alpha$, consistent with $\varpi_\gamma=0$. For an isotropic temperature, the cycle integrals give
\begin{equation}
	\frac{\mathcal{E}\mathcal{D}_\Theta}{4\pi^2}=\gamma^2R^2l_2^2=\frac{\gamma^2}{4k}.
\label{eq:cycle.iso}
\end{equation}
The dissipation per coherent oscillation in the weak-noise limit, $\pi^2\gamma^2/k$, thus lies below the overdamped value $4\pi^2$ whenever $\gamma<2\sqrt{k}$. For the Gaussian profile with $\ell^2=\gamma\sqrt{k}$ and $h_0=2e\gamma\sqrt{k}$, the orbit has $R^2=(1+\ln2)\gamma\sqrt{k}$ and $h'(R)=-2R$, and the stability condition reduces to $\gamma^2+2(1-\ln2)k>0$, which holds for every $\gamma$. Because $h$ is bounded and decays at large radius, the motion far from the origin is that of a damped harmonic oscillator, and deterministic trajectories from random initial conditions converge to the cycle. Here, $h_0$ and $\ell$ are chosen for the given $\gamma$ and kept fixed when the friction is varied, which leaves $\varpi_\gamma=0$ unchanged. Along this family, the weak-noise cost $\pi^2\gamma^2/k$ vanishes as $\gamma\to0$; since the radius and the slowest transverse relaxation rate vanish as well, the weak-noise limit must be taken first, at fixed $\gamma$. Inertia thus allows highly coherent oscillations at low dissipation when the frequency is insensitive to friction.

\subsection{Violations of the conventional TUR by velocity-resolved currents}\label{sec:violation}
The conventional TUR, $\Var(J_\tau)/\E[J_\tau]^2\ge2/\Sigma_\tau$, holds for currents in steady-state overdamped diffusions and Markov jump processes~\cite{Barato.2015.PRL,Gingrich.2016.PRL,Horowitz.2020.NP}, but it is not guaranteed for underdamped dynamics~\cite{Vu.2019.PRE.UnderdampedTUR,Fischer.2020.PRE}. We now show that weighting the displacement by a suitable function of the velocity violates it by arbitrarily large factors, even in the simplest driven underdamped dynamics, whereas the friction-response TUR remains valid. Comparing the two bounds also reveals why the conventional TUR fails: the fluctuations of currents are bounded by the friction response of their mean, not by the mean itself.

We consider a particle on a ring driven by a constant force $f>0$,
\begin{equation}
	\dd{x_t}=v_t\dd{t},\quad
	\dd{v_t}=(f-\gamma v_t)\dd{t}+\sqrt{2\gamma T}\dd{W_t}.
\label{eq:violation.model}
\end{equation}
In the steady state, the position is uniformly distributed, and the velocity is independent of it and Gaussian with mean $\vbar=f/\gamma$ and variance $T$. The irreversible local mean velocity is therefore the constant $\nu^{\irr}=-f$, and the entropy production rate reads $\sigma_{\st}=f^2/(\gamma T)$. We characterize the driving strength by $\kappa\equiv\vbar^2/T=\sigma_{\st}/\gamma$, the entropy produced per velocity relaxation time. We consider velocity-resolved currents
\begin{equation}
	J_\tau=\int_0^\tau\psi(v_t)\circ\dd{x_t}=\int_0^\tau\Phi(v_t)\dd{t},
\label{eq:violation.current}
\end{equation}
where $\Phi(v)=v\psi(v)$. These currents have the form~\eqref{eq:und.J} with $\phi=\Phi$ and $\bw=0$. For an even function $\psi$, the rate $\Phi$ is odd under velocity reversal, so that $J_\tau$ is a time-antisymmetric current. It records the displacement weighted by a function of the instantaneous velocity, which can be evaluated from velocity-resolved trajectories; for $\psi=1$, it reduces to the net displacement $\int_0^\tau v_t\dd{t}$ along the ring.

Because $\Phi$ depends only on the velocity, which is an Ornstein-Uhlenbeck process, the long-time problem can be solved exactly in the eigenbasis of this process. On functions of $v$ alone, the backward generator reduces to $\Lg^\dagger=(f-\gamma v)\partial_v+\gamma T\partial_v^2$. Its eigenfunctions are the probabilists' Hermite polynomials of the normalized velocity~\cite{Risken.1989}, $\varphi_n(v)\equiv\He_n((v-\vbar)/\sqrt T)$, which satisfy
\begin{equation}
	\Lg^\dagger\varphi_n=-\gamma n\varphi_n,~
	\partial_v\varphi_n=\frac{n\varphi_{n-1}}{\sqrt T},~
	\ev{\varphi_n\varphi_m}_{\st}=n!\,\delta_{nm}.
\label{eq:violation.basis}
\end{equation}
We expand the rate in terms of the eigenfunctions as
\begin{equation}
	\Phi(v)=\sum_{n=0}^{\infty}a_n\varphi_n(v),\quad a_n=\frac{\ev{\Phi\varphi_n}_{\st}}{n!}.
\label{eq:violation.Hermite}
\end{equation}
The mean rate is $\jr_{\st}=a_0$, and the Poisson equation~\eqref{eq:general.poisson} is solved mode by mode, $\Psi=\sum_{n=1}^\infty a_n\varphi_n/(\gamma n)$. The long-time amplitude is the velocity gradient $\partial_v\Psi=\sum_{n=1}^\infty a_n\varphi_{n-1}/(\gamma\sqrt T)$. Inserting it into Eq.~\eqref{eq:general.DJdiff} with $\bw=0$ and $\D=\gamma T$, and using the orthogonality in Eq.~\eqref{eq:violation.basis}, we obtain
\begin{equation}
	D_J=\frac{1}\gamma\sum_{n=1}^\infty a_n^2(n-1)!.
\label{eq:violation.DJ}
\end{equation}

The friction response of the mean rate can also be obtained in closed form. At fixed $f$ and $T$, the stationary velocity distribution depends on the friction only through its mean $\vbar=f/\gamma$, with $\gamma\partial_\gamma\vbar=-\vbar$, so that $\gamma\partial_\gamma\jr_{\st}=-\vbar\ev{\partial_v\Phi}_{\st}$. Since $\ev{\partial_v\varphi_n}_{\st}=\delta_{n1}/\sqrt T$ by Eq.~\eqref{eq:violation.basis}, we obtain
\begin{equation}
	\gamma\partial_\gamma\jr_{\st}=-\frac{\vbar a_1}{\sqrt T}=-\sqrt\kappa\,a_1.
\label{eq:violation.response}
\end{equation}
Thus, every coefficient $a_n$ with $n\ge1$ contributes to the diffusivity~\eqref{eq:violation.DJ}, but only $a_1$ enters the friction response. The reason is that the friction shifts the stationary Gaussian without changing its width, and such a shift changes, to first order, only the mean of $\varphi_1$.

We now express both TURs in the long-time limit in terms of the Hermite coefficients. The conventional TUR in its long-time form requires the uncertainty factor $\Qc\equiv D_J\sigma_{\st}/\jr_{\st}^2$ of a current with $\jr_{\st}\ne0$ to satisfy $\Qc\ge1$. Note that $\Var_0(J_\tau)$ and $\Rgamma-\tau\gamma\partial_\gamma\jr_{\st}$ remain bounded for the stationary preparation. Together with Eqs.~\eqref{eq:violation.DJ} and \eqref{eq:violation.response} and $\sigma_{\st}=\gamma\kappa$, the uncertainty factor of the friction-response TUR~\eqref{eq:und.TUR} in the limit $\tau\to\infty$ reads, for $\gamma\partial_\gamma\jr_{\st}\ne0$, i.e., $a_1\ne0$,
\begin{equation}
	\Qc_{\mathrm{fr}}\equiv\frac{D_J\sigma_{\st}}{(\gamma\partial_\gamma\jr_{\st})^2}=1+\sum_{n=2}^\infty\frac{(n-1)!\,a_n^2}{a_1^2}\ge1.
\label{eq:violation.Qfr}
\end{equation}
The first mode alone saturates the friction-response bound, and each higher mode adds fluctuations without adding response. By definition, the two uncertainty factors are related by
\begin{equation}
	\Qc=\Qc_{\mathrm{fr}}\,(\gamma\partial_\gamma\ln|\jr_{\st}|)^2.
\label{eq:violation.Qrelation}
\end{equation}
Since $\Qc_{\mathrm{fr}}\ge1$, the conventional factor obeys $\Qc\ge(\gamma\partial_\gamma\ln|\jr_{\st}|)^2$, and it can fall below unity only if $|\gamma\partial_\gamma\ln|\jr_{\st}||<1$. For the displacement, $\psi=1$, the only nonzero coefficients are $a_0=\vbar$ and $a_1=\sqrt{T}$, and $\Qc=\Qc_{\mathrm{fr}}=1$. The following two examples reduce the logarithmic friction sensitivity of the mean algebraically and exponentially, respectively.

\subsubsection{Algebraic suppression}
A quadratic function already suffices to make $\Qc$ arbitrarily small. We take
\begin{equation}
	\psi(v)=1+c\qty(\frac{v^2}{T}-1),
\label{eq:violation.algebraicweight}
\end{equation}
with a dimensionless coefficient $c$, so that $c=0$ corresponds to the displacement. Writing $v=\vbar+\sqrt T\,\He_1$ and using $\He_1^2=\He_2+1$ and $\He_1^3=\He_3+3\He_1$, we find that the only nonzero Hermite coefficients of $\Phi=(1-c)v+cv^3/T$ are
\begin{align}
	a_0&=\vbar(1+2c+c\kappa),&a_1&=\sqrt T(1+2c+3c\kappa),\notag\\
	a_2&=3c\vbar,&a_3&=c\sqrt T.
\label{eq:violation.algebraiccoeff}
\end{align}
Equation~\eqref{eq:violation.DJ} then gives, without truncation,
\begin{align}
	D_J&=\frac{T}{\gamma}\qty[(1+2c+3c\kappa)^2+c^2(9\kappa+2)],\\
	\Qc&=\frac{(1+2c+3c\kappa)^2+c^2(9\kappa+2)}{(1+2c+c\kappa)^2}.
\end{align}
Minimizing $\Qc$ with respect to $c$ at a given driving strength yields
\begin{equation}
	c_{\mathrm{opt}}=-\frac{2\kappa}{(\kappa+2)(6\kappa+1)},\quad
	\Qc_{\min}=\frac{9\kappa+2}{4\kappa^2+9\kappa+2}.
\label{eq:violation.algebraicmin}
\end{equation}
Near equilibrium, $\Qc_{\min}=1-2\kappa^2+O(\kappa^3)$, so that the violation is weak. Far from equilibrium, $\Qc_{\min}\simeq9/(4\kappa)$, and the conventional bound is violated by the factor $1/\Qc_{\min}\simeq4\kappa/9$, which grows without bound. The optimal coefficient $c_{\mathrm{opt}}$ is negative, so that $\psi$ decreases for high velocities and compensates the growth of the displacement rate with $v$. The rate $\Phi(v)=v\psi(v)$ then becomes nearly flat around the typical velocity, so that velocity fluctuations about $\vbar$ barely change it. The diffusivity is therefore strongly reduced, while the mean rate remains of the same order. This mechanism requires a weight that resolves the velocity, which is not available in the overdamped case. 

For the same current, Eq.~\eqref{eq:violation.response} gives the friction response $\gamma\partial_\gamma\jr_{\st}=-\vbar(1+2c+3c\kappa)$ at fixed $f$, $T$, and $c$, and Eq.~\eqref{eq:violation.Qfr} yields
\begin{equation}
	\Qc_{\mathrm{fr}}=1+\frac{c^2(9\kappa+2)}{(1+2c+3c\kappa)^2}\ge1,
\label{eq:violation.algebraicfr}
\end{equation}
which equals $1/\Qc_{\min}$ at $c=c_{\mathrm{opt}}$. 

\begin{figure}[t]
\centering\includegraphics[width=1.0\linewidth]{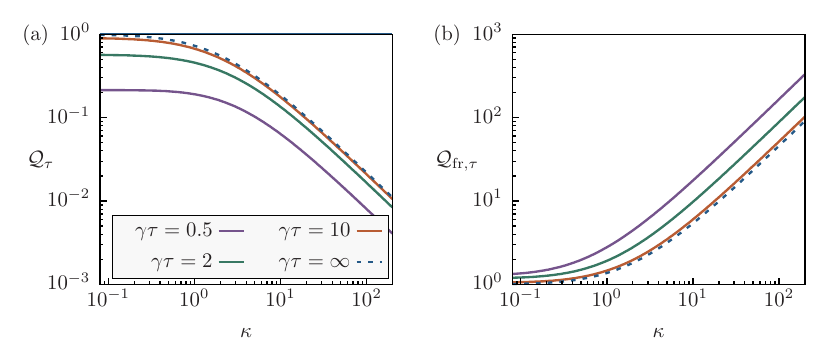}
\caption{Finite-time uncertainty factors for the algebraic velocity-resolved current, with $c=c_{\mathrm{opt}}$ chosen to minimize the long-time conventional factor at each driving strength. (a) Conventional uncertainty factor $\Qc_\tau$ for $\gamma\tau=0.5$, $2$, $10$, and the long-time limit. (b) Preparation-subtracted friction-response factor $\Qc_{\mathrm{fr},\tau}$ for the same currents and observation times. The responses are evaluated with the stationary reference preparation and the current coefficients held fixed. The gray horizontal line marks unity. All curves are exact evaluations of the three-mode Hermite expansion.}
\label{fig:algebraic}
\end{figure}

Finite observation times can be treated exactly in the same basis. Since $\mathcal{P}_s\varphi_n=e^{-\gamma ns}\varphi_n$, the conditional future is $G_t=a_0(\tau-t)+\sum_{n=1}^\infty a_nh_n(\tau-t)\varphi_n$, with $h_n(s)\equiv(1-e^{-\gamma ns})/(\gamma n)$. With $b_n\equiv n!\,a_n^2$ and for the stationary state, we obtain
\begin{align}
	\Var(J_\tau)&=2\sum_{n=1}^\infty b_n\left[\frac{\tau}{\gamma n}-\frac{1-e^{-\gamma n\tau}}{(\gamma n)^2}\right],\\
	\Var_0(J_\tau)&=\sum_{n=1}^\infty b_nh_n(\tau)^2,\\
	\Vd&=2\sum_{n=1}^\infty b_n\gamma n\int_0^\tau h_n(s)^2\dd{s},\\
	\Rgamma&=-\sqrt{\kappa}a_1[\tau-h_1(\tau)].
\label{eq:violation.finitetime}
\end{align}
For the algebraic weight, the sums contain only $n=1,2,3$, with $b_1=T(1+2c+3c\kappa)^2$, $b_2=18c^2\kappa T$, and $b_3=6c^2T$.

We compare the finite-time factors of the conventional TUR and the friction-response TUR,
\begin{equation}
	\Qc_\tau\equiv\frac{\Var(J_\tau)\Sigma_\tau}{2\E[J_\tau]^2},\quad \Qc_{\mathrm{fr},\tau}\equiv\frac{\Vd\Sigma_\tau}{2\Rgamma^2},
\end{equation}
with $\E[J_\tau]=\jr_{\st}\tau$ and $\Sigma_\tau=\sigma_{\st}\tau$. These values tend to $\Qc$ and $\Qc_{\mathrm{fr}}$ as $\tau\to\infty$, respectively. At short times, $J_\tau\simeq\Phi(v_0)\tau$ has a variance of order $\tau^2$, so that
\begin{equation}
	\Qc_\tau\simeq\frac{\gamma\tau}{2}\,\frac{(1+2c+3c\kappa)^2+6c^2(3\kappa+1)}{(1+2c+c\kappa)^2}\to0.
\label{eq:violation.shortQ}
\end{equation}
The friction-response factor instead tends to the finite value $(4/3)[1+c^2(36\kappa+18)/(1+2c+3c\kappa)^2]$, in agreement with Eq.~\eqref{eq:und.Qsub} for $k=1$. Figure~\ref{fig:algebraic} shows both factors at $c=c_{\mathrm{opt}}$: as $\tau$ increases, $\Qc_\tau$ rises monotonically to $\Qc_{\min}$, whereas $\Qc_{\mathrm{fr},\tau}$ decreases monotonically to $1/\Qc_{\min}$ and thus stays above unity at all times.

\subsubsection{Exponential suppression}
Next, we demonstrate that a smooth, bounded weight can suppress the conventional factor exponentially. For $\alpha>0$, we take
\begin{equation}
	\psi(v)=\frac{\erf(\alpha v/\sqrt T)}{v},\quad
	\Phi(v)=\erf(\alpha v/\sqrt T),
\label{eq:violation.erfweight}
\end{equation}
where $\psi(0)=2\alpha/\sqrt{\pi T}$ by continuity, so that the even function $\psi$ is smooth. For large $\alpha$, the rate approaches $\operatorname{sgn}(v)$, and $J_\tau$ approaches the time spent moving forward minus the time spent moving backward. Let $\lambda\equiv2\alpha^2/(1+2\alpha^2)\in(0,1)$ and $\eta\equiv\sqrt{\lambda\kappa/2}$. Since $v/\sqrt T$ is Gaussian with mean $\sqrt\kappa$ and unit variance, the identity $\E[\erf(a+bY)]=\erf(a/\sqrt{1+2b^2})$ for a standard Gaussian variable $Y$ gives $\jr_{\st}=\erf(\eta)$. The mean rate approaches unity at strong driving, while the fluctuations of the bounded rate become very small.

Applying Gaussian integration by parts, $\ev{\Phi\varphi_n}_{\st}=T^{n/2}\ev{\partial_v^n\Phi}_{\st}$, we obtain the Hermite coefficients
\begin{equation}
	a_n=\frac{(-1)^{n-1}}{n!}\frac{2}{\sqrt\pi}e^{-\lambda\kappa/2}\qty(\frac{\lambda}{2})^{n/2}H_{n-1}(\eta)
\label{eq:violation.erfcoeff}
\end{equation}
for $n\ge1$, where $H_n$ is the physicists' Hermite polynomial of degree $n$. Substituting them into Eq.~\eqref{eq:violation.DJ} and shifting $n\to n+1$ yields
\begin{equation}
	D_J=\frac{2\lambda}{\pi\gamma}e^{-\lambda\kappa}
	\sum_{n=0}^\infty\frac{\lambda^nH_n(\eta)^2}{2^n n!(n+1)^2}.
\label{eq:violation.erfseries}
\end{equation}
This series can be written as a single integral. To this end, we write $1/(n+1)^2=-\int_0^1s^n\ln s\dd{s}$, use Mehler's formula, $\sum_{n=0}^\infty (r/2)^nH_n(\eta)^2/n!=(1-r^2)^{-1/2}\exp[2\eta^2r/(1+r)]$ for $|r|<1$, and substitute $r=\lambda s$; with $2\eta^2=\lambda\kappa$, this gives
\begin{equation}
	D_J=\frac{2}{\pi\gamma}\int_0^\lambda
	\frac{\ln(\lambda/r)}{\sqrt{1-r^2}}
	\exp\!\left[-\frac{\lambda\kappa}{1+r}\right]\dd{r}.
\label{eq:violation.erfintegral}
\end{equation}

At strong driving, the integrand is concentrated near the upper limit $r=\lambda$, where the exponent $-\lambda\kappa/(1+r)$ is largest and the logarithmic prefactor vanishes. Expanding the integrand there gives
\begin{align}
	D_J&\simeq\frac{\ell(\lambda)}\gamma\kappa^{-2}e^{-\lambda\kappa/(1+\lambda)},\\
	\Qc&\simeq\ell(\lambda)\kappa^{-1}e^{-\lambda\kappa/(1+\lambda)},
\label{eq:violation.erfasympt}
\end{align}
as $\kappa\to\infty$ at fixed $0<\lambda<1$. Here, $\ell(\lambda)\equiv 2(1+\lambda)^4/(\pi\lambda^3\sqrt{1-\lambda^2})$. Since $\sigma_{\st}=\gamma\kappa$, the conventional uncertainty factor decays exponentially with the dissipation rate.

The friction-response factor behaves in the opposite way. Equation~\eqref{eq:violation.response} with $a_1=(2\lambda/\pi)^{1/2}e^{-\lambda\kappa/2}$, or direct differentiation of $\erf(\eta)$ with $\gamma\partial_\gamma\eta=-\eta$, gives
\begin{equation}
	\gamma\partial_\gamma\jr_{\st}
	=-\sqrt{\frac{2\lambda\kappa}{\pi}}e^{-\lambda\kappa/2},
\label{eq:violation.erfresponse}
\end{equation}
and Eq.~\eqref{eq:violation.Qfr} becomes
\begin{equation}
	\Qc_{\mathrm{fr}}
	=1+\sum_{n=1}^\infty\frac{\lambda^nH_n(\eta)^2}{2^n n!(n+1)^2}\ge1.
\label{eq:violation.erffr}
\end{equation}
Equation~\eqref{eq:violation.Qrelation} shows where the exponential suppression comes from. The squared logarithmic sensitivity $(\gamma\partial_\gamma\ln|\jr_{\st}|)^2=(2\lambda/\pi)\kappa e^{-\lambda\kappa}/\erf(\eta)^2$ decays as $\kappa e^{-\lambda\kappa}$, faster than $\Qc$ itself, whereas Eqs.~\eqref{eq:violation.erfasympt} and \eqref{eq:violation.Qrelation} give $\Qc_{\mathrm{fr}}\simeq[\pi\ell(\lambda)/(2\lambda)]\kappa^{-2}e^{\lambda^2\kappa/(1+\lambda)}$, which grows exponentially. Once the particle almost always moves forward, the mean rate saturates at unity and barely responds to the friction. The conventional factor ignores this insensitivity, whereas the friction-response factor accounts for it. 

These examples show that, in underdamped dynamics, the precision of a current depends on how it resolves the velocity. Reference~\cite{Fischer.2020.PRE} conjectured that, in driven one-dimensional diffusion, the precision of currents $J_\tau=\int_0^\tau\theta(x_t)v_t^n\dd{t}$ with a single odd power $n$ is bounded by that of free diffusion. Equations~\eqref{eq:violation.algebraicmin} and \eqref{eq:violation.erfasympt} show that no universal lower bound extends to arbitrary rates that are odd in the velocity: already for the driven free particle, which is the reference system of that conjecture, a quadratic or a bounded velocity weight lowers $\Qc$ far below unity at strong driving. The origin of the violations also differs from that of the pendulum clocks of Ref.~\cite{Pietzonka.2022.PRL}, which combine underdamped oscillations with a ticking mechanism to achieve precise directed motion at low dissipation. Here, the dynamics is linear and Gaussian, and the gain in precision comes entirely from the nonlinear velocity weighting of the measured current.

\section{Overdamped Langevin dynamics}\label{sec:over}

\subsection{Dynamics, observables, and variance identity}
We now turn to overdamped dynamics, in which the velocity relaxes much faster than the observed time scales. The dynamics is described by the Langevin equation
\begin{equation}
	\dd{\bx_t}=\bb_t(\bx_t)\dd{t}+\sqrt{2\D}\dd{\bW_t},
\label{eq:od.dynamics}
\end{equation}
where $\bb_t$ is the drift and $\D$ is a constant positive-definite diffusion matrix. For a scalar mobility $\mu$ and a mechanical force $\bF_t$, one has $\bb_t=\mu\bF_t$ and, in a homogeneous thermal environment, $\D=\mu\T$. In the overdamped limit of Eq.~\eqref{eq:und.dynamics}, $\mu=1/\gamma$. The probability density $p_t(\bx)$ evolves according to the Fokker-Planck equation $\partial_tp_t=-\nabla\cdot\bj_t$, where $\bj_t=\bb_tp_t-\D\nabla p_t$ is the probability current, and the forward and backward generators are
\begin{equation}
\begin{gathered}
	\Lg_t p=-\nabla\cdot(\bb_t p)+\D:\nabla\nabla p,\\
	\Lg_t^\dagger a=\bb_t\cdot\nabla a+\D:\nabla\nabla a.
\end{gathered}
\end{equation}
The local mean velocity and the entropy production therefore read
\begin{align}
	\bnu_t&=\bj_t/p_t=\bb_t-\D\nabla\ln p_t,\label{eq:od.nu}\\
	\Sigma_\tau&=\int_0^\tau\ev{\bnu_t^\top\D^{-1}\bnu_t}_t\dd{t}.\label{eq:od.entropy}
\end{align}

In contrast to the underdamped case, a position current directly samples the noise that drives the state. We consider the observables
\begin{equation}
	J_\tau=\int_0^\tau\phi_t(\bx_t)\dd{t}+\int_0^\tau\bw_t(\bx_t)\circ\dd{\bx_t},
\label{eq:od.J}
\end{equation}
whose effective rate is $\Phi_t=\phi_t+\bw_t\cdot\bb_t+\D:\nabla\bw_t$. Since every coordinate is noisy, the diffusion realization of Sec.~\ref{sec:general.realizations} applies with $\bq=\bx$ and no deterministic coordinates. The martingale and its amplitude read
\begin{equation}
	\dd{M_t}=\bu_t^\top\sqrt{2\D}\dd{\bW_t},\quad\bu_t=\bw_t+\nabla G_t,
\label{eq:od.M}
\end{equation}
respectively, and the dynamical variance becomes
\begin{gather}
	\Var(J_\tau)-\Var_0(J_\tau)=2\int_0^\tau\ev{\bu_t^\top\D\bu_t}_t\dd{t}.
\label{eq:od.variance}
\end{gather}
This identity holds at finite times, without any stationarity or long-time assumption. As discussed in detail below, Eq.~\eqref{eq:od.variance} extends the stationary overdamped relation obtained in Ref.~\cite{Chun.2026.arxiv} to finite times.

\subsection{Fluctuation-response relations}
\label{sec:over.injection}

Force, mobility, and temperature changes modify different terms of Eq.~\eqref{eq:od.dynamics}, but each affects the mean of an observable only through the probability current that it injects at fixed density. This source current puts all three on the same footing. For a perturbation in the direction $(\delta\bb_t,\delta\D)$, i.e., $\bb_t\to\bb_t+\eps\delta\bb_t$ and $\D\to\D+\eps\delta\D$ with a uniform $\delta\D$, the source current~\eqref{eq:general.source} and the response~\eqref{eq:general.diffresponse} read
\begin{equation}
\begin{gathered}
	\delta\bj_{t,\mathrm{src}}=\delta\bb_t p_t-\delta\D\nabla p_t,\\
	\delta\E[J_\tau]=\int_0^\tau\inprod{\bu_t}{\delta\bj_{t,\mathrm{src}}}\dd{t}.
\end{gathered}
\label{eq:od.response}
\end{equation}
The martingale amplitude that generates the fluctuations in Eq.~\eqref{eq:od.variance} is thus also the susceptibility of the mean to an injected current: the local response $\vb*{R}_{\mathrm{src}}(t,\bx)\equiv\delta\E[J_\tau]/\delta\bj_{t,\mathrm{src}}(\bx)$ equals $\bu_t(\bx)$. Equation~\eqref{eq:od.variance} therefore expresses the preparation-subtracted variance entirely through local responses,
\begin{equation}
	\Var(J_\tau)-\Var_0(J_\tau)=2\int_0^\tau\ev{\vb*{R}_{\mathrm{src}}^\top\D\,\vb*{R}_{\mathrm{src}}}_t\dd{t},
\label{eq:od.FRR}
\end{equation}
which is the overdamped form of Eq.~\eqref{eq:general.FRRdiff} and the counterpart of Eq.~\eqref{eq:und.FRR}. In contrast to the underdamped case, a current can be injected along every direction of position space, because every coordinate is noisy.

When the injected currents are generated by physical parameters, Eq.~\eqref{eq:od.FRR} becomes a finite-time generalization of the overdamped fluctuation-response relation of Ref.~\cite{Chun.2026.arxiv}, which was obtained for stationary processes in the long-time limit. Following Ref.~\cite{Chun.2026.arxiv}, we consider a one-dimensional system with a position-dependent mobility $\mu(x)$, temperature $T(x)$, and force $f(x)$, and define the process by its current $j=\mu(fp-T\partial_xp)$, which corresponds to the anti-It\^o convention used there. The same process has the It\^o drift $\mu f+\partial_xD$ with $D=\mu T$, so that Eq.~\eqref{eq:general.source} applies. Let $R_\theta(t,x)\equiv\delta\E[J_\tau]/\delta\theta_t(x)$ be the response of the mean to a local, instantaneous change of $\theta\in\{f,\mu,T\}$ at fixed observable and initial density. A change of $\mu$ or $T$ alters both $D$ and the It\^o drift, but the terms with $\partial_x\delta D$ cancel in Eq.~\eqref{eq:general.source}, and the injected source current is the variation of $j$ at fixed density, $N_\theta\delta\theta$, with $N_f=\mu p_t$, $N_\mu=j_t/\mu$, and $N_T=-\mu\partial_xp_t$. Each response is therefore the martingale amplitude multiplied by a known factor, $R_\theta=N_\theta u$, and Eq.~\eqref{eq:od.FRR} becomes
\begin{equation}
	\Var(J_\tau)-\Var_0(J_\tau)=2\int_0^\tau\int p_tD\left[\frac{R_\theta(t,x)}{N_\theta(t,x)}\right]^2\dd{x}\dd{t},
\label{eq:od.localFRR}
\end{equation}
provided that $N_\theta\ne0$ wherever $p_t>0$. Where $N_\theta$ vanishes, as $N_\mu$ does in the absence of a probability current, the amplitude must be obtained from another parameter, such as the force, for which $N_f=\mu p_t>0$. For a process started in its stationary state and in the long-time limit $\tau\to\infty$, the preparation variance remains bounded, and $R_\theta/N_\theta$ approaches its stationary value. Dividing Eq.~\eqref{eq:od.localFRR} by $\tau$ then gives the long-time variance relation of Ref.~\cite{Chun.2026.arxiv}. Equation~\eqref{eq:od.localFRR} thus extends that relation to finite times and to arbitrary initial densities. At finite times, the preparation variance is in general nonzero and must be subtracted from the variance. The same holds for covariances between different observables. Since $R_\theta^\alpha=N_\theta u^\alpha$ for every observable $J_\tau^\alpha$ and every parameter $\theta$, polarizing Eq.~\eqref{eq:od.localFRR}, as for the covariance matrix $\Cm-\Cm_0=[\Bform(J^\alpha,J^\beta)]$ in Eq.~\eqref{eq:general.covariance}, gives
\begin{equation}
	\Cov(J_\tau^\alpha,J_\tau^\beta)-\Cov_0(J_\tau^\alpha,J_\tau^\beta)=2\int_0^\tau\int p_tD\frac{R_\theta^\alpha R_{\theta'}^\beta}{N_\theta N_{\theta'}}\dd{x}\dd{t},
\label{eq:od.localcov}
\end{equation}
where the responses of the two observables may be taken with respect to different parameters. In the stationary long-time limit, Eq.~\eqref{eq:od.localcov} divided by $\tau$ reduces to the covariance relation of Ref.~\cite{Chun.2026.arxiv}.

For simplicity, we return to the position-independent diffusion matrix of Eq.~\eqref{eq:od.dynamics}; the results below extend straightforwardly to a position-dependent mobility that multiplies the whole current, with $\partial_{\ln\mu}$ denoting a uniform rescaling. A uniform change of mobility has half the entropy production as its cost, for any initial density. Rescaling $\mu\to(1+\eps)\mu$ at fixed force and temperature injects the source current $\delta\bj_{t,\mathrm{src}}=\bj_t$, i.e., $\bg_t=\bnu_t$, whose cost is $(1/2)\int_0^\tau\ev*{\bnu_t^\top\D^{-1}\bnu_t}_t\dd{t}=\Sigma_\tau/2$. Hence, we obtain
\begin{equation}
	2(\partial_{\ln\mu}\E[J_\tau])^2\le\Vd\Sigma_\tau,
\label{eq:od.mobility}
\end{equation}
which is the overdamped counterpart of the friction-response TUR~\eqref{eq:und.TUR}, since $\gamma\partial_\gamma=-\partial_{\ln\mu}$. More generally, the mobility can be perturbed nonuniformly. If the mobility multiplies the whole current, $\bj=\mu(\bF p-\T\nabla p)$, a perturbation profile $\delta\ln\mu=\psi_t(\bx)$ injects the source current $\psi_t\bj_t$, i.e., $\bg_t=\psi_t\bnu_t$, with the cost $(1/2)\int_0^\tau\ev*{\psi_t^2\bnu_t^\top\D^{-1}\bnu_t}_t\dd{t}\le\psi_{\mathrm{max}}^2\Sigma_\tau/2$, where $\psi_{\mathrm{max}}\equiv\sup_{t,\bx}|\psi_t(\bx)|$. Its response, $\delta\E[J_\tau]=\int_0^\tau\ev*{\psi_t\bu_t\cdot\bnu_t}_t\dd{t}$, therefore obeys $2(\delta\E[J_\tau])^2\le\psi_{\mathrm{max}}^2\Vd\Sigma_\tau$, which reduces to Eq.~\eqref{eq:od.mobility} for $\psi_t=1$. Bounding $\Vd$ from above by $\Var(J_\tau)$ gives the response TUR of Ref.~\cite{Chun.2026.arxiv}, whereas subtracting $\Var_0(J_\tau)$ sharpens it at finite times. Force and temperature perturbations similarly lead to response inequalities, but with their own costs, which are not the entropy production. 

The mobility-response TUR~\eqref{eq:od.mobility} can always be saturated at finite times, like its underdamped counterpart. Since the uniform mobility change has the source $\bg_t=\bnu_t$, equality requires $\bu_t\propto\D^{-1}\bnu_t$. By Eq.~\eqref{eq:od.nu}, $\D^{-1}\bnu_t=\D^{-1}\bb_t+\nabla\Lambda_t$ with $\Lambda_t\equiv-\ln p_t$. In contrast to the underdamped case, the required amplitude is not a gradient when the drift is nonconservative, but its non-gradient part $\D^{-1}\bb_t$ can be carried by the current weight. A saturating observable can be constructed as
\begin{equation}
\begin{gathered}
	\bw_t^{\mathrm{opt}}=\D^{-1}\bb_t+\nabla\Lambda_\tau,\\
	G_t^{\mathrm{opt}}=\Lambda_t-\Lambda_\tau,\\
	\Phi_t^{\mathrm{opt}}=-\partial_tG_t^{\mathrm{opt}}-\Lg_t^\dagger G_t^{\mathrm{opt}},\\
	\phi_t^{\mathrm{opt}}=\Phi_t^{\mathrm{opt}}-\bw_t^{\mathrm{opt}}\cdot\bb_t-\D:\nabla\bw_t^{\mathrm{opt}}.
\end{gathered}
\label{eq:od.optimal}
\end{equation}
Hence, $\bu_t^{\mathrm{opt}}=\bw_t^{\mathrm{opt}}+\nabla G_t^{\mathrm{opt}}=\D^{-1}\bb_t+\nabla\Lambda_t=\D^{-1}\bnu_t$. As in the underdamped case, all quantities are computed from the reference process and kept fixed when the mobility is changed. Equations~\eqref{eq:od.variance} and \eqref{eq:general.diffresponse} then give $\Vd^{\mathrm{opt}}=2\Sigma_\tau$ and $\partial_{\ln\mu}\E[J_\tau^{\mathrm{opt}}]=\Sigma_\tau$, so that Eq.~\eqref{eq:od.mobility} holds with equality, and
\begin{equation}
	\Sigma_\tau=\sup_{\phi,\bw}\frac{2(\partial_{\ln\mu}\E[J_\tau])^2}{\Var(J_\tau)-\Var_0(J_\tau)}.
\label{eq:od.entropyvariational}
\end{equation}
For pure currents, $\phi_t=0$, the function $G_t$ is fixed by $\bw_t$, and the condition $\bw_t+\nabla G_t\propto\D^{-1}\bnu_t$ cannot be met in general; Sec.~\ref{sec:over.tur} discusses when it can.

\subsection{Improved finite-time TUR and its equality conditions}\label{sec:over.tur}
The mobility bound~\eqref{eq:od.mobility} requires changing the mobility. For currents, the same response can instead be obtained by controlling the timing, which yields a finite-time TUR for time-dependent driving. Let the dynamics follow an external protocol, $\Lg_t=\Lg_{\lambda(\upsilon t)}$, where $\upsilon$ sets its speed, and consider pure time-antisymmetric currents, $\phi_t=0$, whose weights depend on time, if at all, only through the protocol; the initial density $p_0$ is held fixed. Under the change $(\tau,\upsilon)\mapsto((1+\eps)\tau,\upsilon/(1+\eps))$, the rescaled time $s=t/(1+\eps)\in[0,\tau]$ sees the same protocol $\lambda(\upsilon s)$ and the same weights, while the drift and the diffusion matrix are both multiplied by $1+\eps$. Because a current $\int\bw_t\circ\dd{\bx_t}$ does not depend on the time parametrization, the current $J_{(1+\eps)\tau}$ accumulated under the slower protocol has the same distribution as the current $J_\tau$ under the original protocol with the mobility multiplied by $1+\eps$. The derivative $(\tau\partial_\tau-\upsilon\partial_\upsilon)\E[J_\tau]$ is therefore the response to the uniform mobility change of Eq.~\eqref{eq:od.mobility}, whose source current is $\bj_t$, and Eq.~\eqref{eq:general.diffresponse} gives
\begin{equation}
	(\tau\partial_\tau-\upsilon\partial_\upsilon)\E[J_\tau]=\Rc[\bnu]=\int_0^\tau\ev{\bu_t\cdot\bnu_t}_t\dd{t}.
\label{eq:od.clock}
\end{equation}
With $\bu_t=\bw_t+\nabla G_t$ and $p_t\bnu_t=\bj_t$, the right-hand side splits into $\int_0^\tau\inprod{\bw_t}{\bj_t}\dd{t}$ and $\int_0^\tau\inprod{\nabla G_t}{\bj_t}\dd{t}$. The first term is $\E[J_\tau]$, because the mean rate of a current is $\inprod{\bw_t}{\bj_t}$. In the second term, an integration by parts and the continuity equation $\partial_tp_t=-\nabla\cdot\bj_t$ give $\inprod{\nabla G_t}{\bj_t}=\inprod{G_t}{\partial_tp_t}$. Hence,
\begin{equation}
	(\tau\partial_\tau-\upsilon\partial_\upsilon)\E[J_\tau]=\E[J_\tau]+\int_0^\tau\inprod{G_t}{\partial_tp_t}\dd{t}.
\end{equation}
The second term arises because the density evolves faster, which changes the expected future accumulation. For a stationary process, $\partial_tp_t=0$, and the response reduces to the mean, as in the conventional TUR.

Combining Eq.~\eqref{eq:od.clock} with the variance identity~\eqref{eq:od.variance} and applying the Cauchy-Schwarz inequality yields the improved finite-time TUR
\begin{equation}
	\frac{\Var(J_\tau)-\Var_0(J_\tau)}{((\tau\partial_\tau-\upsilon\partial_\upsilon)\E[J_\tau])^2}\ge\frac{2}{\Sigma_\tau}.
\label{eq:od.TUR}
\end{equation}
Replacing $\Var(J_\tau)-\Var_0(J_\tau)$ by $\Var(J_\tau)$ gives the TUR obtained in Ref.~\cite{Koyuk.2020.PRL} for time-dependent driving, with the same conventions for the initial density and the protocol; Eq.~\eqref{eq:od.TUR} sharpens it by the preparation variance. For pure currents, Eqs.~\eqref{eq:od.mobility} and \eqref{eq:od.TUR} contain the same response and differ only operationally: there, it is measured by changing the mobility, whereas here it is obtained by changing the observation time and the protocol speed. The two responses differ for observables with a residence part, because the time change also stretches the integration measure of $\phi_t$, whereas a mobility change leaves the instantaneous weight fixed. In general,
\begin{equation}
	\Rc[\bnu]=(\tau\partial_\tau-\upsilon\partial_\upsilon)\E[J_\tau]-\int_0^\tau\ev{\phi_t}_t\dd{t},
\label{eq:od.mixedclock}
\end{equation}
and the last term vanishes for the pure currents used in Eq.~\eqref{eq:od.TUR}.

We now determine the saturation condition of Eq.~\eqref{eq:od.TUR}. Let $\varsigma=\Rc[\bnu]/\Sigma_\tau$, and decompose the martingale amplitude into a part aligned with the irreversible current and a remainder,
\begin{equation}
	\bu_t=\varsigma\,\D^{-1}\bnu_t+\bu_t^{\perp}.
\label{eq:od.udecomp}
\end{equation}
Since $\Rc[\bnu]=\int_0^\tau\ev*{\bu_t\cdot\bnu_t}_t\dd{t}=\varsigma\,\Sigma_\tau+\int_0^\tau\ev*{\bu_t^{\perp}\cdot\bnu_t}_t\dd{t}$, the remainder satisfies $\int_0^\tau\ev*{\bu_t^{\perp}\cdot\bnu_t}_t\dd{t}=0$. The cross term in Eq.~\eqref{eq:od.variance} therefore integrates to zero, and
\begin{equation}
	\Vd-\frac{2\Rc[\bnu]^2}{\Sigma_\tau}=2\int_0^\tau\ev{(\bu_t^{\perp})^\top\D\bu_t^{\perp}}_t\dd{t}.
\label{eq:od.satdecomp}
\end{equation}
For any driving and any initial density, equality in Eq.~\eqref{eq:od.TUR} thus holds if and only if $\bu_t^{\perp}=0$ on the support of $p_t$ at all times, i.e., if the martingale of $J_\tau$ is proportional to the martingale part of the stochastic entropy production, $\dd{M_t}\propto \bnu_t^\top\D^{-1/2}\dd{\bW_t}$.

For pure currents (i.e., $\phi_t=0$), $G_t$ is fixed by the weight, so that the condition $\bu_t^{\perp}=0$ constrains $\bw_t$. Since $\D^{-1}\bnu_t=\D^{-1}\bb_t-\nabla\ln p_t$, the condition reads $\nabla(G_t+\varsigma\ln p_t)=\varsigma\D^{-1}\bb_t-\bw_t$. The weight must therefore have the form $\bw_t=\varsigma\D^{-1}\bb_t-\nabla \Upsilon_t$, where $\Upsilon_t$, which is fixed by $\bw_t$ only up to a function of time, can be chosen such that $G_t= \Upsilon_t-\varsigma\ln p_t$. Inserting this into the backward equation~\eqref{eq:general.backward}, with $\Phi_t=\bw_t\cdot\bb_t+\D:\nabla\bw_t$, gives $\partial_t\Upsilon_t+\varsigma s_t=0$, where
\begin{equation}
	s_t(\bx)\equiv\bnu_t^\top\D^{-1}\bnu_t-2\partial_t\ln p_t.
\label{eq:od.satcond}
\end{equation}
To see this, note that $\Phi_t=\varsigma(\bb_t^\top\D^{-1}\bb_t+\nabla\cdot\bb_t)-\Lg_t^\dagger \Upsilon_t$ for a constant $\D$, so that the terms $\Lg_t^\dagger \Upsilon_t$ cancel in the backward equation and lead to $\partial_t\Upsilon_t+\varsigma[\bb_t^\top\D^{-1}\bb_t+\nabla\cdot\bb_t-(\partial_t+\Lg_t^\dagger)\ln p_t]=0$. The Fokker-Planck equation gives $(\partial_t+\Lg_t^\dagger)\ln p_t=-\nabla\cdot\bb_t+2\D:\nabla\nabla\ln p_t+\nabla\ln p_t^\top\D\nabla\ln p_t$, and with $\bnu_t=\bb_t-\D\nabla\ln p_t$, the term inside the bracket equals $s_t$. The function $s_t(\bx)$ is the expected rate of the stochastic entropy production, $\int_0^t\bnu_r^\top\D^{-1}\circ\dd{\bx_r}-\int_0^t\partial_r\ln p_r(\bx_r)\dd{r}$, from the state $\bx$: in It\^o form, its drift is $\bnu_t^\top\D^{-1}\bb_t+\nabla\cdot\bnu_t-\partial_t\ln p_t$, which equals $s_t$ by the Fokker-Planck equation $\partial_tp_t=-\nabla\cdot(p_t\bnu_t)$. With the terminal condition $G_\tau=0$, i.e., $\Upsilon_\tau=\varsigma\ln p_\tau$, we obtain $\Upsilon_t=\varsigma(\ln p_\tau+\int_t^\tau s_r\dd{r})$, and the saturating weight is
\begin{equation}
	\bw_t^{\mathrm{opt}}=\varsigma\qty(\D^{-1}\bb_t-\nabla\ln p_\tau-\int_t^\tau\nabla s_r\dd{r}).
\label{eq:od.wopt}
\end{equation}
This weight enters Eq.~\eqref{eq:od.TUR} only if it depends on time through the protocol. For a \emph{strictly monotonic} protocol, any time-dependent weight is a function of the protocol, $\bw_t(\bx)=\hat{\bw}_{\lambda(\upsilon t)}(\bx)$, which is constructed for the reference process and held fixed when $\tau$ and $\upsilon$ are varied, so that the TUR~\eqref{eq:od.TUR} can always be saturated by a pure current at finite times. The preparation variance of this current is the variance of its conditional mean given the initial state, $\Var_0(J_\tau)=\Var[G_0(\bx_0)]$ with $\bx_0\sim p_0$. Setting $t=0$ in $G_t=\varsigma(\ln p_\tau-\ln p_t+\int_t^\tau s_r\dd{r})$ gives
\begin{equation}
	\Var_0(J_\tau)=\varsigma^2\Var\qty[\ln\frac{p_\tau(\bx_0)}{p_0(\bx_0)}+\int_0^\tau s_r(\bx_0)\dd{r}],
\end{equation}
where $s_r$ is evaluated at the initial state $\bx_0$ for all $r$, not along the trajectory. This variance is in general nonzero, so that the same current does not saturate the TUR of Ref.~\cite{Koyuk.2020.PRL}, which uses the full variance. 

For autonomous dynamics, by contrast, the weight must be time independent, which requires $\nabla s_t=0$ at all times. Writing $\rho_t=\sqrt{p_t}$, one finds $s_t=U-4\D:\nabla\nabla\rho_t/\rho_t$ with $U\equiv\bb^\top\D^{-1}\bb+2\nabla\cdot\bb$. A spatially uniform value, $s_t(\bx)=\bar s_t$, is therefore equivalent to the eigenvalue equation
\begin{equation}
	\hat H\rho_t=\bar s_t\rho_t,\qquad \hat H\equiv-4\D:\nabla\nabla+U.
\label{eq:od.schrodinger}
\end{equation}
For autonomous dynamics, $\hat H$ is a time-independent Schr\"odinger operator~\cite{Risken.1989}, self-adjoint with respect to $\int\dd{\bx}$, and $\rho_t$ is a positive eigenfunction of it with $\int\rho_t^2\dd{\bx}=1$. We assume a connected state space on which $\hat H$ has a discrete spectrum, as on a compact state space or on $\mathbb{R}^d$ with $U(\bx)\to\infty$ as $\|\bx\|\to\infty$. The lowest eigenvalue $s_{\mathrm{gs}}$ of $\hat H$ is the minimum of $\int(4\nabla\rho^\top\D\nabla\rho+U\rho^2)\dd{\bx}$ over functions with $\int\rho^2\dd{\bx}=1$. Since replacing $\rho$ by $|\rho|$ does not change this functional, the corresponding eigenfunction $\rho_{\mathrm{gs}}$ can be chosen nonnegative, and on a connected state space it is strictly positive and unique up to normalization~\cite{Reed.1978}. Every eigenfunction of a higher eigenvalue is orthogonal to $\rho_{\mathrm{gs}}$ and must therefore take both signs, because the integral of a product of two positive functions cannot vanish. The ground state is thus the only positive eigenfunction of $\hat H$. Consequently, $\bar s_t=s_{\mathrm{gs}}$ and $\rho_t=\rho_{\mathrm{gs}}$ for all $t\in(0,\tau]$, where $p_t>0$, and by continuity also at $t=0$. The density $p_t=\rho_{\mathrm{gs}}^2$ thus does not change in time, and the process must be stationary. In the steady state, $s_t=\bnu_{\st}^\top\D^{-1}\bnu_{\st}$, so that a uniform $s_t$ is a spatially uniform local entropy production rate, equal to its mean $\sigma_{\st}$. Then, $G_t=\varsigma\sigma_{\st}(\tau-t)$, so that $\nabla G_t=0$, and the weight~\eqref{eq:od.wopt} reduces to $\bw=\varsigma\D^{-1}\bnu_{\st}$. Finally, $\Rc[\bnu]=\varsigma\Sigma_\tau$ with $\Rc[\bnu]=\E[J_\tau]=\jr_{\st}\tau$ and $\Sigma_\tau=\sigma_{\st}\tau$ fixes $\varsigma=\jr_{\st}/\sigma_{\st}$. Saturation by a pure current with a time-independent weight is thus possible only in a steady state with uniform local entropy production, and the weight $\bw=(\jr_{\st}/\sigma_{\st})\D^{-1}\bnu_{\st}$ is aligned with the irreversible current of entropy production. This is consistent with Ref.~\cite{Busiello.2019.PRE}, where the most precise current of a steady state, the hyperaccurate current, coincides with the entropy production only when the latter saturates the TUR.

\section{Markov jump processes}\label{sec:jump}

\subsection{Dynamics, observables, and variance identity}
Finally, we consider Markov jump processes on a finite state space. The rates $w_{mn}(t)$ of the transitions $n\to m$ define the master equation $\partial_t\vb*p_t=\mathsf{W}_t\vb*p_t$, with $(\mathsf{W}_{t})_{mn}=w_{mn}(t)$ for $m\ne n$ and $(\mathsf{W}_{t})_{nn}=-\sum_{m(\ne n)}w_{mn}(t)$. We assume irreducible dynamics in which every transition has a reverse, $w_{mn}>0$ if and only if $w_{nm}>0$. In terms of the current $j_{mn}$ and the traffic $a_{mn}$ defined in Sec.~\ref{sec:general.realizations}, which measure the net transport along an edge and how often either of its transitions occurs, the entropy production and the dynamical activity read \cite{Schnakenberg.1976.RMP,Shiraishi.2018.PRL,Vu.2023.PRX,Maes.2020.PR}
\begin{align}
	\Sigma_\tau&=\int_0^\tau\sum_{m<n}j_{mn}\ln\frac{w_{mn}p_n}{w_{nm}p_m}\dd{t},\label{eq:jump.entropy}\\
	A_\tau&=\int_0^\tau\sum_{m<n}a_{mn}\dd{t}.\label{eq:jump.dynact}
\end{align}
Under local detailed balance, the logarithmic ratio has its usual thermodynamic interpretation; the variance identity below requires only a well-defined Markov process.

We consider the additive observables~\eqref{eq:general.Jjump}, which accumulate the increment $d_{mn}(t)$ at each transition $n\to m$ and the weight $\phi_n(t)$ during residence in state $n$. The residence term describes occupation times, and the jump term describes particle, chemical, or cycle currents, so that current-current, state-state, and mixed correlations are treated within a single construction. For antisymmetric increments, $d_{mn}=-d_{nm}$, the mean rate is $\sum_np_n\phi_n+\sum_{m<n}d_{mn}j_{mn}$. By Eqs.~\eqref{eq:general.jumpdM} and \eqref{eq:general.jumpamp}, the Doob martingale is driven by the compensated counting processes~\eqref{eq:general.counting},
\begin{equation}
\begin{gathered}
	\dd{M_t}=\sum_{m\ne n}u_{mn}(t)\dd{\widetilde{N}_{mn}(t)},\\
	u_{mn}(t)=d_{mn}(t)+G_t(m)-G_t(n),
\end{gathered}
\label{eq:jump.kernel}
\end{equation}
where the amplitude $u_{mn}$ combines the direct increment of a jump $n\to m$ with the change of the expected future accumulation that the jump causes.

Since the compensated counting processes of distinct transitions are orthogonal and the variance rate of each equals its rate, Eq.~\eqref{eq:general.var} gives the exact finite-time decomposition
\begin{equation}
	\Var(J_\tau)=\Var_0(J_\tau)+\Vd
\label{eq:jump.decomp}
\end{equation}
with
\begin{align}
	\Var_0(J_\tau)&=\sum_np_n(0)\qty(G_0(n)-\E[J_\tau])^2,\label{eq:jump.prepvar}\\
	\Vd&=\int_0^\tau\sum_{m\ne n}w_{mn}p_nu_{mn}^2\dd{t}.\label{eq:jump.dynvar}
\end{align}
For antisymmetric increments, which include currents, residence observables, and their combinations, the amplitude is antisymmetric, $u_{mn}=-u_{nm}$, and the two directions of each edge combine into the traffic,
\begin{equation}
	\Vd=\int_0^\tau\sum_{m<n}a_{mn}u_{mn}^2\dd{t}.
\label{eq:jump.variance}
\end{equation}
Here, $G_0(n)$ is the expected value of $J_\tau$ conditioned on the initial state $n$, so that the preparation variance coincides with the initial variability in the dynamical fluctuation-response relation of Ref.~\cite{Aslyamov.2026.arxiv}. Equations~\eqref{eq:jump.decomp}--\eqref{eq:jump.variance} thus recover that finite-time identity and express its dynamical part through the martingale amplitude.

The dynamical variance~\eqref{eq:jump.dynvar} also applies to observables whose increments are not antisymmetric, such as the dynamical activity, $d_{mn}=1$ for all transitions, and other counting observables. Their martingale amplitudes need not change sign under edge reversal, and their responses must be expressed through independent changes of the two directed rates of each edge. The traffic form~\eqref{eq:jump.variance} and the thermodynamic bounds below apply to antisymmetric increments.

\subsection{Fluctuation-response relations}
We consider transition rates that depend on parameters attached to the transitions, $w_{mn}=w_{mn}(\theta_{mn})$, and perturb them as $\theta_{mn}\to\theta_{mn}+\eps\delta\theta_{mn}$, with arbitrary dependence on time and on the transition, so that $\delta w_{mn}=\partial_\theta w_{mn}\delta\theta_{mn}$, where $\partial_\theta$ denotes the derivative with respect to the parameter of the respective transition; a global parameter corresponds to the same $\delta\theta_{mn}$ for all transitions. By Eq.~\eqref{eq:general.jumpresponse}, the response of an observable with general increments is
\begin{equation}
	\delta\E[J_\tau]=\int_0^\tau\sum_{m\ne n}u_{mn}p_n\delta w_{mn}\dd{t},
\label{eq:jump.responsegen}
\end{equation}
where $p_n\delta w_{mn}$ is the change of the directed flux $n\to m$ at fixed density. If $\theta_{mn}$ affects only the transition $n\to m$, its local response, $R_{\theta,mn}(t)\equiv\delta\E[J_\tau]/\delta\theta_{mn}(t)$, is the martingale amplitude multiplied by the directed flux that a unit change injects, $R_{\theta,mn}=u_{mn}p_n\partial_\theta w_{mn}$. Wherever $\partial_\theta w_{mn}\ne0$, the dynamical variance~\eqref{eq:jump.dynvar} therefore becomes
\begin{equation}
	\Var(J_\tau)-\Var_0(J_\tau)=\int_0^\tau\sum_{m\ne n}\frac{R_{\theta,mn}^2}{w_{mn}p_n(\partial_\theta\ln w_{mn})^2}\dd{t},
\label{eq:jump.localFRRgen}
\end{equation}
which holds for every observable and is the discrete counterpart of Eq.~\eqref{eq:od.localFRR}. For $\theta_{mn}=\ln w_{mn}$, the denominator reduces to the directed flux $w_{mn}p_n$. A parameter that changes both rates of an edge mixes $u_{mn}$ and $u_{nm}$ in its response, so that two independent parameters per edge are needed to reconstruct a general amplitude.

In the following, we focus on observables with antisymmetric increments, which include currents and residence observables; a similar discussion applies to general increments, except that the cost of a kinetic change then becomes the traffic. Since their amplitudes are antisymmetric, $u_{nm}=-u_{mn}$, the two directed terms of each edge in Eq.~\eqref{eq:jump.responsegen} combine, and the response depends on the perturbation only through the source current,
\begin{equation}
	\delta j_{mn}^{\mathrm{src}}=\delta w_{mn}p_n-\delta w_{nm}p_m.
\label{eq:jump.source}
\end{equation}
The response then reads
\begin{equation}
	\delta\E[J_\tau]=\int_0^\tau\sum_{m<n}u_{mn}\delta j_{mn}^{\mathrm{src}}\dd{t}.
\label{eq:jump.response}
\end{equation}
A parameter $\theta_{mn}$ of the edge, which may change both of its rates, injects the source current $\delta j_{mn}^{\mathrm{src}}=N_{\theta,mn}\delta\theta_{mn}$ with $N_{\theta,mn}\equiv p_n\partial_\theta w_{mn}-p_m\partial_\theta w_{nm}$. A single parameter per edge then suffices, $R_{\theta,mn}=N_{\theta,mn}u_{mn}$, and the traffic form~\eqref{eq:jump.variance} gives, wherever $N_{\theta,mn}\ne0$,
\begin{equation}
	\Var(J_\tau)-\Var_0(J_\tau)=\int_0^\tau\sum_{m<n}a_{mn}\qty(\frac{R_{\theta,mn}}{N_{\theta,mn}})^2\dd{t}.
\label{eq:jump.localFRRedge}
\end{equation}

Next, we parametrize the two rates of each edge by the logarithms of their geometric mean and their ratio, $B_{mn}=(1/2)\ln(w_{mn}w_{nm})$ and $F_{mn}=\ln(w_{mn}/w_{nm})$, so that $w_{mn}=e^{B_{mn}+F_{mn}/2}$ and $w_{nm}=e^{B_{mn}-F_{mn}/2}$, with $B_{mn}=B_{nm}$ and $F_{mn}=-F_{nm}$. We call $B_{mn}$ the kinetic parameter and $F_{mn}$ the entropic force of the edge \cite{Owen.2020.PRX,Aslyamov.2024.PRL}. Increasing $B_{mn}$ speeds up both transitions, as a lower activation barrier would, whereas increasing $F_{mn}$ biases their ratio; under local detailed balance, $F_{mn}$ is the entropy flow into the environment per transition $n\to m$. Since $\delta w_{mn}=w_{mn}(\delta B_{mn}+\delta F_{mn}/2)$ and $\delta w_{nm}=w_{nm}(\delta B_{mn}-\delta F_{mn}/2)$, Eq.~\eqref{eq:jump.source} gives
\begin{equation}
	\delta j_{mn}^{\mathrm{src}}=j_{mn}\delta B_{mn}+\frac{1}{2}a_{mn}\delta F_{mn},
\label{eq:jump.BF}
\end{equation}
so that the time-resolved responses are $R_{B,mn}=j_{mn}u_{mn}$ and $R_{F,mn}=a_{mn}u_{mn}/2$. The traffic form~\eqref{eq:jump.variance} then becomes an exact identity in terms of either response,
\begin{align}
	\Vd&=\int_0^\tau\sum_{m<n}\frac{a_{mn}}{j_{mn}^2}R_{B,mn}^2\dd{t}\notag\\
	&=4\int_0^\tau\sum_{m<n}\frac{R_{F,mn}^2}{a_{mn}}\dd{t}.
\label{eq:jump.localFRR}
\end{align}
The kinetic form requires $j_{mn}\ne0$ on every edge with positive traffic; on edges without a net current, as in equilibrium, the amplitude must be obtained from the entropic response. 

Applying Eq.~\eqref{eq:jump.localFRR} to the observable $J_\tau^\alpha+J_\tau^\beta$ and subtracting the identities for $J_\tau^\alpha$ and $J_\tau^\beta$ gives the corresponding identity for the covariance of two observables with antisymmetric increments,
\begin{align}
	&\Cov(J_\tau^\alpha,J_\tau^\beta)-\Cov_0(J_\tau^\alpha,J_\tau^\beta)\notag\\
	&\quad=\int_0^\tau\sum_{m<n}a_{mn}u_{mn}^\alpha u_{mn}^\beta\dd{t}\notag\\
	&\quad=\int_0^\tau\sum_{m<n}\frac{a_{mn}}{j_{mn}^2}R_{B,mn}^\alpha R_{B,mn}^\beta\dd{t}\notag\\
	&\quad=4\int_0^\tau\sum_{m<n}\frac{R_{F,mn}^\alpha R_{F,mn}^\beta}{a_{mn}}\dd{t},
\label{eq:jump.localcov}
\end{align}
where $\Cov_0(J_\tau^\alpha,J_\tau^\beta)=\Cov[G_0^\alpha(z_0),G_0^\beta(z_0)]$ is the preparation covariance. The two responses may also be taken with respect to different parameters; for example, the integrand $2R_{B,mn}^\alpha R_{F,mn}^\beta/j_{mn}$ combines a kinetic response of one observable with an entropic response of the other.

These identities carry over to the frequency domain. In a steady state, the Fourier transforms of the responses to impulsive changes of $B_{mn}$ and $F_{mn}$ are $R_{B,mn}(\omega)=j_{mn}^{\st}u_{\omega,mn}$ and $R_{F,mn}(\omega)=a_{mn}^{\st}u_{\omega,mn}/2$, whose ratio $2j_{mn}^{\st}/a_{mn}^{\st}$ is independent of frequency, as found in Ref.~\cite{Zheng.2026a.arxiv}. Equation~\eqref{eq:general.spectrumlocal.jump} then gives, with the same caveat for the kinetic form,
\begin{align}
	\mathcal{S}_{\dot{J}}(\omega)&=\sum_{m<n}\frac{a_{mn}^{\st}}{(j_{mn}^{\st})^2}|R_{B,mn}(\omega)|^2\notag\\
	&=4\sum_{m<n}\frac{|R_{F,mn}(\omega)|^2}{a_{mn}^{\st}}.
\label{eq:jump.spectrumBF}
\end{align}
The two sums reproduce the kinetic and entropic response identities of Ref.~\cite{Kwon.2026.arxiv}. Reference~\cite{Aslyamov.2026.arxiv} gives the corresponding identity for perturbations of individual directed rates. These identities show that two of the inequalities of Ref.~\cite{Zheng.2026a.arxiv}, Eqs.~(11b) and (15a), are equalities for observables with antisymmetric increments.

The stationary relations follow in the long-time limit, where the martingale amplitude of an observable $J_\tau^\alpha$ approaches $u_{mn}^\alpha=d_{mn}^\alpha+\Psi^\alpha(m)-\Psi^\alpha(n)$. Using Eq.~\eqref{eq:jump.localcov} and the static responses of the mean rates, $\partial_{B_{mn}}\jr_{\st}^\alpha=j_{mn}^{\st}u_{mn}^\alpha$ and $\partial_{F_{mn}}\jr_{\st}^\alpha=a_{mn}^{\st}u_{mn}^\alpha/2$, we obtain
\begin{align}
	\lim_{\tau\to\infty}\frac{\Cov(J_\tau^\alpha,J_\tau^\beta)}{\tau}
	&=\sum_{m<n}a_{mn}^{\st}u_{mn}^\alpha u_{mn}^\beta \notag\\
	&=\sum_{m<n}\frac{a_{mn}^{\st}}{(j_{mn}^{\st})^2}\partial_{B_{mn}}\jr_{\st}^\alpha\,\partial_{B_{mn}}\jr_{\st}^\beta \notag\\
	&=4\sum_{m<n}\frac{\partial_{F_{mn}}\jr_{\st}^\alpha\,\partial_{F_{mn}}\jr_{\st}^\beta}{a_{mn}^{\st}}.
\label{eq:jump.stationaryFRR}
\end{align}
Equation~\eqref{eq:jump.stationaryFRR} reproduces the stationary fluctuation-response relations of Ref.~\cite{Aslyamov.2025.PRL} when both observables are currents ($\phi=0$), those of Ref.~\cite{Ptaszynski.2026.PRE} when both are state observables ($d=0$), and those of Ref.~\cite{Ptaszynski.2026.PRE2} for the correlation between a current and a state observable. At finite times and for arbitrary initial distributions, their counterpart is the time-resolved identity~\eqref{eq:jump.localcov}, in which the preparation covariance is subtracted, as in the dynamical fluctuation-response relation of Ref.~\cite{Aslyamov.2026.arxiv}.

Kinetic and entropic perturbations lead to response inequalities with different costs. For a perturbation with an edge source $g_{mn}$, the response of the mean and the cost are $\Rc[g]=\int_0^\tau\sum_{m<n}u_{mn}g_{mn}\dd{t}$ and $\Ac[g]=\int_0^\tau\sum_{m<n}g_{mn}^2/a_{mn}\dd{t}$, respectively. Applying the Cauchy-Schwarz inequality to the sum over edges and the integral over time, we obtain
\begin{equation}
	\Rc[g]^2\le\Vd\Ac[g],
\label{eq:jump.RAbound}
\end{equation}
which is the jump form of Eq.~\eqref{eq:general.sourcebounds}, with equality attained when $g_{mn}\propto a_{mn}u_{mn}$. The two kinds of perturbation differ only in the source they inject and hence in the cost.

A kinetic profile $\delta B_{mn}=b_{mn}(t)$, at fixed $F_{mn}$, injects $g_{mn}=b_{mn}j_{mn}$ by Eq.~\eqref{eq:jump.BF}, with the cost $\Ac[g]\le b_{\mathrm{max}}^2\int_0^\tau\sum_{m<n}j_{mn}^2/a_{mn}\dd{t}$, where $b_{\mathrm{max}}\equiv\max|b_{mn}(t)|$. Since
\begin{equation}
	\int_0^\tau\sum_{m<n}\frac{j_{mn}^2}{a_{mn}}\dd{t}\le\min\qty(\frac{\Sigma_\tau}{2},A_\tau),
\label{eq:jump.psent.ub}
\end{equation}
it follows that $\Ac[g]\le b_{\mathrm{max}}^2\min(\Sigma_\tau/2,A_\tau)$. An entropic profile $\delta F_{mn}=f_{mn}(t)$, at fixed $B_{mn}$, injects $g_{mn}=f_{mn}a_{mn}/2$, whose cost $\Ac[g]=(1/4)\int_0^\tau\sum_{m<n}f_{mn}^2a_{mn}\dd{t}$ is bounded by the dynamical activity, $\Ac[g]\le f_{\mathrm{max}}^2A_\tau/4$, where $f_{\mathrm{max}}\equiv\max|f_{mn}(t)|$. Equation~\eqref{eq:jump.RAbound} then gives
\begin{equation}
	\Rc[g]^2\le b_{\mathrm{max}}^2\Vd\min\qty(\frac{\Sigma_\tau}{2},A_\tau)
\label{eq:jump.Kbound}
\end{equation}
for kinetic perturbations and
\begin{equation}
	\Rc[g]^2\le f_{\mathrm{max}}^2\Vd \frac{A_\tau}{4}
\label{eq:jump.Ebound}
\end{equation}
for entropic perturbations. Since $\Vd=\Var(J_\tau)-\Var_0(J_\tau)\le\Var(J_\tau)$, replacing $\Vd$ by $\Var(J_\tau)$ recovers the thermodynamic and kinetic response inequalities of Ref.~\cite{Kwon.2025.PRL}. Subtracting $\Var_0(J_\tau)=\Var(G_0(z_0))$ sharpens them whenever the conditional mean $G_0(n)$ varies over the initial states.

\subsection{Improved finite-time TUR and its equality conditions}\label{sec:jump.tur}
A uniform rescaling of all rates, $w_{mn}\to(1+\eps)w_{mn}$, is the kinetic perturbation with $\delta B_{mn}=1$ on every edge. It injects the source $g_{mn}=j_{mn}$ and is equivalent to a rescaling of time. For a pure current, $\phi=0$, under a protocol $\lambda(\upsilon t)$, with any explicit time dependence of the increments entering through the protocol and with $p_0$ fixed, the same time-change argument as in Sec.~\ref{sec:over.tur} therefore gives
\begin{equation}
	\Rc[g]=(\tau\partial_\tau-\upsilon\partial_\upsilon)\E[J_\tau].
\label{eq:jump.clock}
\end{equation}
The cost $\Ac[g]=\int_0^\tau\sum_{m<n}j_{mn}^2/a_{mn}\dd{t}$ of this perturbation is exactly half the pseudo entropy production~\cite{Shiraishi.2021.JSP},
\begin{equation}
	\Sigma_\tau^{\mathrm{ps}}\equiv\int_0^\tau\sum_{m\ne n}\frac{j_{mn}^2}{a_{mn}}\dd{t}\le\min(\Sigma_\tau,2A_\tau).
\label{eq:jump.pseudo}
\end{equation}
Combining Eqs.~\eqref{eq:jump.clock} and \eqref{eq:jump.pseudo} with the response inequality~\eqref{eq:jump.RAbound}, we obtain an improved finite-time TUR that retains the preparation subtraction,
\begin{equation}
	\frac{\Var(J_\tau)-\Var_0(J_\tau)}{((\tau\partial_\tau-\upsilon\partial_\upsilon)\E[J_\tau])^2}\ge\frac{2}{\Sigma_\tau^{\mathrm{ps}}}\ge\frac{2}{\Sigma_\tau}.
\label{eq:jump.TUR}
\end{equation}
The second inequality holds edge by edge: for the directed fluxes $x=w_{mn}p_n$ and $y=w_{nm}p_m$, $(x-y)^2/(x+y)\le(x-y)\ln(x/y)/2$, with equality only if $x=y$. It is therefore strict whenever the entropy production is nonzero. Replacing $\Var(J_\tau)-\Var_0(J_\tau)$ by $\Var(J_\tau)$ and $\Sigma_\tau^{\mathrm{ps}}$ by $\Sigma_\tau$ gives the TUR for time-dependent driving~\cite{Koyuk.2020.PRL}. A relation equivalent to Eq.~\eqref{eq:jump.TUR} was obtained from the dynamical fluctuation-response relation in Ref.~\cite{Aslyamov.2026.arxiv}. The martingale approach makes the gap in the first inequality of Eq.~\eqref{eq:jump.TUR} explicit, which yields the equality conditions derived below.

We now determine when the first inequality in Eq.~\eqref{eq:jump.TUR} is an equality, for any driving and any initial distribution. With $\varsigma=2\Rc[g]/\Sigma_\tau^{\mathrm{ps}}$, we decompose the martingale amplitude into a part aligned with the probability current and a remainder,
\begin{equation}
	u_{mn}=\varsigma\frac{j_{mn}}{a_{mn}}+u_{mn}^\perp.
\label{eq:jump.udecomp}
\end{equation}
By construction, $\int_0^\tau\sum_{m<n}j_{mn}u_{mn}^\perp\dd{t}=0$, so that the cross term in Eq.~\eqref{eq:jump.variance} integrates to zero, and
\begin{equation}
	\Vd-\frac{2\Rc[g]^2}{\Sigma_\tau^{\mathrm{ps}}}=\int_0^\tau\sum_{m<n}a_{mn}(u_{mn}^\perp)^2\dd{t}.
\label{eq:jump.saturation}
\end{equation}
Equality thus holds if and only if $u_{mn}^\perp=0$ on every edge at all times, i.e., if the martingale increment of $J_\tau$ is proportional, with a constant factor, to that of the representing martingale of the rate rescaling, $\dd{S_t}=\sum_{m<n}(j_{mn}/a_{mn})(\dd\widetilde N_{mn}-\dd\widetilde N_{nm})$.

For pure currents, $\phi=0$, the function $G_t$ is fixed by the increments. Because $u_{mn}=d_{mn}+G_t(m)-G_t(n)$, the increments must be $d_{mn}=\varsigma j_{mn}/a_{mn}-[G_t(m)-G_t(n)]$. In the backward equation~\eqref{eq:general.backward}, $\partial_tG_t(n)+\sum_{m(\ne n)}w_{mn}[G_t(m)-G_t(n)]+\Phi_t(n)=0$ with $\Phi_t(n)=\sum_{m(\ne n)}w_{mn}d_{mn}$, the terms with $G_t(m)-G_t(n)$ then cancel, leaving $\partial_tG_t(n)+\varsigma\kappa_t(n)=0$, where
\begin{equation}
	\kappa_t(n)\equiv\sum_{m(\ne n)}w_{mn}\frac{j_{mn}}{a_{mn}}
\label{eq:jump.kappa}
\end{equation}
is the expected rate, from the state $n$, of the observable with increments $j_{mn}/a_{mn}$. With $G_\tau=0$, this gives $G_t(n)=\varsigma\int_t^\tau\kappa_r(n)\dd{r}$ and the saturating increments
\begin{equation}
	d_{mn}^{\mathrm{opt}}(t)=\varsigma\qty[\frac{j_{mn}(t)}{a_{mn}(t)}-\int_t^\tau[\kappa_r(m)-\kappa_r(n)]\dd{r}].
\label{eq:jump.dopt}
\end{equation}
As in Sec.~\ref{sec:over.tur}, these increments enter Eq.~\eqref{eq:jump.TUR} only if they depend on time through the protocol. If the protocol path $\lambda(\upsilon t)$ does not revisit any value on $[0,\tau]$, e.g., for a \emph{strictly monotonic} scalar protocol, any time-dependent increment is a function of the protocol, $d_{mn}(t)=\hat{d}_{mn}(\lambda(\upsilon t))$, which is held fixed when $\tau$ and $\upsilon$ are varied, and the first inequality of Eq.~\eqref{eq:jump.TUR} can be saturated by a pure current at any finite time. The preparation variance of this current, $\Var_0(J_\tau)=\varsigma^2\Var[\int_0^\tau\kappa_r(n_0)\dd{r}]$ with $n_0\sim p_0$, is in general nonzero, so that the same current does not saturate the TUR of Ref.~\cite{Koyuk.2020.PRL}, which uses the full variance.

For autonomous dynamics with time-independent increments, Eq.~\eqref{eq:jump.dopt} must not depend on time, i.e., $\partial_t(j_{mn}/a_{mn})=\kappa_t(n)-\kappa_t(m)$ on every edge at all times. In a steady state, the left-hand side vanishes, so that $\kappa$ takes the same value on all states. Using $w_{mn}p_n=(a_{mn}+j_{mn})/2$ and stationarity, $\sum_{m(\ne n)}j_{mn}=0$, one finds $\kappa(n)=\sigma_n^{\mathrm{ps}}/(2p_n)$, where $\sigma_n^{\mathrm{ps}}\equiv\sum_{m(\ne n)}j_{mn}^2/a_{mn}$ is the contribution of state $n$ to the pseudo entropy production rate $\sigma^{\mathrm{ps}}\equiv\sum_{m\neq n}j_{mn}^2/a_{mn}$. For a current with $\jr_{\st}\ne0$, the first inequality in Eq.~\eqref{eq:jump.TUR} is therefore an equality if and only if
\begin{equation}
	d_{mn}=\frac{2\jr_{\st}}{\sigma^{\mathrm{ps}}}\frac{j_{mn}}{a_{mn}}~\text{and}~
	\sigma_n^{\mathrm{ps}}=\sigma^{\mathrm{ps}}p_n\ \ \forall m, n,
\label{eq:jump.uniform}
\end{equation}
where $\Rc[g]=\jr_{\st}\tau$ and $\Sigma_\tau^{\mathrm{ps}}=\sigma^{\mathrm{ps}}\tau$ fix $\varsigma=2\jr_{\st}/\sigma^{\mathrm{ps}}$. The first condition aligns the increments with the probability current, and the second requires the conditional pseudo entropy production rate $\sigma_n^{\mathrm{ps}}/p_n$ to be the same for all states. Then, $G_t=\jr_{\st}(\tau-t)$ and $\Var_0(J_\tau)=0$, so that the full-variance bound with $\Sigma_\tau^{\mathrm{ps}}$ is saturated as well, whereas the bound with $\Sigma_\tau$ remains strict at nonzero dissipation. When these conditions hold, the saturating current is the most precise current of the system, whose general form was obtained in Ref.~\cite{Busiello.2022.JPA}.

\subsection{Improved finite-time KUR and its equality conditions}\label{sec:jump.kur}
The TUR~\eqref{eq:jump.TUR} concerns observables with antisymmetric increments. Observables that count transitions irrespective of their direction, such as the dynamical activity, are not covered; their precision is instead constrained by the KUR~\cite{Garrahan.2017.PRE,Terlizzi.2019.JPA}. The same uniform rescaling of all rates, $w_{mn}\to(1+\eps)w_{mn}$, yields an improved finite-time KUR for observables with general increments. For a pure jump observable, $\phi=0$, whose increments depend on time only through the protocol, the time-change argument again identifies the response with that to the uniform rescaling, now given by Eq.~\eqref{eq:jump.responsegen} with $\delta w_{mn}=w_{mn}$,
\begin{align}
	\Rc_J&=\int_0^\tau\sum_{m\ne n}w_{mn}p_nu_{mn}\dd{t}\notag\\
	&=(\tau\partial_\tau-\upsilon\partial_\upsilon)\E[J_\tau].
\label{eq:jump.kurresponse}
\end{align}
Since the rescaling multiplies every rate by the same factor, $\delta\ln w_{mn}=1$ on every transition, and its cost is the dynamical Fisher information~\eqref{eq:general.jumpIcost}, which here equals the dynamical activity,
\begin{equation}
	\Icost=\int_0^\tau\sum_{m\ne n}w_{mn}p_n\dd{t}=\int_0^\tau\sum_{m<n}a_{mn}\dd{t}=A_\tau.
\label{eq:jump.activity}
\end{equation}
The fluctuation-response inequality, $\Rc_J^2\le\Vd\Icost$, then gives
\begin{equation}
	\frac{\Var(J_\tau)-\Var_0(J_\tau)}{((\tau\partial_\tau-\upsilon\partial_\upsilon)\E[J_\tau])^2}\ge\frac{1}{A_\tau},
\label{eq:jump.KUR}
\end{equation}
which tightens the KUR in Ref.~\cite{Terlizzi.2019.JPA} by subtracting the preparation variance.
For autonomous dynamics started in the stationary distribution, the response reduces to the mean, and Eq.~\eqref{eq:jump.KUR} sharpens the KUR of Ref.~\cite{Garrahan.2017.PRE}, $\Var(J_\tau)/\E[J_\tau]^2\ge1/A_\tau$, by the preparation variance. For antisymmetric increments, Eq.~\eqref{eq:jump.KUR} follows from Eq.~\eqref{eq:jump.TUR}, because $\Sigma_\tau^{\mathrm{ps}}\le 2A_\tau$; the two costs would coincide only if every edge were unidirectional, $|j_{mn}|=a_{mn}$, which is excluded by the assumption of reversible transitions. The KUR is therefore informative mainly for observables whose increments have a symmetric part, such as the total number of transitions.

The equality conditions follow as in Sec.~\ref{sec:jump.tur}. With $\varsigma=\Rc_J/A_\tau$, we decompose the martingale amplitude into a part that is the same on every transition and a remainder,
\begin{equation}
	u_{mn}=\varsigma+u_{mn}^\perp.
\label{eq:jump.kurdecomp}
\end{equation}
By construction, $\int_0^\tau\sum_{m\ne n}w_{mn}p_nu_{mn}^\perp\dd{t}=0$, so that Eq.~\eqref{eq:jump.dynvar} gives
\begin{equation}
	\Vd-\frac{\Rc_J^2}{A_\tau}=\int_0^\tau\sum_{m\ne n}w_{mn}p_n(u_{mn}^\perp)^2\dd{t}.
\label{eq:jump.kursat}
\end{equation}
Equality thus holds if and only if the amplitude is the same on every transition at all times. Since $u_{mn}=d_{mn}+G_t(m)-G_t(n)$, the increments must be $d_{mn}=\varsigma -[G_t(m)-G_t(n)]$. In the backward equation~\eqref{eq:general.backward}, the terms with $G_t(m)-G_t(n)$ then cancel, as in Sec.~\ref{sec:jump.tur}, leaving $\partial_tG_t(n)+\varsigma\bar{w}_t(n)=0$, where $\bar{w}_t(n)\equiv\sum_{m(\ne n)}w_{mn}(t)$ is the escape rate of the state $n$. With $G_\tau=0$, this gives $G_t(n)=\varsigma\int_t^\tau\bar{w}_r(n)\dd{r}$ and the saturating increments
\begin{equation}
	d_{mn}^{\mathrm{opt}}(t)=\varsigma\qty(1-\int_t^\tau[\bar{w}_r(m)-\bar{w}_r(n)]\dd{r}).
\label{eq:jump.dkur}
\end{equation}
In contrast to Eq.~\eqref{eq:jump.dopt}, these increments depend only on the rates and not on the distribution. They can take either sign, so the saturating observable is a weighted jump observable rather than a nonnegative count. If the protocol path does not revisit any value on $[0,\tau]$, the increments can be written as functions of the protocol, held fixed when $\tau$ and $\upsilon$ are varied, and Eq.~\eqref{eq:jump.KUR} can be saturated at any finite time. The preparation variance of this observable, $\Var_0(J_\tau)=\varsigma^2\Var[\int_0^\tau\bar{w}_r(n_0)\dd{r}]$ with $n_0\sim p_0$, is in general nonzero, so that the same observable does not saturate the KUR with the full variance. For autonomous dynamics with time-independent increments, Eq.~\eqref{eq:jump.dkur} requires $\bar{w}(m)=\bar{w}(n)$ on every edge, i.e., by irreducibility, a uniform escape rate $\bar w$. The saturating observable is then the total number of transitions, $d_{mn}=1$, which is a Poisson process with rate $\bar w$, so that $G_t=\bar{w}(\tau-t)$, $\Var_0(J_\tau)=0$, and $\Var(J_\tau)=\E[J_\tau]=A_\tau=\bar{w}\tau$. In contrast to the TUR, this condition concerns the rates alone and holds for any initial distribution.

\section{Summary and outlook}\label{sec:discussion}
We have developed a unified finite-time theory of fluctuations and responses built around a single object: the best prediction of an additive observable given the trajectory so far. As a trajectory unfolds, this prediction is the value already recorded plus the expected remaining contribution, and it forms a Doob martingale. Its initial value depends only on the initial state, so part of the run-to-run variability is determined before the observation starts. Its increments, by contrast, are generated by the noise during the observation, and their variances add up to the dynamical variance. The covariance of these increments with those of a second martingale, which represents a perturbation of finite cost, gives the linear response of the mean at fixed preparation. The dynamical variance and the response are thus a squared norm and an inner product, and each bound derived here is a Cauchy-Schwarz inequality in which the cost is that of the physical perturbation.

The reason fluctuations and responses share one martingale amplitude is clearest in jump processes, where all randomness enters through individual transitions. A single jump changes the recorded value of the observable directly, and it also changes the value expected over the rest of the trajectory; the sum of these two contributions is the jump of the martingale. Random jumps thus generate the fluctuations, whereas a perturbation of the rates changes how often each jump occurs, and each additional jump changes the expected final value by the same amplitude. For antisymmetric increments, this amplitude is therefore both the martingale increment and the local response of the mean to a probability current injected along that edge. In a diffusion, a small random displacement plays the same role. This shared amplitude is the common structure behind the finite-time identity for jump processes~\cite{Aslyamov.2026.arxiv} and the long-time identity for overdamped diffusion~\cite{Chun.2026.arxiv}; through it, the latter extends to finite times and arbitrary initial distributions.

For underdamped dynamics, the framework gives an exact finite-time fluctuation-response relation, valid for arbitrary initial distributions and any admissible perturbation. Because the noise acts only on the velocities, the martingale amplitude lives in velocity space, and the response is its overlap with the probability current that the perturbation injects there. Thermodynamic bounds then follow once the perturbation is specified. At fixed temperature, a change of the friction coefficient alters both the damping force and the noise strength and injects the irreversible probability current. Its cost is exactly half the entropy production, so the friction response of any fixed observable gives a lower bound on the entropy production. Maximizing over additive observables turns this bound into an equality at every finite time, without assuming a steady state. If the velocity weight may depend on time, a pure-martingale observable saturates even the full-variance bound; for a fixed observable, however, subtracting the preparation variance can make a large difference: in a rotor prepared in equilibrium, the saturating observable with a time-independent velocity weight recovers the full dissipation at all times, including short times, where the bound based on its total variance captures only a vanishing fraction.

In overdamped and jump dynamics, the natural perturbation is a uniform speed-up, through the mobility or through all transition rates, which is equivalent to a rescaling of time. It gives improved TURs for time-dependent driving with the preparation variance subtracted; for jump processes, an equivalent relation was obtained in Ref.~\cite{Aslyamov.2026.arxiv}. For currents in jump processes, the cost is half the pseudo entropy production, bounded by both half the entropy production and the dynamical activity. For observables that count transitions regardless of their direction, the same speed-up gives an improved KUR, with the dynamical activity as the cost. Once the preparation variance is subtracted, both bounds can be saturated at any finite time under strictly monotonic driving: the improved TUR by a suitably weighted current and the improved KUR by a suitably weighted jump observable. For jump currents, the saturated bound is the one with the pseudo entropy production; the bound with the entropy production then remains strict. These observables do not saturate the bounds with the full variance, because their preparation variances are in general nonzero. For time-independent weights, the conditions are more restrictive. In a steady state, a current saturates the improved TUR only if the entropy production rate conditioned on the present state, or its pseudo counterpart for jumps, is the same in every state. The improved KUR, by contrast, is saturated by the total number of transitions whenever all escape rates are equal; since this condition involves only the rates, saturation holds for any initial distribution.

The friction-response TUR also clarifies two consequences of inertia. First, it explains how underdamped currents can violate the conventional TUR. Under strong driving, a particle on a ring moves forward almost all the time, so a current whose bounded weight records the direction of the velocity has a mean rate close to unity that barely responds to friction. Its conventional uncertainty factor decays exponentially with the dissipation rate, whereas its friction-response factor remains above unity. The violation thus reflects the insensitivity of the mean to friction, which the conventional TUR ignores but the friction-response TUR takes into account. Second, for underdamped limit cycles, the friction-response TUR gives a weak-noise lower bound on the entropy production per coherent oscillation, set by the squared friction sensitivity of the frequency. This factor equals unity for the uniformly driven rotor, which saturates the bound, and can be much larger for other cycles, which then require correspondingly more dissipation to sustain the same coherence. Conversely, when the frequency is set by a conservative restoring force and does not depend on friction, the factor vanishes, and inertial oscillators can sustain coherent oscillations at a dissipation far below the overdamped value $4\pi^2$.

The Doob martingale can be defined for any stochastic process~\cite{Doob.1953}, and the variance decomposition holds without the Markov property. What the Markov property provides is that the expected future accumulation depends only on the present state, so that the martingale amplitude becomes a local response that can be computed and interpreted. When the expected future depends on more than the present observed state, as in non-Markovian dynamics with memory~\cite{Kanazawa.2025.arxiv,Funo.2026.arxiv,Dechant.2026.arxiv,Zheng.2026b.arxiv} and in systems with hidden degrees of freedom~\cite{Esposito.2012.PRE,Hartich.2021.PRX,vanderMeer.2022.PRX,Harunari.2022.PRX}, this local construction does not apply directly. For continuously monitored open quantum systems~\cite{Wiseman.2009,Landi.2024.PRXQ,Manzano.2022.QS,Vu.2026.PRL}, it remains open which physical perturbations play the role of the friction change and whether their costs are thermodynamic quantities~\cite{Hasegawa.2020.PRL,Vu.2022.PRL.TUR,Vu.2025.PRXQ,Prech.2025.PRL}, as the costs of friction and mobility changes are half the entropy production here. We leave these extensions to future work.

On the experimental side, a natural next step is to demonstrate dissipation inference from friction responses at finite times. As in TUR-based inference~\cite{Manikandan.2020.PRL,Otsubo.2020.PRE,Vu.2020.PRE,Kim.2020.PRL}, the observable can be optimized numerically from trajectory data, here by maximizing its squared friction response relative to its dynamical variance. Since this maximum equals half the entropy production at any observation time, the optimization yields the entropy production without knowledge of the forces, with an accuracy limited only by the class of observables and the statistics. This scheme requires changing the friction at fixed temperature, which is possible in levitated nanoparticles~\cite{Li.2011.NP,Rondin.2017.NN}, whose damping is tuned by the gas pressure, and in electrical circuits~\cite{Garnier.2005.PRE,Ciliberto.2013.PRL}, whose resistance sets both the damping and the Johnson-Nyquist noise~\cite{Johnson.1928.PR,Nyquist.1928.PR}; RC circuits realize the overdamped dynamics, and an inductance adds inertia.

\begin{acknowledgments}
We thank Hyun-Myung Chun for a fruitful discussion. We also acknowledge the use of ChatGPT Sol 5.6 and Claude Opus 5.5 for assistance with manuscript preparation and analytical checks. TVV was supported by JSPS KAKENHI Grants No.~JP23K13032, No.~JP26K00019, No.~JP26K00022, and No.~JP26H02015. VTV was supported by QACI, FPT Corporation. RB was supported by JSPS KAKENHI Grant No.~JP25KJ0766. KS was supported by JSPS KAKENHI Grants No.~JP26H02015 and No.~JP26H00388.
\end{acknowledgments}

\section*{Data availability}
The data that support the findings of this work are not publicly available. The data are available from the authors upon reasonable request.

\appendix

\section{Conditional covariance of the martingale increments}\label{app:covariance}
We derive Eq.~\eqref{eq:general.Ncov} from Eqs.~\eqref{eq:general.onestep} and \eqref{eq:general.Nf}. Fix a time $t$ and a short time step $\Delta t$. By the Markov property, conditioning on $\Fc_t$ is equivalent to conditioning on $Z_t=Z$. For convenience, we write $a^f_s\equiv(\partial_s+\hLg_s^\dagger)f_s$ for the drift rate subtracted in Eq.~\eqref{eq:general.Nf} and $\E_Z[\cdot]\equiv\E[\cdot\cond Z_t=Z]$. By Eq.~\eqref{eq:general.Nf}, the increment $\Delta N^f=N^f_{t+\Delta t}-N^f_t$ reads
\begin{equation}
	\Delta N^f=f_{t+\Delta t}(Z_{t+\Delta t})-f_t(Z)-\int_t^{t+\Delta t}a^f_s(Z_s)\dd{s}.
\label{eq:app.DeltaN}
\end{equation}
We separate the value of the integrand at the beginning of the step from a remainder,
\begin{align}
	\int_t^{t+\Delta t}a^f_s(Z_s)\dd{s}&=a^f_t(Z)\Delta t+R^f,\\
	R^f&\equiv\int_t^{t+\Delta t}\qty[a^f_s(Z_s)-a^f_t(Z)]\dd{s}.
\label{eq:app.Rf}
\end{align}
With $F^f\equiv f_{t+\Delta t}(Z_{t+\Delta t})$, this gives $\Delta N^f=F^f-c^f-R^f$, where $c^f\equiv f_t(Z)+a^f_t(Z)\Delta t$ is fixed by the initial point $Z$ of the step. The same decomposition applies to $\Delta N^g$. Since $N^f$ is a martingale, $\E_Z[\Delta N^f]=0$, so that the conditional second moment equals the conditional covariance. The constants $c^f$ and $c^g$ do not contribute to it, and
\begin{align}
	\E_Z[\Delta N^f\Delta N^g]&=\Cov_Z[F^f,F^g]-\Cov_Z[F^f,R^g]\notag\\
	&\quad-\Cov_Z[R^f,F^g]+\Cov_Z[R^f,R^g].
\label{eq:app.covsplit}
\end{align}
With $Z'\equiv Z_{t+\Delta t}$, expanding the explicit time dependence gives $F^f=f_t(Z')+\Delta t\,\partial_tf_t(Z')+o(\Delta t)$, and similarly for $F^g$. Hence,
\begin{align}
	\Cov_Z[F^f,F^g]&=\Cov_Z\big[f_t(Z'),g_t(Z')\big]\notag\\
	&\quad+\Delta t\Cov_Z[\partial_tf_t(Z'),g_t(Z')]\notag\\
	&\quad+\Delta t\Cov_Z[f_t(Z'),\partial_tg_t(Z')]+o(\Delta t).
\label{eq:app.covtime}
\end{align}
By Eq.~\eqref{eq:general.onestep} applied to the pairs $(\partial_tf_t,g_t)$ and $(f_t,\partial_tg_t)$, the covariances in the second and third lines are of order $\Delta t$, so these lines are of order $\Delta t^2$; the explicit time dependence therefore drops out. Applying Eq.~\eqref{eq:general.onestep} to $(f_t,g_t)$ in the first line, we obtain
\begin{equation}
	\Cov_Z[F^f,F^g]=\Delta t\hGam_t[f_t,g_t](Z)+o(\Delta t).
\label{eq:app.covFG}
\end{equation}

The Cauchy-Schwarz inequality in time gives $(R^f)^2\le\Delta t\int_t^{t+\Delta t}[a^f_s(Z_s)-a^f_t(Z)]^2\dd{s}$, and hence
\begin{equation}
	\E_Z[(R^f)^2]\le\Delta t\int_t^{t+\Delta t}\E_Z\qty[(a^f_s(Z_s)-a^f_t(Z))^2]\dd{s}.
\end{equation}
From Dynkin's formula, $\E_Z[\varphi_s(Z_s)]-\varphi_t(Z)=\int_t^s\E_Z[(\partial_u+\hLg_u^\dagger)\varphi_u(Z_u)]\dd{u}$, one can show that
\begin{equation}
	\E_Z[\varphi_s(Z_s)]=\varphi_t(Z)+(s-t)[(\partial_t+\hLg_t^\dagger)\varphi_t](Z)+o(s-t).
\end{equation}
Applying this to $\varphi=(a^f)^2$ and $\varphi=a^f$ gives
\begin{align}
	&\E_Z\qty[(a^f_s(Z_s)-a^f_t(Z))^2]\notag\\
	&\quad=(s-t)\qty[\hLg_t^\dagger(a^f_t)^2-2a^f_t\hLg_t^\dagger a^f_t](Z)+o(s-t)\notag\\
	&\quad=(s-t)\hGam_t[a^f_t,a^f_t](Z)+o(s-t),
\end{align}
because the zeroth-order terms and the explicit time derivatives cancel.
Therefore, $\E_Z[(R^f)^2]=O(\Delta t^3)$. Since $\Var_Z[F^f]=O(\Delta t)$, the Cauchy-Schwarz inequality yields
\begin{align}
	|\Cov_Z[F^f,R^g]|&\le\big(\Var_Z[F^f]\E_Z[(R^g)^2]\big)^{1/2}=O(\Delta t^2),\notag\\
	|\Cov_Z[R^f,R^g]|&\le\big(\E_Z[(R^f)^2]\E_Z[(R^g)^2]\big)^{1/2}=O(\Delta t^3),
\label{eq:app.remainders}
\end{align}
and the same bound as in the first line holds for $\Cov_Z[R^f,F^g]$. All three remainder terms in Eq.~\eqref{eq:app.covsplit} are therefore $o(\Delta t)$.

Inserting Eqs.~\eqref{eq:app.covFG} and \eqref{eq:app.remainders} into Eq.~\eqref{eq:app.covsplit}, we obtain
\begin{equation}
	\E\qty[\Delta N^f\Delta N^g\cond\Fc_t]=\hGam_t[f_t,g_t](Z_t)\Delta t+o(\Delta t),
\end{equation}
which is Eq.~\eqref{eq:general.Ncov} in the limit $\Delta t\to\dd{t}$. At a fixed time $t$, $Z_{t^-}=Z_t$ almost surely, because the processes considered here are stochastically continuous: they have no jumps at deterministic times. The right-hand side of the last equation is therefore $\Fc_{t^-}$-measurable up to a null set, and the tower property gives the same expression conditioned on $\Fc_{t^-}$, which is the form of Eq.~\eqref{eq:general.Ncov}. 

\section{Derivation of the backward equation}\label{app:backward}
We derive the backward equation~\eqref{eq:general.backward} by splitting the remaining observable at $t+\Delta t$ as $J_\tau-J_t=(J_{t+\Delta t}-J_t)+(J_\tau-J_{t+\Delta t})$. Conditioning the second term on $\Fc_{t+\Delta t}$ and using the Markov property, we obtain
\begin{align}
	G_t(z)&=\E[J_{t+\Delta t}-J_t\cond z_t=z]\notag\\
	&\quad+\E[G_{t+\Delta t}(z_{t+\Delta t})\cond z_t=z].
\label{eq:app.split}
\end{align}
By Eq.~\eqref{eq:general.additive}, the first term equals $\E[\int_t^{t+\Delta t}\Phi_s(z_s)\dd{s}\cond z_t=z]=\Phi_t(z)\Delta t+o(\Delta t)$, because the increment of the martingale $K$ has zero conditional mean. For the second term, Dynkin's formula gives $\E[G_{t+\Delta t}(z_{t+\Delta t})\cond z_t=z]=G_t(z)+\Delta t[(\partial_t+\Lg_t^\dagger)G_t](z)+o(\Delta t)$. Inserting both expansions into Eq.~\eqref{eq:app.split}, dividing by $\Delta t$, and letting $\Delta t\to0^+$ yields Eq.~\eqref{eq:general.backward}; the terminal condition $G_\tau=0$ holds by definition. 

\section{Representation of admissible perturbations}\label{app:riesz}
We prove the representation theorem of Sec.~\ref{sec:general.response} in the direction not shown there: an admissible perturbation on $\Cclass$ is represented by a unique martingale $S$ as in Eq.~\eqref{eq:general.Sdef}. The proof uses only the square integrability of the observables, the linearity of the response in the observable, and elementary Hilbert-space geometry~\cite{Reed.1972}, which we include for completeness; the Markov property is not needed.

Square-integrable functionals of the trajectory $z_{[0,\tau]}$, i.e., random variables measurable with respect to $\Fc_\tau$, form a Hilbert space with the inner product $\E[\xi\eta]$ (i.e., the average over trajectories of the reference process started from $p_0$) and the norm $\|\xi\|=\E[\xi^2]^{1/2}$. The subspace $\Hdyn=\{\xi:\E[\xi\cond z_0]=0\}$ is closed, and $\Hdyn_{\Cclass}$, the closure of the linear subspace $\{\xi_J:J\in\Cclass\}$, is again a Hilbert space.

Let the perturbation be admissible. We first show that $\Rc_J$ depends on $J$ only through $\xi_J$. Because both $\Rc_J$ and $\xi_J$ are linear in $J$, this is equivalent to $\Rc_J=0$ for every $J\in\Cclass$ with $\xi_J=0$, i.e., for every observable whose value $J_\tau=G_0(z_0)$ is fixed by the initial state along almost every trajectory. Such an observable has $\Vd=\|\xi_J\|^2=0$, and Eq.~\eqref{eq:general.admissible} gives $\Rc_J=0$. The relation $\ell(\xi_J)\equiv\Rc_J$ therefore defines a linear functional on $\{\xi_J:J\in\Cclass\}$ with $|\ell(\xi_J)|\le C\|\xi_J\|$. By continuity, it extends uniquely to $\Hdyn_{\Cclass}$ with the same bound.

We show that there is a unique $\sigma\in\Hdyn_{\Cclass}$ such that
\begin{equation}
	\ell(\xi)=\E[\xi\sigma]\quad\text{for all }\xi\in\Hdyn_{\Cclass}.
\label{eq:app.riesz}
\end{equation}
If $\ell=0$, we set $\sigma=0$. Otherwise, the kernel $\mathscr{K}=\{\xi\in\Hdyn_{\Cclass}:\ell(\xi)=0\}$ is a closed proper subspace, because $\ell$ is continuous and nonzero. Take $\eta\notin\mathscr{K}$ and let $k_*\in\mathscr{K}$ be the element closest to it, $\|\eta-k_*\|=\min_{k\in\mathscr{K}}\|\eta-k\|$, which exists because $\mathscr{K}$ is closed in a Hilbert space. Then, $\eta-k_*$ is orthogonal to $\mathscr{K}$: for any $k\in\mathscr{K}$ and real $\lambda$, $\|\eta-k_*-\lambda k\|^2\ge\|\eta-k_*\|^2$, i.e., $-2\lambda\E[(\eta-k_*)k]+\lambda^2\|k\|^2\ge0$, which can hold for all $\lambda$ only if $\E[(\eta-k_*)k]=0$. The unit vector $e\equiv(\eta-k_*)/\|\eta-k_*\|$ is thus orthogonal to $\mathscr{K}$, and $\ell(e)\ne0$, since otherwise $e\in\mathscr{K}$ would be orthogonal to itself. For any $\xi\in\Hdyn_{\Cclass}$, the combination $\xi-[\ell(\xi)/\ell(e)]e$ lies in $\mathscr{K}$ and is therefore orthogonal to $e$, which gives $\E[\xi e]=\ell(\xi)/\ell(e)$. Equation~\eqref{eq:app.riesz} follows with $\sigma=\ell(e)\,e$. If $\sigma'$ also satisfies it, then $\E[\xi(\sigma-\sigma')]=0$ for all $\xi\in\Hdyn_{\Cclass}$, and the choice $\xi=\sigma-\sigma'$ gives $\sigma'=\sigma$.

We define $S_t\equiv\E[\sigma\cond\Fc_t]$. By the tower property, $S_t$ is a martingale with $S_\tau=\sigma$ and $S_0=\E[\sigma\cond z_0]=0$. For $J\in\Cclass$, Eq.~\eqref{eq:app.riesz} and the polarized isometry~\eqref{eq:general.isometry} give
\begin{equation}
	\Rc_J=\E[(M_\tau-M_0)S_\tau]=\int_0^\tau\E[\dd{M_t}\dd{S_t}],
\label{eq:app.Sdef}
\end{equation}
which is Eq.~\eqref{eq:general.Sdef}. The martingale is unique: another martingale $S'$ with $S'_0=0$ and $S'_\tau\in\Hdyn_{\Cclass}$ that satisfies Eq.~\eqref{eq:general.Sdef} represents the same functional on $\{\xi_J:J\in\Cclass\}$, hence on its closure by continuity, so that $S'_\tau=\sigma$ by the uniqueness of $\sigma$. Since a martingale is determined by its terminal value, $S'_t=\E[S'_\tau\cond\Fc_t]=\E[\sigma\cond\Fc_t]=S_t$ for all $t\in[0,\tau]$.

\section{Martingale representation of generator perturbations}\label{app:cov.gen}
We derive Eq.~\eqref{eq:general.tilt} for a perturbation that leaves the path measure absolutely continuous. Let $P^\eps$ be the path measure of the perturbed dynamics with the same initial density, and assume that it is absolutely continuous with respect to $P=P^0$, i.e., every set of trajectories that has zero probability under $P$ also has zero probability under $P^\eps$; the likelihood ratio $\dd{P^\eps}/\dd{P}$ then exists. Its restriction to the information up to time $t$ defines the density process $\mathcal{Z}_t^\eps=\E[\dd{P^\eps}/\dd{P}\cond\Fc_t]$. Since the initial densities coincide, $\mathcal{Z}_0^\eps=1$, and Eq.~\eqref{eq:general.score} with $q_0=0$ gives $\partial_\eps\mathcal{Z}_t^\eps|_{\eps=0}=S^{\mathrm{dyn}}_t$. For $f$ in a common domain of the generators, $N^f$ of Eq.~\eqref{eq:general.Nf} is a $P$-martingale, whereas its counterpart with the perturbed generator,
\begin{equation}
	N^{f,\eps}_t=f_t(Z_t)-f_0(Z_0)-\int_0^t(\partial_s+\hLg_s^{\eps\dagger})f_s(Z_s)\dd{s},
\end{equation}
is a $P^\eps$-martingale. The two processes are the same function of the path up to the subtracted drift, $N^f_t-N^{f,\eps}_t=\int_0^t[(\hLg_s^{\eps\dagger}-\hLg_s^\dagger)f_s](Z_s)\dd{s}$. Under the change of measure and by Girsanov's theorem for semimartingales~\cite{Jacod.2003},
\begin{equation}
	N^f_t-\int_0^t\frac{\E[\dd{N^f_s}\dd{\mathcal{Z}^\eps_s}\cond\Fc_{s^-}]}{\mathcal{Z}^\eps_{s^-}}
\end{equation}
is a $P^\eps$-martingale. Subtracting $N^{f,\eps}$ gives the $P^\eps$-martingale
\begin{equation}
	D_t\equiv\int_0^t[(\hLg_s^{\eps\dagger}-\hLg_s^\dagger)f_s](Z_s)\dd{s}-\int_0^t\frac{\E[\dd{N^f_s}\dd{\mathcal{Z}^\eps_s}\cond\Fc_{s^-}]}{\mathcal{Z}^\eps_{s^-}}.
\end{equation}
Both terms of $D_t$ are time integrals, so that $D_t$ is a continuous martingale of finite variation with $D_0=0$; such a martingale vanishes identically~\cite{Jacod.2003}. Hence, the two drifts coincide at every time,
\begin{equation}
	[(\hLg_t^{\eps\dagger}-\hLg_t^\dagger)f_t](Z_{t^-})\dd{t}=\frac{\E[\dd{N^f_t}\dd{\mathcal{Z}^\eps_t}\cond\Fc_{t^-}]}{\mathcal{Z}^\eps_{t^-}}.
\end{equation}
Differentiating at $\eps=0$, where $\mathcal{Z}^0=1$ and $\partial_\eps\dd{\mathcal{Z}^\eps_t}|_{\eps=0}=\dd{S^{\mathrm{dyn}}_t}$, gives $[\delta\hLg_t^\dagger f_t](Z_{t^-})\dd{t}=\E[\dd{N^f_t}\dd{S^{\mathrm{dyn}}_t}\cond\Fc_{t^-}]$. The choice $f=H$, for which $N^H=M-M_0$, yields Eq.~\eqref{eq:general.tilt}.

\section{Interpretation of the susceptibility}\label{app:susceptibility}
We provide a physical interpretation of $\chi_{\bg}(\omega)$: it is the Fourier transform of the response of the mean observable $\ev*{\dot{J}_t}\equiv\dd{\E[J_t]}/\dd{t}$ to an impulsive source. We consider time-independent observable fields in the steady state $p_{\st}$ and a perturbation of strength $\eps$ that acts only at $t=0$, with the source current
\begin{equation}
	\delta\bj_{t,\mathrm{src}}=\delta_+(t)\bg p_{\st},
\end{equation}
where $\delta_+(t)$ is a causal Dirac impulse, normalized as $\int_0^\infty\delta_+(t)\dd{t}=1$. Using $\Lg p_{\st}=0$, the first variation of the density, $p_t^\eps=p_{\st}+\eps\delta p_t+O(\eps^2)$, obeys
\begin{equation}
	\partial_t\delta p_t=\Lg \delta p_t-\delta_+(t)\nabla_{\bq}\cdot(\bg p_{\st}).
\label{eq:freq.linearized.FP}
\end{equation}
Since $\delta p_{0^-}=0$, the impulse creates the deviation $\delta p_{0^+}=-\nabla_{\bq}\cdot(\bg p_{\st})$, which then relaxes under the unperturbed dynamics,
\begin{equation}
	\delta p_t=-\Theta(t)\,e^{t\Lg}\nabla_{\bq}\cdot(\bg p_{\st}),
\label{eq:freq.density.response}
\end{equation}
where $\Theta(t)$ is the Heaviside step function.

Since $\delta\Phi_t=\bw\cdot\delta\bb_t+\delta\D_t:\nabla_{\bq}\bw$, we obtain
\begin{align}
	\delta\ev{\dot{J}_t}&=\int\bw\cdot\delta\bj_{t,\mathrm{src}}\dd{\bz}+\int\Phi\delta p_t\dd{\bz}\notag\\
	&=\delta_+(t)\ev{\bw\cdot\bg}_{\st}+\Theta(t)\ev{\bg\cdot\nabla_{\bq}e^{t\Lg^\dagger}\Phi}_{\st}.
\label{eq:freq.rate.response.Phi}
\end{align}
The first equality follows from an integration by parts of $\delta\D_t:\nabla_{\bq}\bw$ and Eq.~\eqref{eq:general.source}. The second equality uses Eq.~\eqref{eq:freq.density.response}, the adjointness $\int h e^{t\Lg}\rho\dd{\bz}=\int(e^{t\Lg^\dagger}h)\rho\dd{\bz}$, and an integration by parts. Since the gradient annihilates constants, $\Phi$ may be replaced by $\Phi-\jr_{\st}$, and we obtain $\delta\ev*{\dot{J}_t}=\mathcal{K}_{\bg}(t)$ with the impulse response
\begin{equation}
	\mathcal{K}_{\bg}(t)\equiv\delta_+(t)\ev{\bw\cdot\bg}_{\st}+\Theta(t)\ev{\bg\cdot\nabla_{\bq}e^{t\Lg^\dagger}(\Phi-\jr_{\st})}_{\st}.
\label{eq:freq.causal.kernel}
\end{equation}
The one-sided Fourier transform of the impulse response is the susceptibility. The first term contributes $\ev{\bw\cdot\bg}_{\st}$ by the normalization of $\delta_+$. In the second term, the mixing condition allows the time integral to be taken inside the average, and the definition of $\Psi_\omega$ gives
\begin{equation}
	\int_0^\infty e^{-i\omega t}\ev{\bg\cdot\nabla_{\bq}e^{t\Lg^\dagger}(\Phi-\jr_{\st})}_{\st}\dd{t}=\ev{\bg\cdot\nabla_{\bq}\Psi_\omega}_{\st}.
\end{equation}
Since $\bu_\omega=\bw+\nabla_{\bq}\Psi_\omega$, we obtain
\begin{equation}
	\int_0^\infty e^{-i\omega t}\mathcal{K}_{\bg}(t)\dd{t}=\ev{\bg\cdot\bu_\omega}_{\st}=\chi_{\bg}(\omega).
\label{eq:freq.susceptibility.overlap}
\end{equation}
By linearity and the time-translation invariance of the steady state, a source current $\delta\bj_{t,\mathrm{src}}=f(t)\bg p_{\st}$ with an arbitrary time profile produces $\delta\ev*{\dot{J}_t}=\int_0^\infty\mathcal{K}_{\bg}(s)f(t-s)\dd{s}$.

\end{document}